\documentclass[iop,apj,tighten]{emulateapj}
\usepackage[breaklinks,colorlinks,urlcolor=blue,citecolor=blue,linkcolor=blue]{hyperref}
\usepackage{cleveref}
\usepackage{enumitem}
\usepackage{graphicx}
\usepackage{bm}
\usepackage{amsmath,amssymb}
\usepackage{color}
\usepackage[toc,page]{appendix}

\newcommand{\Msun}{{\rm M}_\odot}
\newcommand{\Lsun}{{\rm L}_\odot}

\shorttitle{Binary evolution in Active Galactic Nuclei}
\shortauthors{Tagawa et al.}

\renewcommand{\labelenumi}{\theenumi}

\begin{document}
\title{Formation and Evolution of Compact Object Binaries in AGN Disks}

\author{Hiromichi Tagawa\altaffilmark{1}, Zolt{\'a}n Haiman\altaffilmark{2}, and Bence Kocsis\altaffilmark{1}}
\affil{\altaffilmark{1}Institute of Physics, E{\"o}tv{\"o}s University, P{\'a}zm{\'a}ny P.s., Budapest, 1117, Hungary\\
\altaffilmark{2}Department of Astronomy, Columbia University, 550 W. 120th St., New York, NY, 10027, USA
}
\email{E-mail: htagawa@caesar.elte.hu}

\begin{abstract} 
The astrophysical origin of gravitational wave (GW) events discovered
by LIGO/VIRGO remains an outstanding puzzle. In active galactic nuclei
(AGN), compact-object binaries form, evolve, and interact with a
dense star cluster and a gas disk. 
An important question is whether and how binaries 
merge in these environments. To
address this question, 
we have performed 
one-dimensional $N$-body simulations combined with a semi-analytical model 
which includes the formation, disruption,
and evolution of binaries self-consistently. 
We point out that binaries can form in single-single interactions by the dissipation of
kinetic energy in a gaseous medium. This ``gas capture'' binary
formation channel contributes up to $97\,\%$ of gas-driven mergers and
leads to a high merger rate in AGN disks even without pre-existing
binaries. 
We find the merger rate to be in the range $\sim 0.02-60\,\mathrm{Gpc^{-3}yr^{-1}}$. 
The results are insensitive to the assumptions on gaseous hardening
processes: 
we find that once they are formed, binaries merge efficiently 
via binary-single
interactions even if these gaseous processes are neglected.
We find that the average number of mergers per BH is $0.4$, and the
probability for repeated mergers in 30 Myr is $\sim 0.21-0.45$.  
High BH masses due to repeated mergers, high eccentricities, and a significant
Doppler drift of GWs are promising signatures which distinguish this
merger channel from others.  Furthermore, we find that gas-capture
binaries reproduce the distribution of LMXBs in
the Galactic center, including an outer cutoff at $\sim1$ pc due
to the competition between migration and hardening by gas torques.
\end{abstract}

\section{Introduction}

Recent detections of gravitational waves (GWs) have shown evidence for
a high rate of black hole (BH)-BH and neutron star (NS)-NS mergers in
the Universe
\citep{Abbott16a,Abbott16b,Abbott17,TheLIGO17a,TheLIGO17b,TheLIGO18,Zackay19,Venumadhav19,Zackay19b}. However,
the proposed astrophysical pathways to mergers remain highly
debated. Possible compact object merger pathways include isolated
binary evolution
\citep{Dominik12,Kinugawa14,Belczynski16,Breivik16,Belczynski17,Giacobbo18,Bavera19,Spera19}
accompanied by mass transfer
\citep[][]{Pavlovskii17,Inayoshi17,vandenHeuvel17}, common envelope
ejection \citep[e.g.][]{Paczynski76,Ivanova13}, envelope expansion
\citep{Tagawa18}, 
or chemically homogeneous evolution in a tidally
distorted binary \citep{deMink16,Mandel16,Marchant16}, evolution of
triple or quadruple systems
\citep{Silsbee17,Antonini17,Liu17,Liu18,Hoang18,Randall18,ArcaSedda18,Liu18b,Michaely19,Fragione19,Fragione19b},
gravitational capture
\citep{OLeary09,KocsisLevin12,Gondan18a,Rodriguez18b,Zevin19,Rasskazov19,Samsing19},
dynamical evolution in open clusters
\citep{Banerjee17,Banerjee18a,Banerjee18b,Kumamoto18,Rastello18,Bouffanais19}
and dense star clusters
\citep[e.g.][]{PortegiesZwart00,OLeary06,Samsing14,Ziosi14,OLeary16,Rodriguez16,Rodriguez16b,Mapelli16b,Askar17,Fujii17,Zevin19,Zhang19,DiCarlo19},
and dynamical interaction in gas-rich nuclear regions
\citep{McKernan12,McKernan14,Bellovary16,Bartos17,Stone17,McKernan17,Tagawa18b,Leigh18,Yi18,Secunda19,Yang19a,Yang19b,McKernan19,Gayathri19,Tagawa20}.

Twelve low-mass X-ray binary (LMXB) candidates have also recently been
discovered by \citet{Hailey18} within a distance $r\lesssim\,1$~pc of
the Galactic center, with a density distribution of $\propto
r^{-1.5\pm0.3}$, by observing X-ray sources within $\sim4$~pc
presented by \citet{Muno09}. \citet{Generozov18} proposed that the
density profile of these hard binaries can be explained by the tidal
capture mechanism and stellar relaxation processes. Although this model
predicts the radial distribution of LMXBs to be $\propto
r^{-(0.9-1.4)}$, the outer cut-off at $\sim1$ pc remains
unexplained.

\begin{figure}
\begin{center}
  \includegraphics[width=90mm]{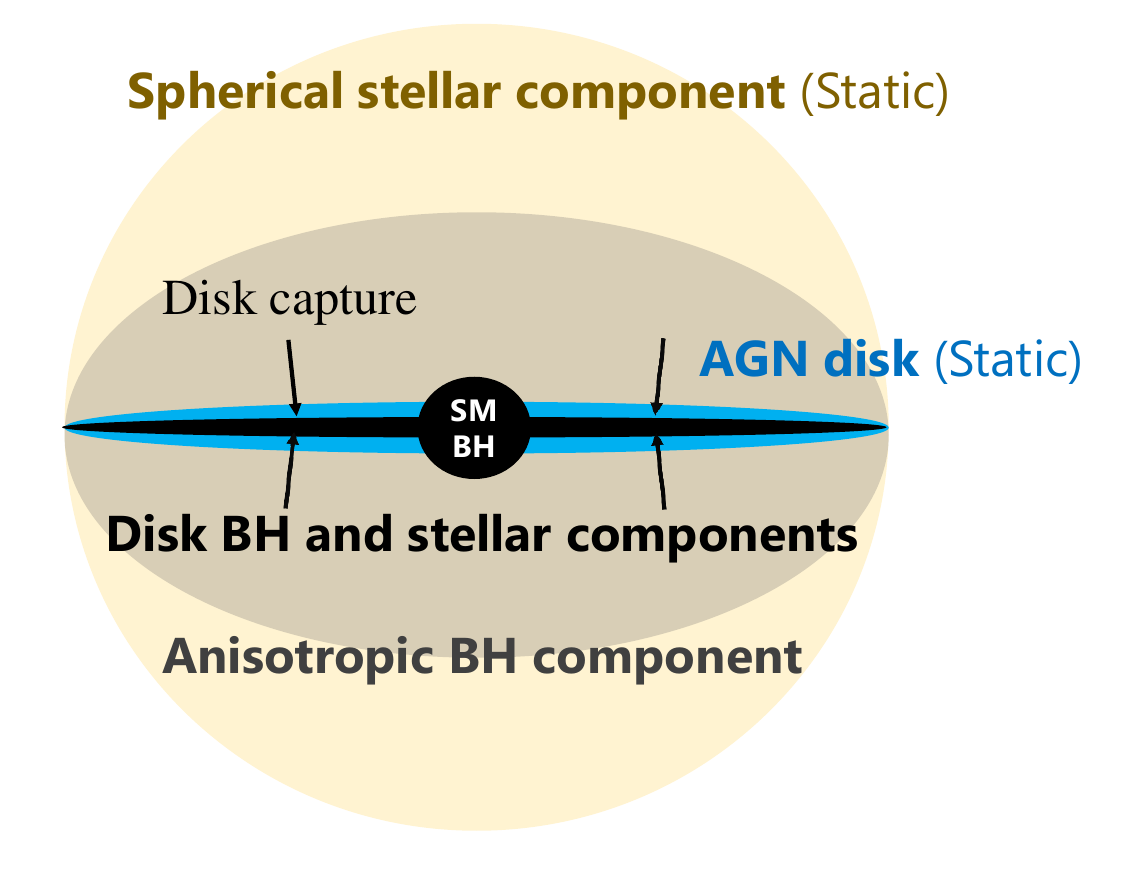}
    \vspace{\baselineskip}
    \caption{Components describing an active galactic nucleus, including
  (1) a central SMBH, (2) a gaseous AGN accretion disk, (3) a
  spherical stellar cluster, (4) an anisotropic (flattened) cluster of
  BHs, and (5) BHs and stars in the AGN disk. We perform numerical
  simulations to follow the evolution of BHs during this AGN phase,
  including the formation and orbital evolution of binaries.  }
\label{fig:components}
\end{center}
\end{figure}

Galactic nuclei are the densest environments of stars and compact
objects in the Universe
\citep{Walcher05,Merritt10,Norris+2014,GallegoCano18}. In the active
phase of a galactic nucleus, a high-density gas disk forms within
0.1--10 pc \citep[][]{Burtscher13} of a central supermassive BH
(SMBH).  In such environments, binaries form and evolve through
interaction with densely populated stars and gas. \citet{Baruteau11}
showed that even when a binary is so massive that it opens a gap
within the accretion disk around a SMBH, it is efficiently hardened via
gas dynamical friction. 
\citet{McKernan12,McKernan14} predicted the formation of intermediate-mass BHs (IMBH) in active galactic nucleus (AGN) disks due to collisions of compact objects. 
\citet{Bartos17} have proposed a pathway for BH-BH mergers in AGN disks, in which binaries are captured in an
accretion disk within $\sim0.01$~pc of the central SMBH due to linear
momentum exchange, and after that, binaries are hardened by gas
dynamical friction of the AGN disk and by type I/II torques of
a circumbinary disk.
\citet{Stone17} proposed another pathway, in which in-situ formed
binaries at $\sim$pc scale evolve via the effects of binary-single
interactions with a stellar disk and via type I/II torques from a
circumbinary disk.
\citet{Leigh18} showed that fewer than ten binary-single interactions
are sufficient to drive hard binaries with a binary separation of
$s\lesssim10$ AU to merger.
\citet{McKernan17,McKernan19} estimated the mass and spin
distributions of the merged BHs, and the merger rate in AGN disks.
\citet{Bellovary16} suggested that BHs accumulate and merge with each
other in migration traps at $20-300$ Schwarzschild radii, where the
torque by the AGN disk changes sign. \citet{Secunda19} and
\citet{Yang19a} modeled the formation of binaries within the migration
traps.
\citet{Just12}, \citet{Kennedy16} and \citet{Panamarev18} discussed the
capture of stars in an AGN disk and their subsequent migration toward the
central SMBH due to the ram pressure of an AGN disk.
\citet{Tagawa20} investigated the distribution of the effective spin parameter for mergers in AGN disks.

Previous studies of compact object mergers in AGN disks have focused
on the role of gas in driving binary mergers assuming pre-existing
binaries in the nuclear star cluster~\citep{Bartos17} or in the disk
itself~\citep{Stone17,McKernan17}, or assuming that binaries form at
migration traps \citep{Bellovary16,Secunda19,Yang19a,Yang19b,Gayathri19}.
In the present study, we examine the formation of binaries during close
two-body encounters in a gaseous medium, where the gas absorbs some of
the initial kinetic energy of the two objects.
\citet{Goldreich02}
have proposed that planetesimal binaries can form in the
Kuiper belt due to the dissipation of the relative velocity between
the planetesimals by dynamical friction in the environment of a
background population of smaller solid bodies. They showed that this
leads to efficient binary formation in the Kuiper belt,
but to our knowledge, the analogous mechanism of binary formation
in an AGN disk has not been previously explored.
Here we include this
``gas capture'' binary formation mechanism, and find that it supplies the majority of
binaries in AGN disks.  We also examine binary formation in dynamical three-body
encounters \citep[e.g.][]{Aarseth76,Binney08}, which have also been
neglected in previous studies of mergers in AGN disks,
but find this mechanism to be less important. 
These mechanisms enable binary formation in AGN disks without migration traps whose existence and properties are poorly understood. 

More generally, in this paper we investigate whether and how binaries
form and merge in AGN disks. 
We combine one-dimensional $N$-body simulations with a semi-analytical model, 
which incorporates the effects of gas dynamical
friction, type I/II migration torques, GW radiation, and several
different types of stellar interaction. We simulate the evolution of
both single and binary objects, and follow their radial position from
the central SMBH, as well as their velocities in time, together with
the evolution of the binaries' separation. 
The other two spatial directions are followed only statistically. 
Our flexible model allows
us to test previous assumptions on whether and how efficiently
binaries may merge in AGN disks \citep[e.g.][and references
  above]{McKernan19}, and to examine the dependence of the merger rate
on the model parameters of the AGN disk and the stellar cluster.

The rest of this paper is organized as follows. In \S~2, we give an
executive summary of our method, and describe the numerical scheme and
the setup of simulations in detail in \S~3. We present our main
results in \S~4. We discuss the implications for the spin and
eccentricity distribution, the merger rate, and a comparison with
LMXBs observed in the Galactic nucleus in \S~5. We summarize our
conclusions in \S~6. For clarity, the variables used
in this paper are listed and defined in Table~\ref{table_notation}.

\begin{figure}
\begin{center}
\includegraphics[width=85mm]{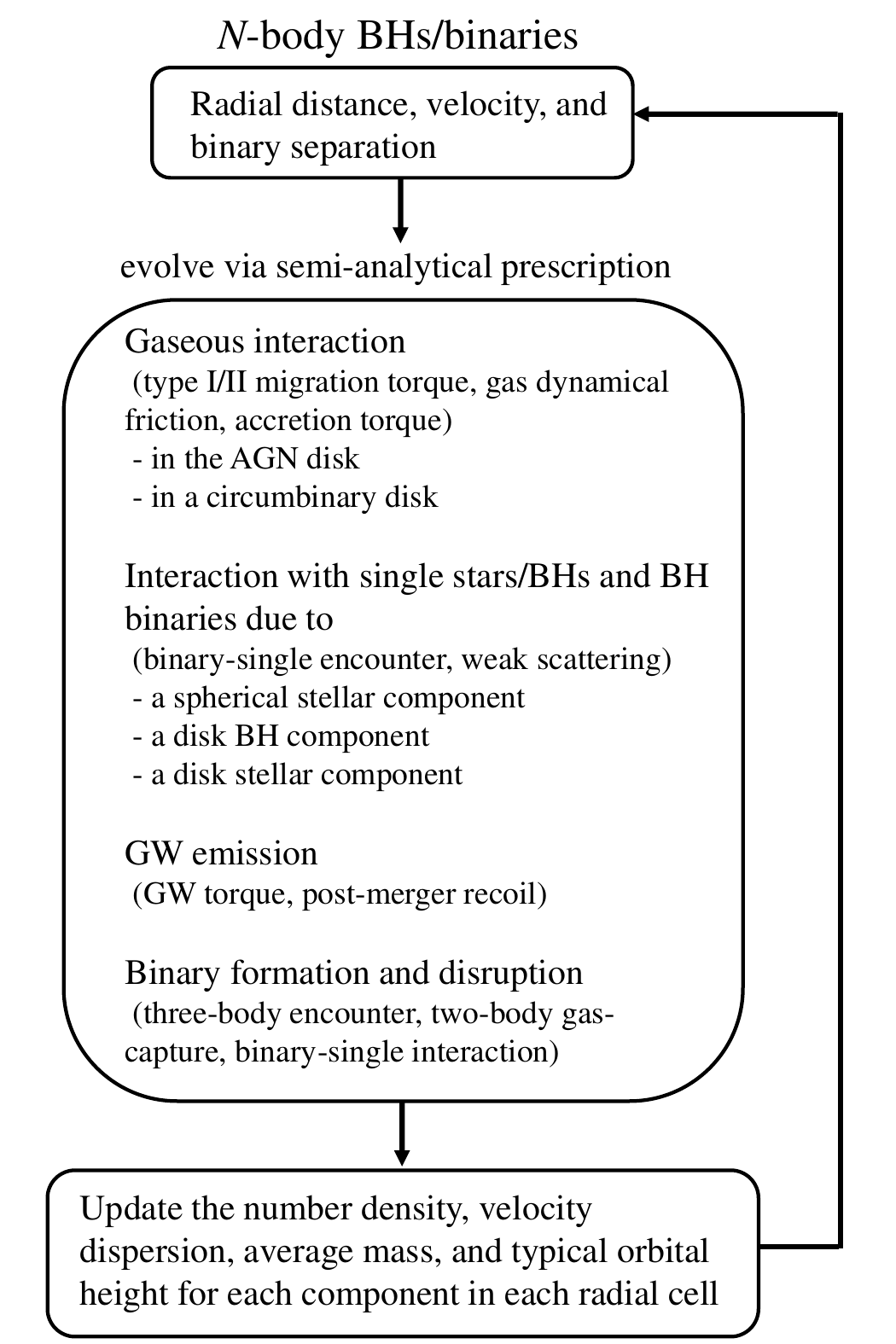}
\vspace{\baselineskip}
\caption{
  Schematic diagram for following the evolution of the BH population
  in our model, including both single and binary BHs.  The $N$-body
  simulation keeps track of these individual objects (starting with
  $2\times 10^4$ BHs and $1.5\times 10^3$ binaries in our fiducial
  model). Single BHs are characterized by their radial position
  ($r_i$) from the central SMBH and by their velocity (${\bf
    v}_i$) relative to the local Keplerian value.  Binaries are
  similarly characterized by their center-of-mass position ($r_j$) and
  velocity (${\bf v}_j$), and additionally by their orbital separation
  ($s_j$).  These variables are updated via semi-analytic
  prescriptions in each ``N-body'' 
  time-step, due to multiple
  processes as listed in the diagram.}
\label{fig:method_diagram}
\end{center}
\end{figure}

\section{Overview of simulations}
\label{sec:overview}

We consider a system describing a galactic nucleus, consisting of the
following five components: (1) a central SMBH, (2) a gaseous accretion
disk (``AGN disk''), (3) a spherical stellar cluster, (4) a flattened
cluster of BHs, and (5) stars and BHs inside the AGN disk, referred to
as the ``disk stellar'' and ``disk BH'' components
($\S$\ref{sec:component}).
Figure~\ref{fig:components} illustrates our setup.

In our fiducial model, we adopt the SMBH mass and the distribution of
stars from observations of the quiescent central region of the
Milky Way at present 
(\S~\ref{sec:bh_stellar_distribution}), which does not have an AGN. 
Also we generate the BH mass distribution using the results of population synthesis models and accounting for an initially mass segregated radial distribution (\S~\ref{sec:bh_stellar_distribution}). 
On the other hand, our model with an
AGN disk represents the conditions during the active phase which is
believed to have existed at an earlier time in the Milky Way's history
\citep[e.g.][]{Wardle08,Su10,Zubovas11,Bland-Hawthorn13}. 

We employ the AGN accretion disk model proposed by \citet{Thompson05}
($\S~$\ref{sec:disk_model}), as adopted in earlier work by
\citet{Stone17}.  This represents a
Shakura-Sunyaev $\alpha$-disk with a constant viscosity parameter
$\alpha$ and accretion rate in the region where it is not
self-gravitating.  The model describes a radiatively efficient, geometrically thin, and optically thick disk and
extends the disk to pc scales with a
constant Toomre parameter in the self-gravitating regime
(see Fig.~\ref{fig:disk_model} below), assuming that it is stabilized by
radiation pressure and supernovae from in-situ star formation. We
assume that stars and BHs form in the disk at the rate required to
stabilize the AGN disk, and some fraction of BHs are initially formed in binaries (\S~\ref{sec:disk_model}).

To follow the time-evolution of the BHs in this system, focusing on
their capture by the disk, and the formation and disruption of BH
binaries in the disk, we run 
one-dimensional $N$-body simulations combined with a semi-analytical method. 
Binaries form in the disk either due to gas
dynamical friction (\S~\ref{sec:gbs}) or due to three-body encounters
(\S~\ref{sec:3bf}), and are disrupted by soft binary-single
interactions. We assume that binaries are disrupted when the binary separation becomes larger than the Hill radius of a binary with respect to the SMBH. 

\begin{figure}
\begin{center}
  \includegraphics[width=90mm]{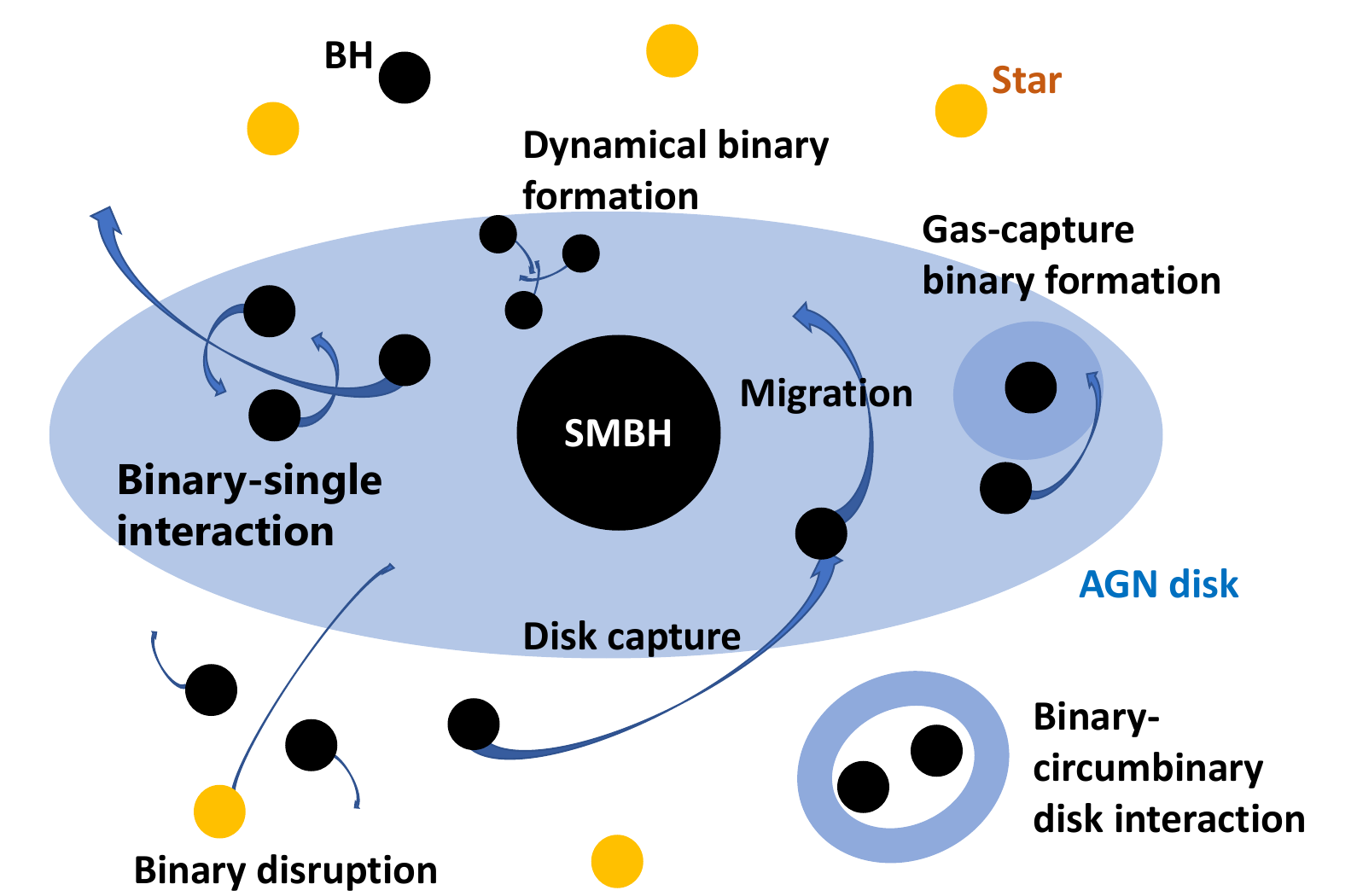}
  \vspace{\baselineskip}
\caption{Schematic diagram illustrating the mechanisms
  affecting the BH population and driving binary formation and
  evolution. See
  \S~\ref{sec:overview} and Fig.~\ref{fig:method_diagram} for an overview and
  \S~\ref{sec:method} for numerical details. }
\label{fig:mechanisms}
\end{center}
\end{figure}

We model the evolution of the orbital separation ($s_j$), the radial
position ($r_j$), and the magnitude of a random velocity relative to
the local Keplerian AGN disk motion ($v_j$) for all binaries labelled
with the binary index $j$. We also track the radial position ($r_i$)
and random velocity ($v_i$) of single compact objects, which represent
stellar-mass BHs.
These quantities evolve according to analytical
formulas as summarized in Figure~\ref{fig:method_diagram} and
illustrated visually in Figure~\ref{fig:mechanisms}. 

Our model includes physical processes due both to the presence
of gas and to multi-body dynamical interactions, as follows.  For the
interaction with gas, the radial positions of all BHs evolve in
response to (i) type I/II migration torques by the AGN disk
($\S$\ref{sec:typeii}), and the velocities of all BHs relative to the local AGN disk decrease due to  (ii) the accretion torque ($\S$\ref{sec:acc}), and
(iii) gas dynamical friction in the AGN disk ($\S$\ref{sec:df}). For
binaries of stellar-mass BHs, the separation evolves due to gas dynamical
friction by the AGN disk and due to type I/II migration torque by a
small circumbinary disk that forms within the Hill sphere of the binary.
We also account for dynamical interactions with single stars and BHs and BH binaries:
the binaries' separations and velocities evolve due to binary-single
interactions ($\S$\ref{sec:bs}), and the
velocities of all BHs additionally evolve due to scattering
($\S$\ref{sec:sss}). 
For dynamical interactions, we neglect the BHs
in the spherical cluster, since they are greatly outnumbered by stars
in the cluster, but we include both stars and BHs in the disk
component.  We also account for GW emission, which reduces the binary
separation rapidly once the binary is sufficiently tight.
For simplicity, eccentricity evolution is ignored and orbits around
the SMBH and binary orbits are both assumed to be circular. 
%{\bf We expect that the assumption of the zero orbital eccentricity less affects results as we consider the anisotropic BH distribution. 
%}

To model the orbits of merged BHs, a recoil velocity is added to the BH remnant
due to anisotropic GW radiation ($\S$\ref{sec:merger_pres}). 
The small mass loss during mergers due to GW radiation is taken into account assuming zero BH spins.

In this study, we ignored several processes for simplicity. 
These include 
exchange of binary components at binary-single interaction, 
the formation and evolution of stellar binaries, 
radial migration of stars due to torque of the AGN disk, 
evolution of compact objects other than BHs, 
the Kozai-Lidov effect of the SMBH or a third stellar-mass object on binaries, 
dynamical relaxation processes, 
counter rotating BHs or stars in the AGN disk \citep{Ivanov15,SanchezSalcedo18}, 
stellar evolution, supernova feedback, binary mass transfer, and 
possible presence of massive perturbers like an SMBH companion and/or IMBHs. A few IMBHs, if present, may efficiently disrupt most BH binaries which may greatly reduce the merger rates \citep{Deme2019}. 

The above model allows us to describe the time-evolution of the binary
BH population in a self-consistent and flexible way.  It extends the
simplified prescriptions of previous studies of stellar-mass BH binary
mergers in AGN disks, and creates a self-consistent one-dimensional $N$-body 
simulation that includes the time-dependent formation,
disruption, and evolution of binaries in AGN.  We use this method to
estimate the contribution of binaries formed during AGN phase. We
confirm previous suggestions~\citep{Secunda19,Yang19b} that repeated
mergers are frequent in AGN disks, although in difference from these
previous works, in our models repeated mergers occur due to efficient
binary formation and evolution processes well outside the ``migration
traps''.

\section{Method}
\label{sec:method}

\begin{table*}
	\caption{
		Notation. 
        	}
\label{table_notation}
\hspace{-0.0mm}
\begin{tabular}{p{2.5cm}|p{5cm}||p{2.5cm}|p{5cm}}
\hline
Symbol&Description&Symbol&Description\\\hline
$i$, $j$, $k$& The index of a single BH, a BH binary, and either of a single BH or a BH binary & 
$v_\mathrm{Kep}(r)$ & The Keplerian velocity at the distance $r$ from the SMBH \\\hline

$l$& The index of a cell, in which physical quantities are stored & 
$M_\mathrm{SMBH}$, ${\bar m}_\mathrm{star}$& The SMBH mass and the average stellar mass\\\hline

$xy$, $z$& The direction of the plane and the angular momentum of the AGN disk & 
$\rho_\mathrm{gas}$, $n_\mathrm{gas}$& The mass and number density of gas \\\hline

$t$, $\Delta t$ & The elapsed time and the time step in simulations & 
$\Sigma_{\mathrm{disk}}$ & The surface density of a gas disk \\\hline

$r$, $r_k$, $r_l$,$\Delta r_l$& The distance, the distance of the $k^{\rm th}$ object, the distance of the geometric center of a cell $l$ from the SMBH, and the width of the $l$th cell &
$h_{\mathrm{disk}}$ & The scale height of a gas disk \\\hline

$s_j$, $v_j$, $v_{\mathrm{bin},j}$& The binary separation, the velocity of the center of mass of binary relative to the local AGN motion, and the relative rotation velocity of binary components in the $j^{\rm th}$ BH binary&
$c_{\mathrm{s}}$& The sound velocity of gas \\\hline

$v_k$, $v_{xy,k}$, $v_{z,k}$& The magnitude of the velocity and the $xy$ and $z$-direction velocity of the center of mass for the $k^{\rm th}$ object relative to the local AGN motion &
$\Omega$& The angular velocity of a gas disk \\\hline

$m_k$, $m_{j_1}$, $m_{j_2}$& The mass of the $k^{\rm th}$ object, and the primary and secondary masses in the $j^{\rm th}$ binary, respectively &
$t_\mathrm{AGN}$& The typical lifetime of AGN disks \\\hline

$h_{k}$& The typical height of orbital motion for the $k^{\rm th}$ object, $h_{k}=v_{z,k}r_k/v_\mathrm{Kep}$ &
${\dot M}_\mathrm{out}$ & The gas accretion rate from the outer radius $r_\mathrm{out}$ \\\hline

$r_{\mathrm{Hill},k}$& The Hill radius of the $k^{\rm th}$ object with respect to the SMBH, $r_\mathrm{Hill}=r_k(m_k/3M_\mathrm{SMBH})^{1/3}$ &
${\dot M}_\mathrm{Edd}$ & The Eddington accretion rate \\\hline

$v_{\mathrm{rel},k}$&The typical relative velocity of a third body relative to the center of mass for the $k^{\rm th}$ object &
$\beta_\mathrm{v}$ & The parameter setting the initial velocity dispersion for BHs \\\hline

$b_{\mathrm{90},k}$& The impact parameter of the $k^{\rm th}$ object at which the direction of particles changes by 90 degree after an encounter, $b_{90,k}=G(m_k+m_\mathrm{c})/v_{\mathrm{rel},k}^2$ &
$\delta_\mathrm{IMF}$& A pawer law exponent for the initial mass function of stars \\\hline

$r_{\mathrm{BHL},k}$& The Bondi-Hoyle-Lyttleton radius for the $k^{\rm th}$ object, $r_{\mathrm{BHL},k}= m_k G/(v_k^2+c_s^2)^{3/2}$  &
$M_\mathrm{star,3pc}$ & The stellar mass within 3 pc from the SMBH\\\hline

${\dot m}_{\mathrm{BHL},k}$& The Bondi-Hoyle-Lyttleton accretion rate for the $k^{\rm th}$ object (Eq.~\ref{eq:macc})&
$\gamma_\rho$ & A power law exponent of the radial profile of the initial BH distribution (Eq.~\ref{eq:bh_density}) \\\hline

$p_{\mathrm{disk},k}$, $p_{\mathrm{c},k}$& The time fraction that the $k^{\rm th}$ object spends within the AGN disk and a background component, respectively, $p_{\mathrm{disk},k}=\frac{2}{\pi}\mathrm{asin}(h_{\mathrm{AGN},l}/h_{k}$), $p_{\mathrm{c},k}=\frac{2}{\pi}\mathrm{asin}(h_{\mathrm{c}}/h_{k}$)  
&$\Gamma_\mathrm{Edd,cir}$ & The Eddington accretion rate onto a BH binary from a circumbinary disk \\\hline

$\rho_\mathrm{c}$, $n_\mathrm{c}$& The mass and number density of a background component &
$m_\mathrm{AM}$ & The efficiency of angular momentum transport in outer regions of the AGN disk \\\hline

$\sigma_\mathrm{c}$& The one-dimensional velocity dispersion of a background component &
$v_\mathrm{GW}$ & The recoil velocity on a merger remnant due to anisotropic GW radiation \\\hline

$m_\mathrm{c}$& The average mass of a background component &
$f_\mathrm{pre}$ & The fraction of the number of pre-existing binaries over the initial BHs number  \\\hline

$h_{\mathrm{c}}$&The average orbital height of a background component & 
$\alpha_\mathrm{SS}$ & The $\alpha$ parameter which gives the efficiency of angular momentum transport in standard thin disks \\\hline

$\rho_{\mathrm{AGN},l}$& The mass density of the AGN disk at $r_l$ &
$\eta_t$& The time step parameter  \\\hline

$h_{\mathrm{AGN},l}$& The height of the AGN disk at $r_l$ &
$N_\mathrm{cell}$ & The number of the cells storing physical quantities  \\\hline

$N_\mathrm{totBH,ini}$, $N_{\mathrm{mer}}$, $N_\mathrm{bin}$, $N_\mathrm{acc}$, $N_\mathrm{mer,SF}$
&The number of initial BHs, mergers, binaries, migrator within $r_\mathrm{in}$, mergers among in-situ formed BHs &

$r_\mathrm{in,BH}, r_\mathrm{out,BH}$& The inner and outer boundaries for $r$ within which BHs are initially distributed   \\\hline

$f_\mathrm{mer,pre}$, $f_\mathrm{mer,gas}$, $f_\mathrm{mer,dyn}$, $f_\mathrm{mer,rep}$
&The fraction of mergers among pre-existing binaries, gas-capture binaries, dynamically formed binaries, and repeated mergers over total mergers &
$r_\mathrm{in}, r_\mathrm{out}$& The inner and outer boundaries for $r$ within which we calculate  \\\hline

$f_\mathrm{BH,mer}$& The number of merger over the initial number of BHs, $f_\mathrm{BH,mer}=N_\mathrm{mer}/N_\mathrm{ini,BH}$&
$f_\mathrm{BH}$, $f_\mathrm{BH,n}$& The fraction of the mass and the number of all stellar-mass BHs over the mass of all stars \\\hline

\end{tabular}
\end{table*}

Here we describe in detail the method and the initial conditions
adopted in this study. Table~\ref{table_notation} lists the definition of variables which appear in this paper.

\subsection{Stellar mass BHs, stars and AGN disk}

In this section we describe the initial condition in the calculations.

\label{sec:initial_stellar_distribution}

\subsubsection{Initial BH and stellar distribution}
\label{sec:bh_stellar_distribution}

We simulate the evolution of $N$-body particles representing
stellar-mass BHs. We assume these are initially distributed according
to 
\begin{align}
\label{eq:bh_density}
\frac{dN_{\rm BH, ini}(r)}{dr}
\propto r ^{\gamma_{\rho}}
\end{align}
where $N_{\rm BH, ini}(r)$ labels the total initial number of BHs within distance $r$ from the central SMBH, and $\gamma_{\rho}$ is
a power-law index.  Theoretically,
$\gamma_{\rho}$ is expected to be between $\sim -0.5$ and $0.25$ for
plausible mass functions for spherically symmetric systems 
\citep{Hopman06,Freitag06,OLeary09,Keshet09,AlexanderHopman2009}.  In our
fiducial model, we adopt $\gamma_{\rho} = 0$ between
$r_\mathrm{in,BH} \leq r \leq r_\mathrm{out,BH}$ where
$r_\mathrm{in,BH}=10^{-4}$~pc and $r_\mathrm{out,BH}=3$~pc. 

We set the total stellar mass within 3 pc to be
\begin{equation}\label{eq:Mstar3pc}
 M_\mathrm{star,3pc}=10^7\,\Msun   
\end{equation}
\citep{Feldmeier14} as the fiducial
value. The minimum and maximum mass for progenitor stars are assumed
to be 0.1 and $140\,\Msun$, respectively.  The BH mass is determined
through the relations between the progenitor mass
($m_{\mathrm{star},i}$) and the BH mass ($m_{\mathrm{BH},i}$) of
\begin{align}\label{eq:mBH}
m_{\mathrm{BH},i}
=
\left\{
\begin{array}{l}
\dfrac{m_{\mathrm{star},i}}{4}~~\mathrm{for}~20\,\Msun<m_{\mathrm{star},i}<40\,\Msun, \\
10\,\Msun~~~\mathrm{for}~40\,\Msun<m_{\mathrm{star},i}<55\,\Msun, \\
\dfrac{m_{\mathrm{star},i}}{13}+5.77\,\Msun \\ \qquad~~~~~~\mathrm{for}~55\,\Msun<m_{\mathrm{star},i}<120\,\Msun, \\
15\,\Msun ~~~~\mathrm{for}~120\,\Msun<m_{\mathrm{star},i}<140\,\Msun, 
\end{array}
\right.
\end{align}
which roughly matches population synthesis simulation results in
\citet{Belczynski10} for their model with solar metallicity and a weak
wind.

Since observational studies \citep{Bartko10,Lu13} suggest a top-heavy
initial mass function (IMF) for stars in the Galactic center region, we
investigate IMFs
\begin{equation}\label{eq:IMF}
 \frac{{\rm d}N}{{\rm d}m_\mathrm{star}} \propto m_\mathrm{star}^{-\delta_\mathrm{IMF}}\,,  
 \qquad 0.1\Msun \leq m_{\rm star} \leq 140\Msun\,,
\end{equation}
with $\delta_\mathrm{IMF}$ in the range $1.7-2.35$.  We set the
fiducial value to be $\delta_\mathrm{IMF}=2.35$, yielding
an average stellar mass ${\bar m}_\mathrm{star}=0.36\,\Msun$  and 
initial number of BHs $N_\mathrm{ini,BH}=2.0\times
10^4$.  For $1.7\leq\delta_\mathrm{IMF}\leq
2.35$, ${\bar m}_\mathrm{star}$ and $N_\mathrm{ini,BH}$ vary
between $0.36-1.78\,\Msun$ and $2.0\times 10^4-1.0\times
10^5$.

The simulation tracks the velocity of particles relative to the local
Keplerian AGN disk in the plane of the disk $v_{xy,k}$ and
perpendicular to it $v_{z,k}$ at the point where the orbit crosses the equatorial plane, where $k$ is the particle index.  The
direction of $v_{xy,k}$ is assumed to be axisymmetrically random in the $xy$ plane.
The $x-$, $y-$, and $z-$ components of the velocity of each BH
relative to the local disk are initially drawn randomly from a
Gaussian distribution with dispersion of $\beta_\mathrm{v}
v_\mathrm{Kep}(r)/\sqrt{3}$ and zero mean. Here
$v_\mathrm{Kep}(r)=\{G[M_{\rm SMBH}+ M_{\rm star}(<r)]/r\}^{1/2}$ is
the Keplerian orbital velocity at the distance $r$ from the central
SMBH, $M_{\rm star}(<r)$ is the stellar mass within $r$, and
$\beta_\mathrm{v}$ is a parameter that determines the initial velocity
dispersion of BHs.  In our fiducial model we set
$\beta_\mathrm{v}=0.2$.  Since this is with respect to the comoving Keplerian frame, it corresponds to a net rotation for the pre-existing BHs component, which is consistent with observational
\citep[e.g.][]{Trippe08,Yelda14,Feldmeier14,Feldmeier15} and
theoretical \citep{Kocsis11b,Szolgyen18} suggestions that massive
stars in the central region of $\lesssim 1\,\mathrm{pc}$ have some
degree of net rotation.

We assume that some fraction of BHs are initially in binaries as
follows.  Spectroscopic observations show that the binary fraction of
O stars in the Galactic field is $\sim0.7$ \citep{Sana12}, but the
binary fraction of OB/WR stars in the Galactic center is estimated to be only $\sim
0.3$ \citep{Pfuhl14}. We define the corresponding initial
``pre-existing'' BH binary fraction $f_\mathrm{pre}$ as the number of
BH binaries over the total number of (single+binary) BHs.
\citet{Belczynski04} found that if the binary fraction of progenitor
stars is $50\%$ (2/3 of stars are in binaries), the binary fraction of
BHs is $\sim 10\,\%$ as a result of stellar evolution due to supernova
kicks and mergers during the common envelope and Roche-lobe overflow
phases.  In the Galactic center, binary disruption due to soft
binary-single interactions may further decrease the BH binary
fraction \citep[see also][]{Stephan16}. We adopt $f_\mathrm{pre}=0.15$ in our fiducial model.

We draw the initial separation of pre-existing binaries randomly from
a log-flat distribution following \citet{Abt83}. The minimum
separation $R_\mathrm{min}$ has large uncertainties since
$R_\mathrm{min}$ is determined by common envelope evolution, which is
not well understood \citep{Ivanova13}. For the fiducial value, we set
$R_\mathrm{min}$ by the sum of the radii of the progenitor binary
components. We compute the stellar radius as \citep{Torres10}
\begin{equation}\label{eq:Rstar}
 R_{\mathrm{star},i}=R_\odot \left(\frac{m_{\mathrm{star},i}}{\Msun}\right)^{1/2}.   
\end{equation}
We set the maximum binary separation $R_\mathrm{max}$ to
$10^5\,R_\odot$ following binary evolution models
\citep[e.g.][]{Belczynski08,Kinugawa14}.  We further assume that
binaries that are soft compared to the local spherical stellar
component (\S \ref{sec:bs}) are promptly disrupted prior to the AGN
phase.  The timescale of the disruption of binaries is $\sim 300$ Myr
at $r_j\lesssim0.01$ pc and $\sim 20$ Myr at $r_j=3$ pc 
(Eq.~7.173 of \citealt{Binney08}), in which we use the velocity dispersion,
the density, and the average stellar mass for the spherical stellar
component in the fiducial setting of our simulations
(Eq.~\ref{eq:quantity_sph}), the Coulomb logarithm is assumed to be
10, each BH binary component is assumed to have a mass of $5\,\Msun$,
and the binary separation is the maximum of the hard-soft boundary
(Eq.~\ref{eq:shs} below) and the separation at which a binary merges
within the Hubble time (Eq.~\ref{eq:sgw} below).  Due to the
disruption of binaries prior to the AGN phase, the binary fraction
at the beginning of the simulation is reduced to $\sim
7\%$. Note that this value varies according to the initial mass
function, the total stellar mass, and the mass of the SMBH (see
\S~\ref{sec:dependence} below).

\subsubsection{Stellar and BH components}
\label{sec:component}

We categorize stars and BHs by whether they reside within or orbit outside of the AGN disk. 
These components are referred to as the disk stellar, the spherical stellar, the disk BH, and the anisotropic BH components (Figure~\ref{fig:components}).  We assume that stars are initially
spherically distributed, and the velocity of stars follow a
Maxwell-Boltzman distribution with no net rotation, while BHs are
initially distributed with some degree of net rotation (see
justification below).  Due to the interaction with the AGN disk and
star formation in the outer regions of the AGN disk, the number of BHs
and stars in the AGN disk gradually increase.  As a result, the BH and
stellar density within the AGN disk rise. Thus the BH and stellar
disk components form during the AGN phase
(Figure~\ref{fig:components}). 
Since the mass outside of the AGN disk is dominated by stars (i.e. the spherical stellar component), 
we ignored the interaction of BHs/binaries with BHs orbiting outside of the AGN disk (i.e the anisotropic BH component) 
in the fiducial model 
for simplicity (but investigate their importance below, in Model~37).

To compute the rates of various density and velocity-dependent
processes (see \S~\ref{sec:bs}--\ref{sec:gbs} below), we calculate
the number density ($n_\mathrm{c}$), the velocity dispersion
($\sigma_\mathrm{c}$), the average mass ($m_\mathrm{c}$), and the
typical orbital height ($h_\mathrm{c}$) for each component in each
radial cell in a grid. 
The spherical radial grid extends from
$r_\mathrm{in}=10^{-4}$ to $r_\mathrm{out}=$5 pc, and is
divided into $N_\mathrm{cell}=120$ cells uniformly on a log scale in
the fiducial model. The dependence of results on $N_\mathrm{cell}$ is
discussed in \S \ref{sec:dependence}.  We neglect the possible effects
related to migration traps, which may exist at $\lesssim
9\times 10^{-5}(M_\mathrm{SMBH}/4\times 10^6\,\Msun)$ pc for the model by \citet{Thompson05} 
\citep{Bellovary16}, 
just outside of the simulated domain.

We set the time-independent physical quantities for the spherical stellar component in each cell to be  
\begin{align}
\label{eq:quantity_sph}
(m_\mathrm{c},n_\mathrm{c},\sigma_\mathrm{c},h_\mathrm{c})_{\mathrm{Sstar},l}=
\left({\bar m}_\mathrm{star},\frac{\rho_\mathrm{Sstar}(r_\mathrm{l})}{{\bar m}_\mathrm{star}},\frac{v_\mathrm{Kep}(r_l)}{\sqrt{3}},\frac{r_l}{\sqrt{2}}  \right)
\end{align}
where ${\bar m}_\mathrm{star}$ is the average stellar mass in the spherical
stellar component determined from the assumed IMF Eq.~\eqref{eq:IMF},
$v_{\mathrm{Kep},l}$ is the Keplerian velocity at the radius $r_l$ of
the geometric center of a cell $l$, and $\rho_\mathrm{Sstar}(r)$ is
the density profile of the spherical stellar component. We adopt
\begin{align}
\label{eq:stellar_density}
\rho_\mathrm{Sstar}(r)=&\left(\frac{M_\mathrm{star,3pc}}{4.3}\right)\,\mathrm{pc}^{-3}\nonumber\\
&\times\left(\frac{r}{0.3\,\mathrm{pc}}\right)^{-0.5}\left[1+\left(\frac{r}{0.3\,\mathrm{pc}}\right)^4\right]^{-0.325},
\end{align}
chosen to match the observed stellar surface density distribution in
the Galactic nucleus \citep{Merritt10,Feldmeier14}, where $M_\mathrm{star,3pc}$ is given by 
Eq.~\eqref{eq:Mstar3pc}.  The typical orbital height
for the spherical stellar component
$(h_\mathrm{c})_{\mathrm{Sstar},l}$ is set to $r_l/\sqrt{2}$ considering a uniform distribution in each spherical shell.

On the other hand, the quantities for the disk BH component in each cell are
initialized as
\begin{align}
\label{eq:quantity_disk}
&(m_\mathrm{c},n_\mathrm{c},\sigma_\mathrm{c},h_\mathrm{c})_{\mathrm{DBH},l}=\nonumber\\
&\quad\left[
\frac{\sum_{k \in \mathrm{DBH},l}m_k}{N_{\mathrm{DBH},l}} ,
\frac{N_{\mathrm{DBH},l}r_l}{V_l h_{\mathrm{c},_\mathrm{DBH},l}}
,\left(
\frac{\sum_{k \in \mathrm{DBH},l}v_k^2}{3 N_{\mathrm{DBH},l}}\right)^{1/2},\right. \nonumber\\
&\quad\;
\left. \frac{r_l}{v_\mathrm{Kep}(r_l)}\left(\frac{\sum_{k \in \mathrm{DBH},l}v_{z,k}^2}{N_{\mathrm{DBH},l}}\right)^{1/2} 
\right], 
\end{align}
which evolve with time, where DBH,$l$ refers to the BH components
which are within the AGN disk in the $l$th cell, $N_{\mathrm{DBH},l}$
is the number of single BHs and BH binaries within the AGN disk in cell $l$, 
$V_l=4\pi r_l^2 \Delta r_l$ is the spatial volume of the $l$th  spherical shell so that $h_l V_l/r_l$ is the spatial volume of the AGN disk in the $l$th cell, 
and $\Delta r_l$ is the width of
the $l$th cell.  
Note that we include the number of BH binaries in $N_{\mathrm{DBH},l}$. For the binaries, 
$m_k$ and $v_k$ refer to the total mass and center of mass velocity
of the $k^{\rm th}$ binary relative to the Keplerian velocity in Eq.~(\ref{eq:quantity_disk}).
The average mass $m_{\mathrm{c},\mathrm{DBH},l}$ and
the velocity dispersion
$\sigma_{\mathrm{c},\mathrm{DBH},l}$ of the disk BH component in a cell $l$ are given
by the the average of $m_k$ and the root mean square of $v_k /\sqrt{3}$ 
of BHs, respectively.

We assume that when $h_{k}<h_{\mathrm{AGN},l}$, the $k^{\rm th}$
object is embedded in the AGN disk, where $h_{k}=v_{z,k}r_k/v_\mathrm{Kep}$ 
is the typical height of orbital motion 
for the $k^{\rm th}$ object
and $h_{\mathrm{AGN},l}$ is the height of the AGN disk 
at $r_l$ ($h_{\mathrm{AGN},l}$ is derived in \S~\ref{sec:disk_model}).  We
assume that the disk BH component rotates in the same sense as the AGN disk, so $\sigma_\mathrm{c}=0$ means that BHs in the disk
co-rotate with the Keplerian gas.

In Model~37, we take into account interactions with BHs outside the AGN disk (i.e. the anisotropic BH component). Each quantity for this new component is calculated in the same way as in Eq.~\eqref{eq:quantity_disk}, but including BHs outside the AGN disk ($h_{k}>h_{\mathrm{AGN},l}$).

\begin{figure}
\includegraphics[width=90mm]{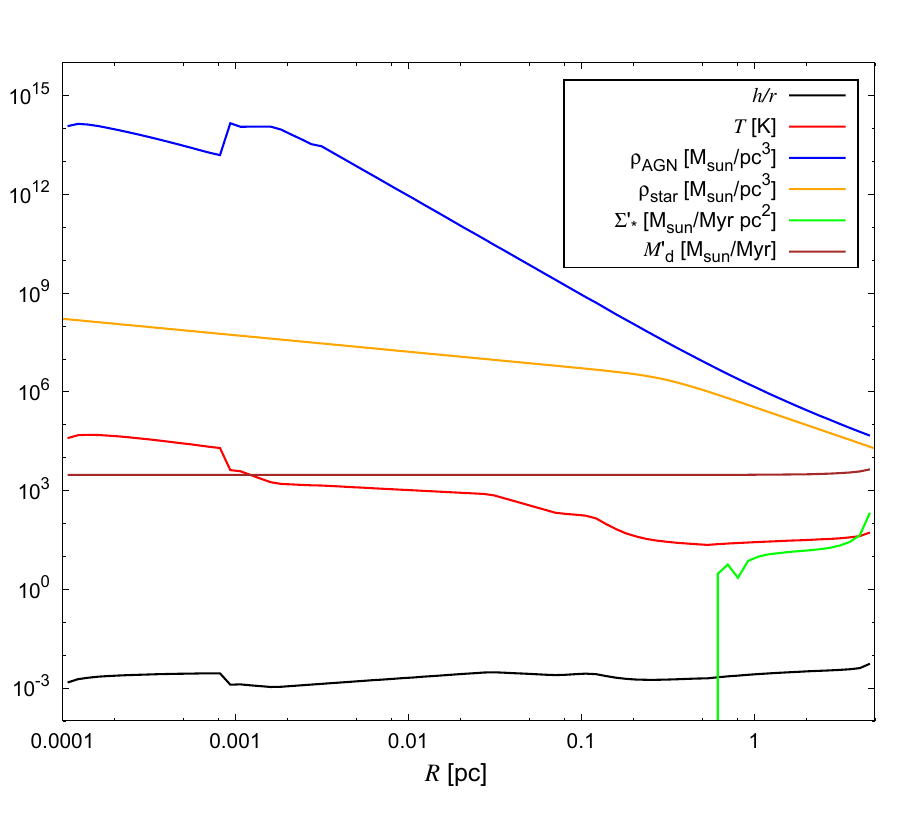}
\caption{Physical quantities for the adopted disk model as a function of the distance $r$ from the SMBH. The colored lines represent the disk height over the distance $h_\mathrm{disk}/r$ (black), the midplane temperature $T$ (red),
the gas density $\rho_{\rm AGN}$ (blue), the background stellar density (orange), the star formation surface density ${\dot \Sigma}_*$ (green), and the accretion rate ${\dot M}_{\rm d}$ (brown). Units are $M_\sun$, Myr, $^\circ$K~and~pc. 
}
\label{fig:disk_model}
\end{figure}

The statistical quantities of the disk stellar component in each cell are calculated as
\begin{align}
\label{eq:quantity_sdisk}
(m_\mathrm{c},n_\mathrm{c},\sigma_\mathrm{c},h_\mathrm{c})_{\mathrm{Dstar},l}
=&({\bar m}_{\mathrm{star}},n_{\mathrm{Dstar},l},\sigma_{\mathrm{c},\mathrm{DBH},l}
\nonumber\\
&\; h_{\mathrm{c},\mathrm{DBH},l})
\end{align}
where $\mathrm{Dstar},l$ refers to the stellar component embedded
within the AGN disk in the $l$th cell.  
We ignore the accretion and migration of stars captured by the AGN disk for simplicity. 
This is a conservative assumption for the merger fraction since it neglects the possibility for migration to increase the stellar density in the inner regions which would facilitate BH mergers by frequent binary-single interactions. 
Since we do not calculate the
evolution of stars, 
we assume that their velocity dispersion and scale height match 
that of the BH disk component, given by Eq.~\eqref{eq:quantity_disk}.
The density of the disk stellar component is calculated from 
the number of the stars by simply assuming that the stars
reside in the same volume as the disk BH component 
($V_l h_{\mathrm{c},\mathrm{DBH},l}/r_l$).
The number of stars in the disk stellar component is
calculated considering three factors.  First we assume that stars form
with rate ${\dot \Sigma}_\mathrm{*}$ in the outer regions of the AGN disk
(Figure~\ref{fig:disk_model}, $\S$ \ref{sec:disk_model}). 
We ignore the evolution of newly formed stars. 
Second we
assume that spherically distributed stars are captured in the AGN disk
at the rate estimated in $\S$3 of \citet{Bartos17}. \citet{Bartos17}
estimated the timescale on which objects are captured in the AGN disk
based on the torque due to Bondi-Hoyle-Lyttleton accretion during
crossing the AGN disk. In our simulation, we calculate the critical
inclination angle of stellar orbits with respect to the AGN disk at
which the alignment timescale (Eq.~11 of \citealt{Bartos17})
becomes the same as the elapsed time. We assume that the inclination
of stars $\cos i$ is distributed uniformly between -1 and 1 as
in \citet{Bartos17}, and derive the fraction of stars whose
inclination is smaller than the critical inclination angle.  
We assume that gas within the Bondi-Hoyle-Lyttleton radius is captured by a star, and we neglect cases in which the stellar radius is larger than the Bondi-Hoyle-Lyttleton radius, which is realized for $v_i\gtrsim 300\,\mathrm{km/s}(m_{\mathrm{star},i}/0.36\,\Msun)^{1/2}$. 
In this way we calculate the number of stars captured in the AGN disk in each
step.  Third, we reduce the number of stars in the disk stellar component in the $l$th cell hosting the
binary by one for each BH binary that experiences a hard binary-stellar
single interaction with objects in the disk stellar component.
This reflects the fact that the recoil kick following a hard BH binary-single interaction with the typical low-mass stars is so large that the interacting
stars are usually kicked out from the AGN disk in the vertical direction.

\subsubsection{AGN disk}
\label{sec:disk_model}

We employ the AGN accretion disk model proposed by \citet{Thompson05}. 
In the fiducial model, we set the SMBH mass to $M_\mathrm{SMBH}=4\times 10^6M_\odot$, and the accretion rate from outer boundary of $r_\mathrm{out}=$
5 pc to be 
${\dot M}_\mathrm{out}=0.1\,{\dot M}_\mathrm{Edd}$, where ${\dot M}_\mathrm{Edd}=L_\mathrm{Edd}/(c^2 \eta_{\rm c})$ is the Eddington accretion rate, $L_\mathrm{Edd}$ is the Eddington luminosity, $c$ is the light speed, and $\eta_{\rm c}$ is the radiative efficiency. We adopt $\eta_{\rm c}=0.1$ assuming a standard thin disk model. 
We adopt the opacity model given by \citet{Bell94} which gives the opacity as a function of temperature and density. Following the fiducial values in \citet{Thompson05}, 
we assume the pressure ratio parameter $\xi = 1$. 
We improve the calculation of the conversion efficiency  $\epsilon$ of star formation to radiation in \citet{Thompson05} by taking into account the limitations due to the AGN and stellar lifetimes (see Appendix \S~\ref{sec:conversion_eff} for details). 
This reduces $\epsilon$ by a factor $\sim 4$ to $\epsilon=1.5\times 10^{-4}$ for the fiducial model, and $\epsilon$ varies between $1.5\times 10^{-4}$ and $7.7\times 10^{-4}$ according to the IMF exponent ($-2.35\leq -\delta_\mathrm{IMF}\leq -1.7$).  
We assume that the efficiency of angular momentum transport due to global torque in the outer region is $m_\mathrm{AM}\sim 0.1-0.2$ 
as suggested by \citet{Thompson05}. 
In the inner region, we adopt the $\alpha$-model for angular momentum transport, in which the alpha parameter is $\alpha_\mathrm{SS}=0.1$ \citep{King07,Bai13}. 
In the transition between the inner and outer regions, the degree of angular momentum transport is adjusted to keep the Toomre parameter at $Q=1$. 

We assume a locally isothermal equation of state to calculate the sound speed of the AGN disk as 
$c_s=(p/\rho_\mathrm{gas})^{1/2}$, where $p$ is the total gas+radiation pressure, 
and $\rho_\mathrm{gas}$ is the local density of a gas disk. 
The viscosity is given by $\nu=r v_r /(d\mathrm{ln}\Omega/d\mathrm{ln} r)$, where $v_r$ is the radial velocity, and $\Omega$ is the angular velocity of the gas disk, 
which is rotating around the enclosed mass of the SMBH and stars (Eq. \ref{eq:stellar_density}). 
In the $\alpha$-prescription, the viscosity $\nu$ is assumed to be proportional to the total pressure, $\nu=\alpha_\mathrm{SS} c_s^2/\Omega$. 
The surface density of the gas disk is calculated using the radial velocity and the gas inflow rate (${\dot M}$) as $\Sigma_\mathrm{disk}= {\dot M}/(2\pi r v_r)$.

The AGN disk properties are shown in Figure~\ref{fig:disk_model}. 
The outer region ($\gtrsim 1$ pc) is stabilized by radiation pressure and supernovae from in-situ formed stars. Therefore the star formation rate is determined by the disk model, which depends on the mass of the SMBH, the accretion rate, and efficiency parameters 
through Eqs.~(C1)-(C12) of \citet{Thompson05}. 
The green line in Figure~\ref{fig:disk_model} represents the surface density of the star formation rate ${\dot \Sigma_\mathrm{*}}$. 
The star formation rate and BH formation rate surface densities are given by $f_\mathrm{star}\times {\dot \Sigma_\mathrm{*}}$ and $f_\mathrm{BH}\times {\dot \Sigma_\mathrm{*}}$, respectively, where $f_\mathrm{star}$ and $f_\mathrm{BH}$ are the ratio of the mass of stars with the mass less than $20\,\Msun$ and all stellar-mass BHs to the mass of all stars at formation, respectively. 
BHs continuously form at this rate in our simulations, and the number of BHs formed within 100 Myr is $\sim7\%$ of the initial number of BHs in the fiducial model. 
The accretion disk model with large size and star formation is motivated by 
several observations, which are early enhancement of metallicity in disks \citep{Artymowicz93,Xu18,Novak19}, 
long-timescale transients in AGN \citep{Graham17}, 
and supernovae found in a vicinity of an AGN \citep{PerezTorres10}.

Based on Eq.~\eqref{eq:mBH}, 
$f_\mathrm{BH}$ varies from 0.016 to 0.092 according to the initial mass function ($1.7\leq \delta_\mathrm{IMF}\leq 2.35$), and $f_\mathrm{BH}=0.016$ in the fiducial model. 
We set the velocity of newly formed BHs relative to the local AGN motion $v_k$ to be the sonic velocity of the AGN disk at their location. This is motivated by numerical simulations suggesting that the scale height of newly born stars within an AGN disk is roughly similar to the thickness of the AGN disk \citep{Nayakshin07}. 
In this study, BHs form immediately during star formation neglecting the lifetime of the progenitor stars for simplicity. 
We assume that the mass distribution and the binary fraction of in-situ formed BHs are the same at formation as those of the pre-existing BHs described above.

\subsection{Formation, destruction, and orbital evolution of binaries}
\label{sec:equations}

In this study we do not follow the evolution of individual stars, but only track their average statistical properties in a grid as explained in \S~\ref{sec:component} for both the spherical and the disk stellar components. However we follow the orbital parameters of individual BHs and BH binaries as follows. 

For clarity we use the indices $i$ and $j$ in this paper exclusively to label a single BH and a BH binary, respectively. We use the index $k$ to denote either a single BH or a BH binary.
For single BHs, we characterise their orbits with their orbital radius around the SMBH $r_i$,  their
magnitude of $z$-direction velocity $v_{z,i}$, and the $xy$-direction velocity $v_{xy,i}$ at $z=0$. 
Here $v_{i}=(v_{xy,i}^2+v_{z,i}^2)^{1/2}$ is the velocity relative to the local Keplerian velocity of a circular orbit with $z=0$. Given $v_{z,i}$, the maximum height of the orbit is $h_i = r_i v_{z,i}/v_{\rm Kep}(r_i)$.
For binaries, we follow their radial position $r_j$, their center of mass velocity components
$v_{xy,j}$ and $v_{z,j}$, and their binary separation $s_j$.
The position, velocity, and separation of BHs evolve via interacting with gas and/or stellar and BH components (Figure~\ref{fig:mechanisms}). 
In this paper, we do not consider the evolution of the binary eccentricity ($e_j$), and assume $e_j=0$ for all binaries.

We incorporate the effects of gas dynamical friction, GW radiation, weak gravitational scattering, binary-single interaction, 
type I/II migration torque, binary formation via three-body encounter and via gas-capture mechanism, binary disruption, and star formation. 

The velocity ${\bf v}_{i}=(v_{xy,i},v_{z,i})$ of a single BH evolves via the equation of motion~\citep{Papaloizou00}
\begin{equation}
\label{eq:dvdt_i}
\frac{{\rm d}{\bf v}_{i}}{\mathrm{d}t}={\bf a}_{\mathrm{acc},i}+{\bf a}_{\mathrm{GDF},i}+{\bf a}_{\mathrm{WS},i},
\end{equation}
where ${\bf a}_{\mathrm{acc},i}$ is the acceleration due to the accretion torque (\S \ref{sec:acc}), 
gas dynamical friction ${\bf a}_{\mathrm{GDF},i}$ (\S \ref{sec:df}), and weak gravitational scattering ${\bf a}_{\mathrm{WS},i}$ (\S \ref{sec:sss}). 

For BH binaries, the center of mass velocity ${\bf v}_j$ changes due to all of these processes and additionally due to binary-single interactions as 
\begin{equation}
\label{eq:dvdt_j}
\frac{\mathrm{d}{\bf v}_{j}}{\mathrm{d}t}={\bf a}_{\mathrm{acc},j}+{\bf a}_{\mathrm{GDF},j}+{\bf a}_{\mathrm{WS},j}+{\bf a}_{\mathrm{BS},j},
\end{equation}
where ${\bf a}_{\mathrm{BS},j}$ is the acceleration due to binary-single interactions (${\bf a}_{\mathrm{BS},j}$.

\citet{Bartos17} proposed that the binary separation decreases due to gas dynamical friction in an AGN disk, type I/II torque from a circumbinary disk, and GW radiation after the capture of close-in binaries in the smooth disk in $\leq 0.01$ pc. 
\citet{Stone17} considered mergers from binaries formed in-situ at the unstable part of the disk at larger radii of $\sim$ pc. \citet{Stone17} find that binary-single interaction with the disk stellar component also harden the binary separation. 
In this study, we incorporate these effects in the evolution of the binary separation as 
\begin{equation}
\frac{\mathrm{d}s_{j}}{\mathrm{d}t}=
\left.\frac{\mathrm{d}s_j}{\mathrm{d}t}\right|_{\mathrm{GW}}+
\left.\frac{\mathrm{d}s_j}{\mathrm{d}t}\right|_{\mathrm{gas}}+
\left.\frac{\mathrm{d}s_j}{\mathrm{d}t}\right|_{\mathrm{BS}},
\end{equation}
where the three terms on the right hand side are the evolution rates for the binary separation due to GW radiation (\S \ref{sec:gw}), 
gaseous torque (\S \ref{sec:typeii} and \S \ref{sec:df}), and binary-single interaction, respectively.  

Single and binary BHs migrate radially toward the SMBH due to the gaseous torque of the AGN disk, given by 
type I or type II torque formulae (\S \ref{sec:typeii}). 
\begin{equation}
\frac{\mathrm{d}r_{k}}{\mathrm{d}t}=
\left.\frac{\mathrm{d}r_{k}}{\mathrm{d}t}\right|_{\mathrm{type\,I/II}},
\end{equation}

Furthermore, binaries form and are disrupted according to
\begin{equation}
\label{eq:dndt}
\frac{\mathrm{d}N_\mathrm{bin}}{\mathrm{d}t}=
\sum_i P_{\mathrm{3bbf},i}+
\sum_i P_{\mathrm{gas},i}+
K_\mathrm{dis}
\end{equation}
where $P_{\mathrm{3bbf},i}$ is the binary formation rate by three-body encounter (\S \ref{sec:3bf}), $P_{\mathrm{gas},i}$ is the binary formation rate by gas-capture mechanism (\S \ref{sec:gbs}), and $K_\mathrm{dis}=\mathrm{d}N_\mathrm{bin}/\mathrm{d}t|_{\mathrm{dis}}<0$ is the binary disruption rate. 
Binaries are disrupted in the simulation when the binary separation $s_{j}$ becomes larger than the Hill radius of a binary with respect to the SMBH 
\begin{equation}\label{eq:rHill}
r_{\mathrm{Hill},j}=r_j \left(\frac{m_j}{3M_\mathrm{SMBH}}\right)^{1/3}\,.    
\end{equation}

The terms in Eqs.~(\ref{eq:dvdt_i})-(\ref{eq:dndt}) are described in the next section, \S \ref{sec:mechanisms}. 
These equations are calculated separately using the local statistical quantities describing the stellar environment in each shell $l$. 

\subsection{Individual Processes}
\label{sec:mechanisms}

Here we describe in detail the prescription adopted for each of the mechanisms included in our simulations. 

\subsubsection{Gravitational wave radiation}
\label{sec:gw}

When the binary separation is small, GW radiation strongly decreases the binary separation. The hardening rate via GW radiation is given as
\begin{eqnarray}
\frac{\mathrm{d}s_j}{\mathrm{d}t}|_{\mathrm{GW}}=-\frac{64}{5}\frac{G^3m_{j_1}m_{j_2}(m_{j_1}+m_{j_2})}{c^5s_j^3}
\end{eqnarray}
assuming zero eccentricity \citep{Peters64}, 
where $m_{j_1}$ and $m_{j_2}$ are the masses of the primary and secondary BH in the $j^{\rm th}$ binary, respectively. 
\subsubsection{Type I and II migration}
\label{sec:typeii}

Objects in a gaseous disk interact gravitationally with nearby gas, resulting in radial migration. 
When the gravitational torque exerted by an object within the disk exceeds the viscous torque of gas, a gap opens in the disk around the object \citep[e.g.][]{Ward97,Crida06}. 
When a gap does not open around the object, the object migrates due to torques from the Lindblad and corotation resonances on the type I migration timescale \citep[e.g.][]{Ward97,Tanaka02,Paardekooper10,Baruteau11} of 
\begin{eqnarray}
\label{eq:typeI}
t_\mathrm{type\,I}\simeq \frac{1}{2f_\mathrm{mig}}
\left(\frac{M_\mathrm{cen}}{M_\mathrm{sat}}\right)
\left(\frac{M_\mathrm{cen}}{\Sigma_\mathrm{disk} r^2}\right)
\left(\frac{h_\mathrm{disk}}{r}\right)^{2}
\Omega^{-1},
\end{eqnarray}
where $f_\mathrm{mig}$ is a dimensionless factor depending on the local temperature and density profiles \citep[see][]{Paardekooper10,Baruteau11}, 
$h_\mathrm{disk}$ is the half thickness of the disk, 
$M_\mathrm{cen}$ and $M_\mathrm{sat}$ are the central and satellite object mass, respectively, 
and $\Sigma_\mathrm{disk}$ is the surface density of the disk.\footnote{It has been argued that $f_\mathrm{mig}$ may change sign in the inner region of the disk and BHs may migrate outward \citep{Bellovary16}. The region where this may occur, around $\sim 40-600\,R_g \sim 8\times 10^{-6}- 10^{-4}\,\mathrm{pc}(M_\mathrm{SMBH}/4e6\,\Msun)$,  is not included in our simulated domain. Here $R_g=GM_\mathrm{SMBH}/c^2$ is the gravitational radius of the SMBH. 
}
We calculate the evolution in the range $r\geq r_\mathrm{in} \sim 10^3 R_g$) where $f_\mathrm{mig}$ remains positive and its variation is not significant. We set $f_\mathrm{mig}=2$ in the fiducial model, which is the typical value found numerically by \citet{Kanagawa18}. 
If a gap opens in a disk around an object, the object migrates due to the torque of the gas approaching the gap boundary on a timescale related to the viscosity. This process is the so-called type II migration \citep[e.g.][]{Lin86,Ward97,Ida04,Edgar07,Haiman09,Duffell14}.

Recent hydrodynamic simulations \citep{Duffell14,Durmann15,Kanagawa18} have shown that even when 
the torque from a BH exceeds the viscous torque of gas, 
the gas is able to pass through the gap. 
\citet{Duffell14} and \citet{Kanagawa18} show that the migration timescale even for massive migrator is given by the type I migration timescale with a reduced gas surface density in the gap ($\Sigma_\mathrm{disk,min}$), 
\begin{equation}
\label{eq:typeI_II}
t_\mathrm{type\,I/II}=\frac{\Sigma_\mathrm{disk}}{\Sigma_\mathrm{disk,min}}t_\mathrm{type\,I}.
\end{equation}
\citet{Fung14} and \citet{Kanagawa15} show that
\begin{equation}
\label{eq:typeI_II_Sigma}
\Sigma_\mathrm{disk,min}=\Sigma_\mathrm{disk} /(1+0.04K),
\end{equation} 
where 
$K=(M_\mathrm{sat}/M_\mathrm{cent})^2(h_\mathrm{disk}/r)^{-5}\alpha_\mathrm{eff}^{-1}$, 
and $\alpha_\mathrm{eff}=\nu /(c_s h_\mathrm{disk})$ is the effective $\alpha$ parameter.

We calculate the migration rate of stellar-mass BHs within the AGN disk as
\begin{eqnarray}
\label{eq:drdt_typeii}
\frac{\mathrm{d}r_{k}}{\mathrm{d}t}|_\mathrm{type\,I/II}=
-\frac{ r_{k}}{t_\mathrm{type\,I/II}}p_{\mathrm{disk},k}
\end{eqnarray} 
where $p_{\mathrm{disk},k}$ is the fraction of time that the $k^{\rm th}$ object spends in the disk along its orbit around the SMBH, which we calculate as
\begin{equation}\label{eq:pdisk}
p_{\mathrm{disk},k} = 
\left\{
\begin{array}{cc}
    1 & \mathrm{for}~h_{k}<h_\mathrm{\rm disk}(r_k)\,, \\[1.5ex]
\dfrac{2}{\pi} {\rm arcsin}\left[\dfrac{h_{\rm disk}(r_k)}{h_{k}}\right]  & \mathrm{otherwise}\,.   
\end{array}
\right.
\end{equation}
assuming that the time spent inside the AGN disk is approximated by the ratio of the scale height of the disk and the BH's orbit. 
To calculate the migration rate, we substitute $h_{\mathrm{AGN},l}$ into $h_\mathrm{disk}$ in Eq.~(\ref{eq:typeI}). To reduce the computational cost, we use quantities at the center of each cell $r_l$ in Eqs.~(\ref{eq:typeI})-(\ref{eq:typeI_II}). 
For comparison, we also investigate cases in which migration by the type I/II torque does not operate, since the migration of BHs is not well understood due to the complexity of the effects of $N$-body migrators \citep{Broz18}, feedback from BHs \citep[e.g.][]{delValle18,Regan19}, and inhomogeneities in the turbulent accretion disk \citep{Laughlin2004,Baruteau10}.

Due to the torques exerted by the BHs, 
the gas density is reduced near each BH according to Eq.~\eqref{eq:typeI_II_Sigma}. We take into account this reduction when we use gas dynamical friction (\S \ref{sec:df}), gas accretion (\S \ref{sec:acc}), and the gas-capture mechanism (\S \ref{sec:gbs}). 
In these mechanisms, 
when a BH is in the AGN disk, 
we use the local gas density of the AGN disk around each BH as 
\begin{equation}
\rho_\mathrm{gas}=\rho_{\mathrm{AGN},l}\Sigma_\mathrm{disk,min}/\Sigma_\mathrm{disk}, 
\end{equation} 
where $\rho_{\mathrm{AGN},l}$ is the unperturbed density of the AGN disk in the $l$th cell hosting a BH (\S \ref{sec:disk_model}). 
When a BH orbits outside of the AGN disk ($h_{{z},k}>h_{\mathrm{AGN},l}$), we use $\rho_\mathrm{gas}=\rho_{\mathrm{AGN},l}=\Sigma_{\mathrm{disk},l}/(2h_{\mathrm{disk},l})$.

After gas is captured within the Hill's sphere of a binary, 
a circumbinary disk forms. This disk exerts a torque on the binary and changes its separation similar to Type I/II migration.   
The naive expectation, detailed in several semi-analytic treatments (see, e.g, \citealt{Haiman09} and references therein) has been that this process hardens the binary, on a time-scale related to the viscous time-scale in the disk, but modified by a factor involving the ratio of the mass of the binary and the disk.  On the other hand, recent hydrodynamical simulations have found that an equal-mass binary is softened, rather than hardened by the presence of a circumbinary gas disk \citep{Miranda17,Tang17,Munoz19,Moody19}.   At the present time, these simulations have many limitations: they treat equal-mass binaries on prescribed orbits, and disks with a specific viscosity parameter $\alpha$, isothermal equation of state, and a temperature that is generically chosen (for numerical reasons) to be much higher than expected in real disks.  
They do not include radiation and cooling.  Most of them are 2D, and study circular binaries. As a result, it remains unclear how generic these simulations results are, and whether they are applicable to real systems.  For example, recent studies have found that unequal-mass binaries~\citep{Duffell19}, as well as binaries embedded in cooler disks~\citep{Tiede+20} are generally hardened by the circumbinary disk.

Here we consider several prescriptions (Models~3-6). 
In the fiducial model, 
we follow the conclusions of the earlier semi-analytic studies, and  assume that the binary can be hardened by the torque of a circumbinary disk,
and the hardening timescale is given as Eq.~(\ref{eq:drdt_typeii}) by substituting $s_j$ into $r_k$, 
\begin{eqnarray}
\label{eq:dsdt_typeii}
\frac{\mathrm{d}s_{j}}{\mathrm{d}t}|_\mathrm{type\,I/II}=
-\frac{ s_{j}}{t_\mathrm{type\,I/II}}p_{\mathrm{disk},k}.
\end{eqnarray} 
Here, we substitute the BH binary component masses for $M_\mathrm{cen}$ and $M_\mathrm{sat}$, and use the angular velocity of the binary in $\Omega$ in Eq.~(\ref{eq:drdt_typeii}). 
If the Toomre parameter of the circumbinary disk satisfies $Q>1$, it is stable against gravitational fragmentation. In this case, we calculate the surface density of a circumbinary disk using Eq.~(14) in \citet{Goodman04}, in which we assign the opacity consistently following \citet{Bell94}, and the disk temperature is given by the maximum of Eq.~(13) in \citet{Goodman04} and the temperature of the AGN disk at the position $r_k$. 
We assume that the gas pressure dominates over the radiation pressure for the circumbinary disk, which is a valid approximation for the radial range of the disk ($\gg \sim 10^9~\mathrm{cm}$) we are interested in.  
When $Q<1$, we reduce the surface density of gas disk around a binary 
to satisfy $Q=1$.

We assume the accretion rate onto a binary or a single BH is given by the minimum of the Bondi-Hoyle-Lyttleton rate (Eq.~\ref{eq:macc}) times $p_{\mathrm{disk},k}$ (Eq.~\ref{eq:pdisk}) and Eddington limited accretion, 
\begin{equation}
\label{eq:macc_binary}
\dot m_{\mathrm{acc},k}=\mathrm{min}\left({\dot m}_{\mathrm{BHL},k}p_{\mathrm{disk},k},\frac{\Gamma_\mathrm{Edd,cir}L_{\mathrm{Edd},k}}{\eta_{\rm c} c^2}\right)
\end{equation}
where $\Gamma_\mathrm{Edd,cir}$ is the Eddington ratio,
$L_{\mathrm{Edd},k}$ is the Eddington luminosity for a binary, and $\eta_{\rm c}$ is the energy conversion efficiency. We set $\Gamma_\mathrm{Edd,cir}=1$ and $\eta_{\rm c}=0.1$. 
The Eddington limited accretion rate is motivated by regulation due to strong radiation pressure acting on dust grains for gas of around solar metallicity \citep{Toyouchi19}, by a bipolar jet \citep{Regan19}, 
and also by inefficient angular momentum transfer in circumbinary disks \citep{Sugimura18,Inayoshi18}.

We assume that the type I/II migration torque generated by a circumbinary disk surrounding the binary operates to shrink the binary separation only when the Bondi-Hoyle-Lyttleton radius $r_{\mathrm{BHL},k}$ is larger than the binary separation and otherwise the circumbinary disk does not exert a torque on the binary. 
Furthermore, we assume that a circumbinary disk is always aligned with the binary. This is justified since the alignment timescale of the disk with the binary is roughly given by 
the viscous timescale at the binary separation \citep[e.g.][]{Ivanov99,Moody19}. This timescale is 
$r^2/\nu \sim \alpha_\mathrm{SS}^{-1}(h_\mathrm{disk}/r)^{-2}\Omega^{-1}\sim 10^5$ yr \citep{Moody19} for $s_j \sim $ AU, which is shorter than the evolution timescale of binaries by type I/II torque from circumbinary disks ($\gtrsim$ Myr).

\subsubsection{Gas dynamical friction}
\label{sec:df}
When an object has a non-zero velocity relative to the ambient gas, 
gas dynamical friction reduces the relative velocity. 
In this manner, gas dynamical friction hardens binaries \citep{Escala04,Kim07,Baruteau11} and 
damps the velocity dispersion of BHs and stars \citep{Papaloizou00,Tanaka02} in the disk. 
For simplicity, we adopted the formulation for deceleration by gas dynamical friction derived by \citet{Ostriker99} as 
\begin{align}
\label{eq:gdf}
a_{\mathrm{GDF},k}(v_k)&=-\frac{4\pi G^2 m_k \rho_\mathrm{gas}p_{\mathrm{disk},k}}{ v_k^2}f(v_k/c_s),\nonumber\\
\mathrm{where}~
f(x) &= 
\left\{
\begin{array}{ll}
\frac{1}{2} \mathrm{ln}\left(\frac{1+x}{1-x}\right)-x & \mathrm{for}~~0 < x < 1\,,\\
\frac{1}{2}\,\ln\left(x^2 -1 \right)+\ln\Lambda_\mathrm{gas} & \mathrm{for}~~x>1\,,~~
\end{array}
\right.
\end{align}
where $\ln\Lambda_\mathrm{gas}$ is the Coulomb logarithm for gas. We set $\ln\Lambda_\mathrm{gas}=3.1$ referring the results in \citet{Chapon13}. 
For both binary hardening and velocity damping, 
we use the sonic velocity $c_s$ from the disk model of \citet{Thompson05} (\S \ref{sec:disk_model}) at the geometric center of the cell $l$ hosting BHs. 
Although these formulae apply for linear motion, they remain approximately correct for circular motion, despite the strong curvature and possible interaction of the pair of ``wakes'' in the binary case~\citep{Kim07}. 
For binary hardening, we assume that gas dynamical friction operates while the binary is captured to the AGN disk ($h_{k}<h_\mathrm{AGN}$).

The AGN disk capture timescale on which the initial supersonic velocity of an object $v_{\mathrm{ini},k}$ decays due to crossing the disk due to gas dynamical friction is
\begin{align}\label{eq:dt_dec}
t_\mathrm{capAGN} &\equiv \int_0^{v_{\mathrm{ini},k}} \frac{1}{a_{\mathrm{GDF},k}(v_{k})p_{\mathrm{disk},k}}{\rm d}v_{k}\nonumber\\
&\simeq \frac{v_{\mathrm{ini},k}^4 r_k}{16 \sqrt{3} \pi G^2 \rho_\mathrm{gas} m_k h_\mathrm{AGN} v_\mathrm{Kep}}\nonumber\\
&\sim 22\,\mathrm{Myr}\left(\frac{v_{\mathrm{ini},k}}{0.2v_\mathrm{Kep}}\right)^4
m_{k,10}^{-1}M_{\rm SgrA}^{3/2}
r_{1\rm pc}^{-3/2}
\nonumber\\
&\quad\times\left(\frac{h/r}{0.01}\right)^{-1}\rho_{\mathrm{gas},6}^{-1},
\end{align}
where we introduced the abbreviated labels $M_{\rm SgrA}=M_{\rm SMBH}/(4\times10^6\,\Msun)$, $m_{k,10}=m_i/(10\,\Msun)$, $r_{1\rm pc}=r_k/\rm pc$
$h/r=h_{\mathrm{AGN}}(r_k)/r_k$, $\rho_{\mathrm{gas},6} = \rho_\mathrm{gas}/(10^6\,\Msun\mathrm{pc}^{-3})$ (cf. Figure~\ref{fig:disk_model}),
where we assumed $f\sim1$ in $a_{\mathrm{GDF}}$ for simplicity (see Eq.~\ref{eq:gdf}), 
and ignore the contribution of the stellar mass as $v_\mathrm{Kep}\sim (GM_{\rm SMBH}/r_k)^{1/2}$, 
which underestimates $t_\mathrm{capAGN}$ by $\sim 2$ at $r_k= \mathrm{pc}$. 

Similarly we define the gas dynamical friction hardening timescale of a binary as
\begin{align}
\label{eq:df_timescale}
t_\mathrm{GDF,HS} &\equiv \int_0^{v_\mathrm{HS}} {\rm d}v/a_\mathrm{GDF}
\nonumber\\
&\simeq \frac{v_{\mathrm{HS}}^3}{12 \pi G^2 \rho_\mathrm{gas} m_k}\nonumber\\
&\sim 7\,\mathrm{Myr}
\left(\frac{m_\mathrm{star}}{0.36\,\Msun}\right)^{3/2}
m_{k,10}^{-5/2}M_{\rm SgrA}^{3/2}
r_{1\rm pc}^{-3/2}\rho_{\mathrm{gas},6}^{-1},
\end{align}
where $v_\mathrm{HS}=({\bar m}_\mathrm{star}/m_k)^{1/2} v_\mathrm{Kep}(r) $ is the orbital velocity of a binary of mass $m_k$ around its center of mass at which the binary is at the hard-soft boundary compared to the spherical stellar component.

During the hardening of binaries, 
it is not obvious whether type I/II torques by a circumbinary disk or gas dynamical friction by the AGN disk is a better description. 
\citet{Baruteau11} showed that a binary in an AGN disk is hardened roughly on the timescale of gas dynamical friction. 
\citet{Derdzinski18} find that the gas captured by a rapidly migrating BH within an AGN disk has almost no rotation with respect to the migrator, which suggests that gas torques may operate in the manner of the dynamical friction (although this was demonstrated only for a single specific binary + disk configuration). 
Since this issue has not been settled yet, we investigate several different prescriptions. 
In the fiducial model, 
we investigate the cases in which 
the binary hardening rate due to gas interaction is  given by
\renewcommand{\theenumi}{(\roman{enumi})}
\begin{enumerate}
    \item the maximum value of the gas dynamical friction and the type I/II torque, 
    \item only the gas dynamical friction, 
    \item only the type I/II torque, or
    \item zero (no hardening by gas interaction). 
\end{enumerate}

\citet{Park17} have shown that gas dynamical friction does not operate 
when 
the Bondi-Hoyle-Lyttleton radius ($r_{\mathrm{BHL},k}=Gm_k/(c_s^2+v_k^2)$) is smaller than the size of a HII sphere ($r_{\mathrm{HII},k}=(3Q_{\mathrm{ion},k}/4\pi \alpha_\mathrm{rec,B}n_\mathrm{gas})^{1/3}$), where $\alpha_\mathrm{rec,B}$ is the case-B recombination coefficient for H (evaluated at $T=10^4\,\mathrm{K}$), 
and $Q_{\mathrm{ion},k}$ is the ionizing photon number flux from the $k^{\rm th}$ BH. 
This is caused by radiation feedback from a BH 
which diminishes the wake created by gas dynamical friction. 
The condition $r_{\mathrm{BHL},k} > r_{\mathrm{HII},k}$ can be rewritten as  
\begin{align}
\left(\frac{n_\mathrm{gas}}{2\times 10^8\,\mathrm{cm}^{-3}}\right)^2
\left(\frac{m_k}{10\,\Msun}\right)^2
\left(\frac{v_k}{10\,\mathrm{km/s}}\right)^{-6} > 1
\end{align}
if we assume the Eddington accretion rate with the 
radiative efficiency 
$\eta_{\rm c} = 0.1$,
$c_s \ll v_k$, and $Q_\mathrm{ion}\sim L_k /h\nu_\mathrm{p}$ with $h \nu_\mathrm{p} \sim 13.6$ eV, where $h$ is Planck's constant, $\nu_\mathrm{p}$ is the average photon's frequency, and $L_k$ is the luminosity from the $k^{\rm th}$ BH. 
Also when the velocity of a BH exceeds twice the sound velocity in the HII region ($v_k \gtrsim 50\,\mathrm{km/s}$), gas dynamical friction is recovered \citep{Park13,Park17}. 
We set these conditions, $r_{\mathrm{BHL},k}> r_{\mathrm{HII},k}$ or $v_k > 50\,\mathrm{km/s}$, as criteria for gas dynamical friction to operate 
(also see a recent update considering dust emission by \citealt{Toyouchi20}). 
In this work, we ignore kinetic feedback on gas dynamical friction, whose effects are highly debated \citep{Gruzinov19,LiXinyu19,Regan19,Takeo20}.

\subsubsection{Accretion torque}
\label{sec:acc}

As a BH crosses the AGN disk, it captures gas from the disk. 
The velocity of the BH decreases to satisfy conservation of momentum.   
The capture rate of gas on the $k^{\rm th}$ BH during passing the AGN disk is given by the Bondi-Hoyle-Lyttleton rate  
\begin{eqnarray}
\label{eq:macc}
{\dot m}_{\mathrm{BHL},k}=4\pi  r_{\mathrm{w},k} r_{\mathrm{h},k} \rho_\mathrm{gas} (c_s^2 + v_{k}^2)^{1/2}, 
\end{eqnarray}
where 
\begin{eqnarray}
r_{\mathrm{w},k}=\mathrm{min}(r_{\mathrm{BHL},k},r_{\mathrm{Hill},k},r_{\mathrm{shear},k}) 
\end{eqnarray}
and 
\begin{eqnarray}
r_{\mathrm{h},k}=\mathrm{min}(r_{\mathrm{w},k},h_{\mathrm{AGN},l})
\end{eqnarray} 
are the width and height of the gas bound to the $k^{\rm th}$ BH, and 
\begin{eqnarray}
r_{\mathrm{shear},k}=\frac{G m_k}{(r_{\mathrm{Hill},k}\Omega)^2}
\end{eqnarray} 
is the typical radius over which gas motion changes due to the Keplerian shear of the AGN disk. 
Since ${\bf v}_k$ represents the relative velocity compared to the local rotating motion of the AGN disk, 
the acceleration by gas accretion torque is 
\begin{eqnarray}
\label{eq:acc}
{{\bf a}}_{\mathrm{acc},k}=- {\bf v}_k \frac{{\dot m}_{\mathrm{BHL},k}}{m_k}p_{\mathrm{disk},k}.
\end{eqnarray}

Although gas is considered to be captured at the rate of Eq.~(\ref{eq:macc}), 
the accretion rate onto the BH may be smaller than this value due to radiation feedback and the inefficiency of angular momentum transport. 
When we calculate the type I/II torque by a circumbinary disk, the accretion rate onto the binary is limited by the Eddington accretion rate (Eq.~\ref{eq:macc_binary}). 

We adopt the gas accretion rate onto single BHs and binary BHs using Eq.~\eqref{eq:macc_binary}. 
On the other hand, when BHs are in binaries, we apportion the gas accretion rate between the binary components as 
\begin{align}
\label{eq:macc_BHs}
{\dot m}_{\mathrm{acc},j_2}&=\min\left[\frac{\lambda}{1+\lambda}{\dot m}_{\mathrm{BHL},j}p_{\mathrm{disk},j},\frac{\Gamma_\mathrm{Edd,cir}L_{\mathrm{Edd},j_2}}{\eta_{\rm c} c^2}\right]\\
{\dot m}_{\mathrm{acc},j_1}&=\min
\left[{\dot m}_{\mathrm{acc},j}-{\dot m}_{\mathrm{acc},j2},\frac{\Gamma_\mathrm{Edd,cir}L_{\mathrm{Edd},j_1}}{\eta_{\rm c} c^2}\right]
\end{align}
where $\lambda$ is the ratio of the accretion rate onto the $j_2^{\rm th}$ BH over that onto $j_1^{\rm th}$ BH. 
We adopt $\lambda$ given by the fitting formula in Eq. (1) of \citet{Kelley19}, based on the results of earlier hydrodynamical simulations \citep{Farris14}.

\subsubsection{Binary-single interaction}
\label{sec:bs}

After a binary-single interaction, the binary separation changes depending on the hardness of the binary \citep[e.g.][]{Heggie75,Hills75,Binney08}. We make two types of prescriptions for binary-single interaction according to whether a binary is hard, which satisfies $E_{\mathrm{b},j}>3m_\mathrm{c}\sigma_\mathrm{c}^2/2$, or soft, where $E_{\mathrm{b},j}=Gm_{j_1}m_{j_2}/(2s_j)$ is the binding energy of the $j^{\rm th}$ binary. 

For soft binary-single interactions, we employ the prescription for the average softening rate derived in \citet{Gould91}, 
\begin{equation}
\label{eq:hard-bs}
\frac{\mathrm{d}s_j}{\mathrm{d}t}|_{\mathrm{BS}}=
\frac{16}{3}\frac{G n_\mathrm{c} m_\mathrm{c} s_j^2}{m_{j}\sigma_\mathrm{c} ^3}
\mathrm{ln}\Lambda \left(E_{\mathrm{b},j}-\frac{3}{2}m_\mathrm{c}\sigma_\mathrm{c} ^2 \right), 
\end{equation}
where $\Lambda=v_{\mathrm{bin},j}^2/\sigma_\mathrm{c}^2$ is the Coulomb factor, and $v_{\mathrm{bin},j}$ is the relative velocity of components in the $j^{\rm th}$ binary. 
We set $v_{\mathrm{bin},j}=(Gm_{j}/s_j)^{1/2}$ assuming $e_j=0$. 
This equation assumes an isotropic Maxwell-Boltzmann distribution for the velocity of the background objects, which is approximately justified for the interaction with the spherical stellar component. 
We do not account for soft binary-single interactions with the disk BHs and stellar components since most interactions with these components are not soft, due to the low relative velocity  ($\sigma_\mathrm{c}<v_\mathrm{Kep}h_\mathrm{AGN}/r\lesssim Gm_j/r_{\mathrm{Hill},j}$).

On the other hand, for hard binary-single interactions, 
the hardening rate and the kick velocity after a binary-single encounter are given following \citet{Leigh18}.  The binary hardening rate is given by
\begin{equation}
    E_{\mathrm{af},j}=E_{\mathrm{be},j}-(\Delta K_j+\Delta K_\mathrm{c}),
\end{equation}
where $E_{\mathrm{be},j}$ and $E_{\mathrm{af},j}$ are the binding energies of the $j^{\rm th}$ binary, i.e. $G m_{j_1} m_{j_2} / (2s_j)$, before and after the binary-single interaction, and
$\Delta K_j$ and $\Delta K_\mathrm{c}$
are the changes in the kinetic energy of the $j^{\rm th}$ binary and the escaping third body, respectively. 
Since $\Delta v_{\mathrm{BS},j}$ and $\Delta v_{\mathrm{BS,c}}$ are on order of $v_{\mathrm{bin},j}$ and $v_j$ is typically smaller than $v_{\mathrm{bin},j}$, we approximate $\Delta K_j=\frac{1}{2}m_j \Delta v_{\mathrm{BS},j}^2$ and $\Delta K_\mathrm{c}=\frac{1}{2}m_{\rm c} \Delta v_{\mathrm{BS,c}}^2$, where $\Delta v_{\mathrm{BS},j}$ and $\Delta v_{\mathrm{BS,c}}$ are the kick velocities onto the $j^{\rm th}$ binary and the third body, respectively.
Due to the conservation of linear momentum, $\Delta {\bf v}_{\mathrm{BS},j}=-(m_{\rm c}/m_j)\Delta {\bf v}_{\mathrm{BS,c}}$. 
Here $\Delta v_{\mathrm{BS,c}}$ is set to be the mode of the probability distribution for the kick velocity, 
which is determined by the energy and the total angular momentum $L$ for the three-body system. Following \citet{Leigh18}, we sample $L^2$ uniformly from 0 to $(11.5/18)\,L_\mathrm{max}^2$ and set it to zero for the spherical and the disk components, respectively, where $L_\mathrm{max}$ is the maximum angular momentum of the three-body system (Eq.~7.27 in \citealt{Valtonen06}). 
The assumption of the low angular momentum for the disk components is due to the low $v_{\mathrm{rel},j}$, and this assumption increases $\Delta v_{\mathrm{BS},c}$ for the disk components by a factor $\sim 1.6$ compared to $\Delta v_{\mathrm{BS},c}$ for the spherical cases with the typical angular momentum value of $(2/9)\,L_\mathrm{max}^2$ \citep{Leigh18}. 
To determine the energy of the three-body system, 
we set $r_{\mathrm{Hill},j}$ and $v_{\mathrm{rel},j}$ to the initial third body position and velocity relative to the binary center of mass, respectively. 
Hence, when a hard binary-single interaction occurs, the binary is hardened as 
\begin{equation}
\frac{\mathrm{d}s_j}{\mathrm{d}t}|_{\mathrm{BS}}=
\left(\frac{1}{E_{\mathrm{bf},j}}-\frac{1}{E_{\mathrm{af},j}}\right)\frac{Gm_{j_1}m_\mathrm{j_2}}{2\Delta t}, 
\end{equation}
and a binary center of mass receives 
the kinetic energy $\Delta K_j$ due to a recoil kick. 
For the interaction with the disk BH component, we randomly choose the third body $k'$ which is captured by the AGN disk ($h_{k'}<h_\mathrm{AGN}$) and resides in the same cell with the binary $j$, and set $m_c=m_{k'}$. We assign the third body a recoil kick given by $\Delta K_c$. Even when the $k'$th object is binary, we treat as interaction with a single object with the mass $m_{k'}$. 
We assume that the direction of this kick velocity is 
random and isotropically distributed.
\citet{Geller19} verified that the binary evolution due to encounters calculated by this semi-analytical approach matches the results from direct $N$-body simulations.

The timescale for the occurrence of a binary-single encounter is given by \citep[e.g.][]{Binney08}
\begin{equation}
\label{eq:t_hardbs}
t_{\mathrm{BS},j}=1/(n_\mathrm{c} \sigma_\mathrm{coll} v_{\mathrm{rel},j}p_{\mathrm{c},j}),
\end{equation}
where 
\begin{equation}
v_{\mathrm{rel},j}=\mathrm{max}(\sqrt{3}\sigma_\mathrm{c},v_{j},v_{\mathrm{rel,mig},j},v_{\mathrm{shear},j})
\end{equation}
represents the typical velocity of the third body relative to the center of mass for the binary $j$, $\sigma_\mathrm{coll}$ is the cross section, $v_{\mathrm{rel,mig},j}$ is the migration velocity of the binary relative to the third body, 
$v_{\mathrm{shear},j}$ is the shear velocity between the center of mass for the binary $j$ and the third body, 
 $p_{\mathrm{c},j}$ is the fraction of time that the $j^{\rm th}$ object spends within the scale height of each component along its orbit around the SMBH. 
We set $v_{\mathrm{shear},j}=p_\mathrm{uni}v_\mathrm{Kep}r_{\mathrm{Hill},j}/r_j$ assuming that the difference between the SMBH and the binary $j$ distance and the third object is $p_\mathrm{uni}r_{\mathrm{Hill},j}$, where $p_\mathrm{uni}$ is a random number uniformly distributed between 0 and 1. 
In Eq.~\eqref{eq:t_hardbs}, we define $p_{\mathrm{c},j}$ using
\begin{align}
\label{eq:pc_disk}
p_{\mathrm{c},k} \equiv 
\left\{
\begin{array}{cc}
    1 & \mathrm{for}~h_{k}<h_\mathrm{c}\,, \\[1.5ex]
\dfrac{2}{\pi} {\rm arcsin}\left[\dfrac{h_{\rm c}}{h_{k}}\right]  & \mathrm{otherwise}\,.   
\end{array}
\right.
\end{align}
For the interaction with the spherical stellar component, we always set $p_{\mathrm{c},j}=1$. 
We set the migration velocity of a third body in the disk BH and stellar components by the average of the migration velocity for BHs in a cell hosting BHs $j$, while a third body in the spherical stellar component has no migrating motion. 
Here, $n_{\rm c}$ and $\sigma_{\rm c}$ are the number density and velocity dispersion of objects in the cell of the binary, including the spherical stellar component, the disk BH component, and the disk stellar component.\footnote{We calculate the binary-single interaction rate separately for the three components.} 
The cross section is approximated by
\begin{equation}
\sigma_\mathrm{coll} = b_{xy,j} b_{z,j} 
\end{equation}
where $b_{xy,j}$ and $b_{z,j}$ are the effective maximum impact parameters of objects approaching from different directions such that they approach the binary center of mass within a binary separation $s_j$ at closest approach. This is given approximately by
\begin{align}\label{eq:bxy}
b_{xy,j} &= \min\left\{s_j\sqrt{1+2\frac{b_{90,j}}{s_j}},r_{\mathrm{Hill},j}\right\}\,,\\
b_{90,j} &= \frac{G (m_{j}+m_\mathrm{c})}{v_{\mathrm{rel},j}^2} \label{eq:b90}\,,\\
b_{z,j} &= \min(b_{{xy},j}, h_{\mathrm{eff}})\,,\label{eq:bz}\\
h_{\rm eff}&=\max (h_{z,j},h_{\rm c})\,,\label{eq:heff}
\end{align}
where the square root term accounts for gravitational focusing by the binary's center of mass calculated in the limit that it dominates over the gravity of the SMBH inside the binary's Hill sphere. We conservatively neglect binary-single interactions with objects that have a larger impact parameter than $r_{\mathrm{Hill},j}$.\footnote{Note further that $s_j<r_{\mathrm{Hill},j}$ holds for stable binaries.} In Eq.~\eqref{eq:bz}, we also limit the maximum impact parameter due to the lack of objects moving at the elevation above $h_{\rm eff}$.

When the binary is embedded in the AGN disk, 
it interacts with both the spherical stellar and the disk stellar and BH components, and 
otherwise interacts only with the spherical stellar component. 
We reduce the number of stars in the disk stellar component in the $l^{\rm th}$ cell hosting the binary by one for each BH
binary that experiences a hard binary-stellar single interaction with objects in the disk stellar component.

\subsubsection{Weak gravitational scattering}
\label{sec:sss}

Weak gravitational scattering is the velocity exchange due to encounters between single objects or between the scattering of the center of mass of a binary with a single object.
Based on the Fokker-Planck approximation for an infinite homogeneous medium (Eq.~7.92 in \citealt{Binney08}), we assume that the mean acceleration due to weak gravitational scattering is
\begin{align}
\label{eq:ss_scatter2}
\bm{a}_{\mathrm{WS},k}=
&-p_{\mathrm{c},k}
\frac{4\pi G^2(m_k+m_\mathrm{c}) m_\mathrm{c} n_{\mathrm{c}} \ln\Lambda'}{\sigma_\mathrm{c}^2}
     g(X_k)\, \bm{\hat{v}}_{k} \nonumber\\
&+ p_{\mathrm{c},k} \left[
\frac{4\sqrt{2}\pi G^2 m_\mathrm{c}^2 n_{\mathrm{c}} 
\ln\Lambda'}{\Delta t \sigma_\mathrm{c}}
    \frac{\mathrm{erf}(X_k)}{X_k}
    \right]^{1/2}
    \,\bm{\hat{n}} 
  \end{align}
where $p_{\mathrm{c},k}$ sets the fraction of time that the object spends within a scale height for each component, respectively (Eq.~\ref{eq:pc_disk}), the first term corresponds to dynamical friction, and the second term represents Brownian motion\footnote{The change in the velocity due to this term after a timestep $\Delta t$ is $\Delta v_{k} \propto a_{\mathrm{WS},k}\Delta t \propto \Delta t^{1/2}$ as expected for diffusion. This term is equivalent to $v_{k}^{-1}\langle \Delta v_{k}^2 \rangle/\Delta t$, but this representation is numerically more stable for $v_{k}\ll \sigma_{\rm c}$.}, $\bm{\hat{v}}_{k}\equiv\bm{v}_{k}/v_{k}$,  $\bm{\hat{n}}$ is a unit vector in a random direction\footnote{Note that we assume an isotropic diffusion and averaged Eq. (7.92) in \citet{Binney08} over the direction.}, $X_k\equiv|\bm{v}_k-\bm{v}_\mathrm{c}|/(\sqrt{2}\sigma_\mathrm{c})$, $v_\mathrm{c}$ is the mean velocity of the medium in the comoving Keplerian frame\footnote{For the BH and stellar disk components $v_\mathrm{c}=0$, while for the stellar spherical component $v_\mathrm{c}=v_{\rm Kep}(r_k)$.}, 
$\mathrm{erf}(X)$ is the error function, 
\begin{equation}
    g(X) = \frac{1}{2X^2}\left[ \mathrm{erf}(X) - \frac{2X}{\sqrt{\pi}}e^{-X^2} \right]\,.
\end{equation}
and $\ln\Lambda'$ is the Coulomb logarithm. We set $\Lambda'=h_\mathrm{c}/b_{90,k} $ following \citet{Papaloizou00}, where $b_{90,k}$ given by Eq.~\eqref{eq:b90} is the impact parameter at which the direction of particles change by 90 degree during the encounter. 
Eq.~\eqref{eq:ss_scatter2} is valid when $\Lambda'>1$. 
On the other hand, 
when $h_\mathrm{c}<b_{90,k} $, the scattering is approximately confined within a two-dimensional plane. In this case, we assume that the direction of the acceleration is along the $xy$-plane, 
and set the acceleration according to the Fokker-Planck results for objects confined to an infinite homogeneous 2D medium (Kocsis, Tagawa, Haiman, in prep.)

\begin{align}
\bm{a}_{\mathrm{WS},k}=
p_{\mathrm{c},k}\left\{D[\Delta v_{\parallel}] \bm{\hat{v}}_{k} + 
\sqrt{\frac{D[\Delta v_{\perp}^2]+D[\Delta v_{\parallel}^2] }{\Delta t} }\bm{\hat{n}}_{xy}\right\},
\end{align}
where $\bm{n}_{xy}$ is a unit vector in a random direction in the $xy$ (AGN disk) plane and the dynamical friction and Brownian motion terms are
\begin{align}
    \label{eq:dv_3}
    D[\Delta v_{\parallel}] 
    =& - (2\pi)^{3/2} G\Sigma_{\rm c} x^{1/2} e^{-x}[I_0(x)+I_1(x)]
    \\
    \label{eq:dvdv_3}
    D[\Delta v_{\perp}^2] =& D[\Delta v_{\parallel}^2] = (2\pi)^{3/2}  \frac{G \Sigma_{\rm c} \sigma_{\rm c} m_{\rm c}}{m+m_{\rm c}}
      x e^{-x}
      \nonumber\\&\times
    \left[
     \left(1+\frac{1}{2x}\right)\mathrm{I}_0(x) + \mathrm{I}_1(x)
    \right]
    \,,
\end{align}
where $x=v_{\rm c}^2/(4\sigma_{\rm c}^2)$ and $I_0(\cdot)$ and $I_1(\cdot)$ are modified Bessel functions. In practice we resort to the approximations in Eqs.~\eqref{eq:dv_app_12}--\eqref{eq:dv_app_34} of  Appendix \S~\ref{sec:scattering_twod}.

To simply incorporate the effect of the recoil on a third body due to dynamical friction by the disk BH component, we give the same kinetic energy ($\sum_{k\in l}{\bm K}_{\mathrm{WS-DF},k}/N_{\mathrm{DBH},l}$) to the BHs which are in the AGN disk in the same cell as 
\begin{equation}
\label{eq:back_ws}
   {\bf a}_{\mathrm{WS},k}=\frac{1}{\Delta t}
    \left( \frac{\sum_{k\in l}{\bm K}_{\mathrm{WS-DF},k}}{m_k N_{\mathrm{DBH},l}} \right)^{1/2}
\end{equation}
where ${\bm K}_{\mathrm{WS-DF},k}$ is the kinetic energy added onto $k^{\rm th}$ objects due to the dynamical friction by scattering with the disk BH component.

\subsubsection{Binary formation via three-body interaction}
\label{sec:3bf}

If three bodies make a close encounter, a binary can form dynamically \citep[e.g.][]{Aarseth76,Binney08}. 
This process can be efficient in very dense or low velocity dispersion stellar environments. 

For the formation rate of binaries we extend the rough derivation of Eq. (7.111) in \citet{Binney08} to account for the limitations of a disk geometry and the destructive effect of the SMBH. We assume that binaries can form only when the three bodies undergo a strong encounter. We require to have an encounter between an object $i$ and a second object with impact parameter less than 
\begin{equation}\label{eq:bi}
b_i=\min( b_{{90},i}, r_{\mathrm{Hill},i})    
\end{equation}
in a region where there is a third object. Here $b_{{90},i}$ is given by (Eq.~\ref{eq:b90}).\footnote{
A binary can also form if $b_i>b_{{90},i}$, but in this case the initial binary separation is typically large, and the binaries are soon disrupted by a soft binary-single interaction \citep{Aarseth76}. For simplicity, we ignore such soft-binary formation. }
For a given BH $i$, the mean rate of strong single-single scattering encounters is given by $p_{\mathrm{c},i} n_\mathrm{c} b_i b_{z,\mathrm{eff},i} v_\mathrm{rel,i}$, where $b_{z,\mathrm{eff},i}=\min(b_i,h_{\mathrm{eff},i})$ and $h_{\mathrm{eff},i}$ is given by Eq.~\eqref{eq:heff}.
In each encounter, the probability that a single third BH lies within a distance $b_i$ is of order $n_\mathrm{c} b_i^2 b_{z,i}$ \citep{Binney08}, where $b_{z,i}=\min(b_i,h_{\mathrm{c}})$. 
Thus the mean rate for two single bodies to come within a distance $b_i$ to the $i^{\rm th}$ BH is 
\begin{eqnarray}
\label{eq:three_body}
\Gamma_{\mathrm{3bbf},i}
=\frac{1}{3} p_{\mathrm{c},i} n_\mathrm{c}^2 b_i^3 b_{z,\mathrm{eff},i} b_{z,i}v_{\mathrm{rel},i}
\end{eqnarray}
\citep{Binney08}. 
The factor of $1/3$ in Eq.~(\ref{eq:three_body}) compensates for triple-counting each three-body pair.

To get a rough understanding of the rate of binary formation per BH via three-body encounters, let us evaluate the order of magnitude of Eq.~\eqref{eq:three_body}, assuming that the mass of objects in the nuclear star cluster is $M_{\rm NSC}$ within an effective radius $r_{\rm NSC}$, 
with the density profile in Eq.~\eqref{eq:bh_density}.
We approximate the BH density as 
\begin{align}n_\mathrm{BH}=n_0 \left( \frac{r}{r_\mathrm{NSC}}\right)^{\gamma_\rho - 2 },\quad
n_0 = f_0 \frac{\gamma_{p}}{4\pi} \frac{\eta_\mathrm{n,BH} M_{\rm NSC}}{r_\mathrm{NSC}^{3}}\frac{r}{h}, 
\label{eq:bh_density2}
\end{align}
where $\gamma_\rho=0$, $f_{\mathrm{DBH}}$ 
is the fraction of BHs in the AGN disk\footnote{Note that all BHs have their own orbital height in the simulation, which changes in time. Here $f_{\mathrm{DBH}}$ labels the objects whose height is at most that of the AGN disk, which is often higher than $10\%$ of the total population.
}, 
$v_\mathrm{rel}\sim \sqrt{3}v_\mathrm{Kep}(r)\,h/r$ and for simplicity we approximated all height-related quantities with the disk scaleheight $h$ assuming that $h \equiv h_c\sim h_{z,i}\sim h_{\mathrm{eff},i}$ in the radial cell of the object $i$. 
Furthermore, we use the empirical $M_{\rm SMBH}-\sigma$, $M_{\rm NSC}-\sigma-r_{\rm NSC}$ relations given below in Eqs.~\eqref{eq:Msmbh-sigma}, \eqref{eq:Mnsc-sigma}, \eqref{eq:Mnsc-r} and the number of BHs per unit stellar mass is assumed to be  $\eta_\mathrm{n,BH}=0.002\,\Msun^{-1}$ in Eq.~\eqref{eq:etaBH}. This leads to
\begin{align}
    {\Gamma}_{\mathrm{3bbf},i} &\sim  n_\mathrm{c}^2 \min(b_{90,i}^3,r_{\mathrm{Hill},i}^3) \min(b_{90,i}^2,r_{\mathrm{Hill},i}^2, h^2) v_{\mathrm{rel},i}
    \nonumber\\ 
    &=
    \min({\Gamma}^{(1)}_{\mathrm{3bbf},i},{\Gamma}^{(2)}_{\mathrm{3bbf},i},{\Gamma}^{(3)}_{\mathrm{3bbf},i},{\Gamma}^{(4)}_{\mathrm{3bbf},i})
\end{align}
where the four cases are 
\begin{align}
\label{eq:three_1}
   {\Gamma}^{(1)}_{\mathrm{3bbf},i} &\sim 
   n_\mathrm{BH}^2 
   b_{90,i}^5 v_{\mathrm{rel},i}=
   n_\mathrm{BH}^2 
   \frac{2^5 G^5 m_i^5}{v_{\mathrm{rel},i}^9}
    \nonumber \\
    &\sim 7\,\mathrm{Myr}^{-1} 
    r_{1\rm pc}^{1/2}
      \left(\frac{h/r}{0.01} \right)^{-11}
    m_{i,10}^5 M_{\rm SgrA}^{-3.85}
    f_{\mathrm{DBH}}^2\,,
    \nonumber \\
    &\qquad\mathrm{if}~b_{90,i} \leq \min(h,r_{\mathrm{Hill},i})\,,\\
\label{eq:three_2}
    {\Gamma}^{(2)}_{\mathrm{3bbf},i} &\sim n_\mathrm{BH}^2
    r_{\mathrm{Hill},i}^5 v_{\mathrm{rel},i} 
    \nonumber \\
    &\sim 0.2\,\mathrm{Myr}^{-1} 
    r_{1\rm pc}^{1/2}
  \left(\frac{h/r}{0.01} \right)^{-1}
   m_{i,10}^{5/3}
    M_{\rm SgrA}^{-0.51}
    f_{\mathrm{DBH}}^2\,,
    \nonumber \\
    &\qquad\mathrm{if}~r_{\mathrm{Hill},i} \leq h \leq b_{90,i}\,,\\
\label{eq:three_3}
    {\Gamma}^{(3)}_{\mathrm{3bbf},i} &\sim n_\mathrm{BH}^2
     r_{\mathrm{Hill},i}^3 h^2 v_{\mathrm{rel},i} 
    \nonumber \\
    &\sim 0.2\,\mathrm{Myr}^{-1} 
    r_{1\rm pc}^{1/2}
    \left(\frac{h/r}{0.01} \right)
   m_{i,10}
   M_{\rm SgrA}^{0.153}
   f_{\mathrm{DBH}}^2\,,
        \nonumber \\
    &\qquad\mathrm{if}~ h\leq r_{\mathrm{Hill},i} \leq b_{90,i}\,,\\
\label{eq:three_4}
    {\Gamma}^{(4)}_{\mathrm{3bbf},i} &\sim 
       n_\mathrm{BH}^2 
   b_{90,i}^3 h^2 v_{\mathrm{rel},i}=
    n_\mathrm{BH}^2
      \frac{2^3 G^3 m_i^3}{v_{\mathrm{rel},i}^5} h^2 
    \nonumber \\
    &\sim 2\,\mathrm{Myr}^{-1} 
    r_{1\rm pc}^{1/2}
    \left(\frac{h/r}{0.01} \right)^{-5}
    m_{i,10}^3
    M_{\rm SgrA}^{-5.85}
    f_{\mathrm{DBH}}^2\,,
        \nonumber \\
    &\qquad\mathrm{if}~h\leq b_{90,i} \leq r_{\mathrm{Hill},i}\,,
\end{align}
where we used the abbreviated labels defined under Eq.~\eqref{eq:dt_dec}.
Thus the three-body binary formation rates are greatly increased in the AGN disk since gas reduces the velocity dispersion of the particles. This leads to a strong dependence on $h/r$ in Eq.~\eqref{eq:three_1}. However, since 
\begin{align}
\frac{r_\mathrm{Hill}}{r} = 0.0094\,\frac{m_{i,10}^{1/3}}{M_{\rm SgrA}^{1/3}}\,, \quad
\frac{b_{90}}{r} \simeq 0.02\,\frac{m_{i,10}}{M_{\rm SgrA}} \left(\frac{h/r}{0.01} \right)^{-2}\,,
\end{align}
and typically $h/r<h_\mathrm{AGN}/r\sim 0.01$--$0.001$, 
Eq.~\eqref{eq:three_3} represents the most typical case of the binary formation rate via three-body encounters after the BHs are captured by the AGN disk. 
Eqs.~\eqref{eq:three_1}--\eqref{eq:three_4} show that the binary formation rate is high compared to the AGN lifetime; BHs are expected to form binaries outside of $0.01\,\rm pc$. 

We form a binary due to this dynamical three-body binary formation mechanism in the simulation with probability $P_{\mathrm{3bbf},i}=\Gamma_{\mathrm{3bbf},i}\Delta t$ using Eq.~\eqref{eq:three_body}. We assume that the initial separation of the newly formed binary is $s_i\equiv b_i$ (Eq.~\ref{eq:bi}). This approximation is justified by the fact that in reality the distribution scales with $s^{9/2}$ for $s\lesssim b_{90}$ and it is exponentially suppressed above $b_{90}$ \citep[see Eq. 7.175 in][]{Binney08}.

When the binary formation criteria is satisfied for some BH $i$, 
we search for a binary counterpart $i'$ in the same cell $l$. 
The binary mass is given by $m_i+m_{i'}$, and the binary has a velocity of ${\bf v}_{j}={\bf v}_\mathrm{kick}+{\bf v}_\mathrm{cen}$,
where ${\bf v}_\mathrm{cen}=(m_i {\bf v}_i+m_{i'} {\bf v}_{i'}+m_\mathrm{c} {\bf \sigma}_\mathrm{c}\bm{\hat{n}} )/(m_i+m_{i'}+m_\mathrm{c})$ is the center of mass velocity of the three body system, 
and 
$\bm{v}_\mathrm{kick}=[m_\mathrm{c}/(m_i+m_{i'}+m_\mathrm{c})](Gm_i m_{i'}/d)\bm{\hat{v}}_\mathrm{kick}$
is the kick velocity due to the binary formation, where $\bm{\hat{n}}$ and $\bm{\hat{v}}_\mathrm{kick}$ are unit vectors drawn randomly from the isotropic distribution.

We also consider cases in which the third body is a member of the disk stellar component. 
In this case, we substitute $(n_\mathrm{c})_{\mathrm{DBH},l} (n_\mathrm{c})_{\mathrm{Dstar},l}$ into $n_\mathrm{c}^2$ in Eq.~(\ref{eq:three_body}), which further increases the binary formation rate compared to Eqs.~\eqref{eq:three_1}--\eqref{eq:three_4}.
We assume that all newly formed binaries are BH-BH binaries, 
and due to the steep scaling of the formation rate with mass we do not consider the formation of BH-stellar binaries or stellar-stellar binaries in this study.
Formation of BH-stellar or stellar-stellar binaries may further increase the rate of BH-BH mergers since such binaries are plausibly exchanged to BH-BH binaries after several binary-single interactions with BHs.

\subsubsection{Gas-capture binary formation}
\label{sec:gbs}

Generally, if any form of dissipation removes a sufficient amount of energy during the passage of two bodies inside their mutual Hill sphere, the two objects may not travel back to infinity following the encounter, and a binary forms.
In the AGN disk, gas dynamical friction serves as a dissipation mechanism. 
\citet{Goldreich02} confirmed that 
the fraction of two bodies passing within their Hill radius which becomes bound ($P_{\mathrm{cap},i}$, C in \citealt{Goldreich02}) is roughly coincident with 
the fractional decrease in the binding energy 
during a passage through the mutual Hill sphere 
(D in \citealt{Goldreich02}). 
Assuming that the latter is approximated as $t_{\mathrm{pass},i}/t_{\mathrm{GDF},i}$, 
where 
\begin{equation}
t_{\mathrm{pass},i}=\frac{r_{\mathrm{Hill},i}}{v_{\mathrm{rel},i}}
\end{equation}
is the crossing timescale across the Hill radius, and 
\begin{equation}
t_{\mathrm{GDF},i}=\frac{v_{\mathrm{rel},i}}{a_{\mathrm{GDF},i}(v_{\mathrm{rel},i})}
\end{equation}
is the timescale of damping of the relative velocity between two bodies by gas dynamical friction (Eq.~\ref{eq:gdf}). 
Hence the probability for binary formation by the gas-capture mechanism during a single BH encounter, is 
\begin{equation}
P_{\mathrm{cap},i}=\mathrm{min}(1,t_{\mathrm{pass},i}/t_{\mathrm{GDF},i}), 
\end{equation}
The binary formation timescale due to the gas-capture mechanism is 
\begin{equation}
\label{eq:t_cap}
t_{\mathrm{cap},i}=\frac{t_{\mathrm{enc},i}}{P_{\mathrm{cap},i}}\,, 
\end{equation}
where 
\begin{equation}\label{eq:t_enc}
t_{\mathrm{enc},i}^{-1}=p_{\mathrm{c},i} n_\mathrm{DBH} r_{\mathrm{Hill},i} z_{\mathrm{Hill},i} v_{\mathrm{rel},i}
\end{equation}
defines the rate of objects within the disk black hole component to enter the Hill sphere of BH $i$, where
\begin{equation}
z_{\mathrm{Hill},i}=\min(r_{\mathrm{Hill},i}, h_{\mathrm{eff},i}) 
\end{equation}
is the maximum height within the Hill's sphere, $h_{\mathrm{eff},i}$ is given by Eq.~\eqref{eq:heff}
Thus the binary formation rate for the $i^{\rm th}$ BH is $\Gamma_{\mathrm{gas},i}=\frac12 t_{\mathrm{cap},i}^{-1}=\frac12 t_{\mathrm{enc},i}^{-1} P_{\mathrm{cap},i}$, where the $1/2$ factor compensates for double-counting each two-body pair. 
The probability of binary formation for the $i^{\rm th}$ BH within the simulation timestep is $P_{\mathrm{gas},i}=\Delta t/(2 t_{\mathrm{cap},i})$.
Since the gas-capture mechanism is caused by gas dynamical friction, we assume that it operates only within the AGN disk (\S \ref{sec:df}) 
and when radiation feedback is inefficient 
(\S \ref{sec:df}). 
We set the initial separation to be $s_j=r_{\mathrm{Hill},i}$. 
As in the case of the three-body binary formation, 
after the binary formation criteria are satisfied for some BH $i$, we search for a binary counterpart $i'$ in the same cell $l$ in the AGN disk. The binary mass is given by $m_i+m_{i'}$, and its velocity by $v_{j}=v_i$ since the velocity of a counterpart $i'$ is dissipated during the passing time. 

\begin{table*}
\begin{center}
\caption{Fiducial values for the model parameters}
\label{table_parameter}
\hspace{-5mm}
\begin{tabular}{c|c}
\hline 
Parameter & Fiducial value \\
\hline\hline
Mass of the central SMBH & $M_\mathrm{SMBH}=4\times 10^6\,\Msun$ \\\hline
Gas accretion rate from the outer radius & ${\dot M}_\mathrm{out}=0.1\,{\dot M}_\mathrm{Edd}$\\\hline
Fraction of pre-existing binaries & $f_\mathrm{pre}=0.15$ \\\hline
Power-law exponent for the initial density profile for BHs & $\gamma_{\rho}=0$ \\\hline
Parameter setting the initial velocity anisotropy for BHs & $\beta_\mathrm{v}=0.2$ \\\hline
Efficiency of angular momentum transport in the $\alpha$-disk & $\alpha_\mathrm{SS}=0.1$ \\\hline
Stellar mass within 3 pc &$M_\mathrm{star,3pc}=10^7\,\Msun$\\\hline 
Initial mass function slope & $\delta_\mathrm{IMF}=2.35$\\\hline
Angular momentum transfer parameter in the outer disk &$m_\mathrm{AM}=0.15$\\\hline
Accretion rate in Eddington units onto stellar-mass BHs &$\Gamma_\mathrm{Edd,cir}=1$\\\hline
Numerical time-step parameter &$\eta_t=0.1$\\\hline
Number of radial cells storing physical quantities &$N_\mathrm{cell}=120$\\\hline
Maximum and minimum $r$ for the initial BH distribution& $r_\mathrm{out,BH}=3$ pc, $r_\mathrm{in,BH}=10^{-4}$ pc \\\hline
\end{tabular}
\end{center}
\end{table*}

Let us estimate the rough timescale for gas-capture binary formation as in $\S~\ref{sec:3bf}$. 
The capture fraction $P_{\mathrm{cap},i}=\mathrm{min}(1,t_{\mathrm{pass},i}/t_{\mathrm{GDF},i})$ is $\sim 1$ in the ranges we calculate ($10^{-4}\,\mathrm{pc}<r<5\,\mathrm{pc}$) as
\begin{align}
\label{eq:gas_capture_time_ratio}
\frac{t_{\mathrm{pass},i}}{t_{\mathrm{GDF},i}} \sim 9.9\times 10^5 r_{1\rm pc}^{4}
    \left(\frac{h/r}{0.01} \right)^{-4}
    m_{i,10}
    M_{\rm SgrA}^{-4/3}
    \rho_\mathrm{\rm gas,6}
    \,,
\end{align}
where $\rho_\mathrm{\rm gas,6}\equiv\rho_\mathrm{gas}/(10^6\,\mathrm{\Msun\,pc^{-3}})\propto r_{1\rm pc}^{-3}$ in the fiducial model according to Figure~\ref{fig:disk_model}, 
and we assume $f\sim 1$ in $a_\mathrm{GDF}$. 
The rate of gas-capture binary formation per BH is roughly estimated as 
\begin{align}
\label{eq:gas_binary_rate}
\Gamma_{\mathrm{cap},i}&\sim 
t_{\mathrm{cap},i}^{-1}\sim t_{\mathrm{enc},i}^{-1}
\sim 7\,\mathrm{Myr}^{-1}  
    r_{1\rm pc}^{-1/2}
    \frac{h/r}{0.01}
    m_{i,10}
    M_{\rm SgrA}^{0.494}
    f_0
    \,,
\end{align}
where we use $h<r_\mathrm{Hill}$. 
By comparing Eq.~\eqref{eq:three_3} with Eq.~\eqref{eq:gas_binary_rate}, $\Gamma_{\mathrm{3bbf},i}\propto r_{1\rm pc}^{1/2}$, $\Gamma_{\mathrm{cap},i}\propto r_{1\rm pc}^{-1/2}$, and $\Gamma_{\mathrm{cap},i}\sim 30 \,\Gamma_{\mathrm{3bbf},i}$ at $r=1\,\mathrm{pc}$, 
we conclude that the rate of gas capture binary formation dominates the rate of three-body binary formation.

If the center of mass of a binary makes a close encounter with a third object, it may undergo a gas-capture interaction to form a hierarchical triple. 
In such a case, the Kozai-Lidov effect may facilitate the merger of the inner binary \citep[e.g.][]{Silsbee17,Liu18}. Additionally, perturbations caused by density inhomogeneities may lead to a close binary-single encounter, which may contribute to the hardening of the binary. We conservatively neglect these merger pathways in the simulation. This subject merits future investigation.

\subsubsection{Time-step}

We use a shared time-step 
\begin{eqnarray}
\label{eq:time_step}
\Delta t=\eta_t\, \mathrm{min}_k\Big(
\frac{s_k}{{\rm d}s/{\rm d}t|_{\mathrm{GW},k}},
\frac{s_k}{{\rm d}s/{\rm d}t|_{\mathrm{typeII},k}},\nonumber\\ 
\frac{s_k}{{\rm d}s/{\rm d}t|_{\mathrm{GDF},k}},
t_{\mathrm{BS},k}, t_{\mathrm{GBS},k},\nonumber\\ 
\frac{r_k}{\mathrm{d}r_k / \mathrm{d}t|_\mathrm{typeI/II}}
\Big),~~~~~
\end{eqnarray}
where $\eta_t=0.1$ is a constant in the fiducial model,
but is varied below.

\subsubsection{Merger prescription}
\label{sec:merger_pres}

Since $N$-body particles represent stellar-mass BHs, a merger occurs  when the separation of a binary becomes smaller than the sum of the innermost stable circular orbits ($6Gm_j/c^2$) of the binary components.
During a BH merger, the merged remnant recoils due to anisotropic GW emission. The GW recoil kick velocity depends on the mass ratio, spin magnitude and spin direction of the merged BHs \citep[e.g.][]{Baker07,Campanelli07,Koppitz07,Herrmann07}. 
In the fiducial model, we assume that the BH spins are zero, 
and set the recoil velocity to
\begin{equation}
    v_\mathrm{GW}=8,830 \,\mathrm{km/s}\,\frac{q^2(1-q)}{(1+q)^5}
\end{equation}
whose functional form and normalization are derived by Post-Newtonian predictions and fits to numerical relativity simulations, respectively, 
where $q$ is the mass ratio of merged BHs \citep{Baker07}. 
This formula gives a maximum kick velocity of $v_\mathrm{GW} \simeq 160\,\mathrm{km/s}$ at $q\sim 0.4$. However, since the kick velocity is very sensitive to the BH spins, we also investigate cases with constant kick velocity of 
$v_\mathrm{GW}=400$, 600, and 1000 km/s in Models~26, 27, and 28, respectively. 
We add the recoil velocity to the velocity of the merged remnant $v_i$, and the direction is assumed to be random and isotropically distributed.  
We set the mass loss due to GW radiation at mergers assuming zero spins for BHs as 
\begin{eqnarray}
m_\mathrm{GW}=m_j (1-0.2\eta) 
\end{eqnarray}
\citep{Tichy08}, where $\eta=q/(q+1)^2$ is the symmetric mass ratio.

\begin{figure*}
\includegraphics[width=170mm]{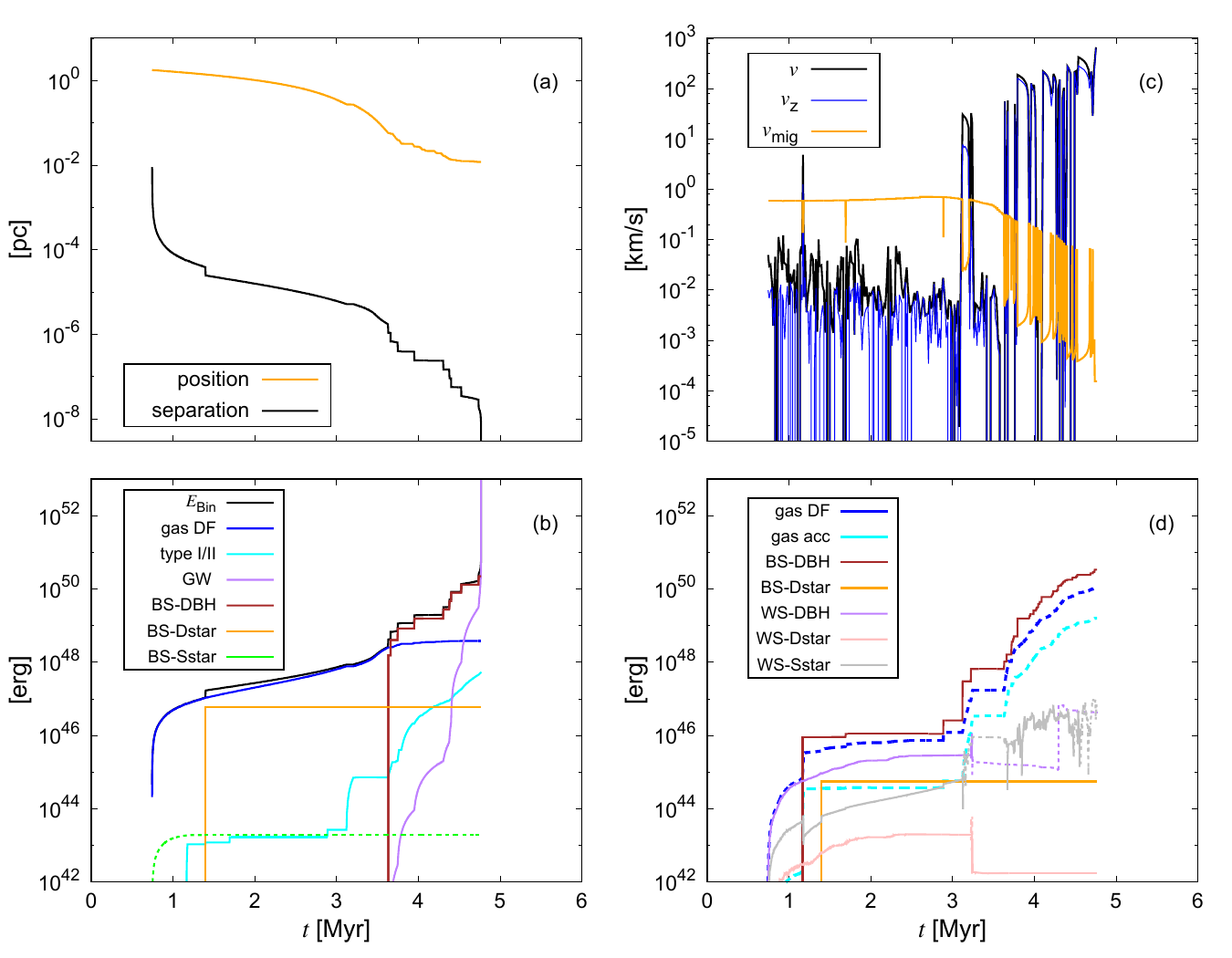}
\caption{
Time evolution of the physical quantities describing a typical
binary that formed in the fiducial model by the gas-capture mechanism. 
(a): The evolution of the binary separation (black) and the radial distance from the SMBH (orange). 
(b): Black line shows the evolution of the binding energy of the binary. The other lines show the cumulative contributions of
gas dynamical friction (blue), 
type I/II torque (cyan), 
gravitational wave radiation (purple), 
and binary-single interactions with the disk BH (brown) and stellar (orange) and spherical stellar (green) components. 
Negative values are indicated
by dashed curves in panels (b) and (d). 
(c): The evolution of the velocity of the center of the binary relative to the local motion of the AGN disk $v_j$ (black), the $z$-direction velocity $v_{z,j}$ (blue), and the migration velocity (orange). 
(d): The cumulative change in the binary center of mass's kinetic energy due to 
dynamical friction by gas (dashed blue), 
gas accretion torque (dashed cyan), 
binary-single interactions with the disk BH (brown) and disk stellar (orange) components, 
and weak gravitational scattering with disk BHs (purple), disk stars (pink), and the spherical stellar (gray) components. 
}
\label{fig:tot_cont}
\end{figure*}

\section{Results}
\label{sec:results}

We start this section by describing an illustrative example of a binary that formed and merged in our fiducial model (\S~\ref{sec:evolution_binaries}). We then present the demography and various physical properties of the entire binary population in the fiducial model (\S~\ref{sec:number-evolution} and \ref{sec:contribution}), and discuss how the most important quantities change when model parameters are varied (\ref{sec:dependence}).

\subsection{Formation and evolution of binaries}
\label{sec:evolution_binaries}

We first present the evolution of a binary in the fiducial model (labeled as Model~1 in Table~\ref{table_results}, whose parameter values are listed in Table~\ref{table_parameter}) in Figure~\ref{fig:tot_cont}.  The binary in this figure forms at $0.74$ Myr with an initial separation of $8.9\times 10^{-3}$ pc at $r_j = 1.8$ pc by the gas-capture mechanism. 
The masses of binary components are $9.9\,\Msun$ and $9.1\,\Msun$. 
In the early phase, the binary separation decreases 
significantly due to 
gas dynamical friction of the AGN disk (blue line in panel (b)). 
The binary also migrates toward the SMBH due to type I/II torque of the AGN disk (orange line in panel (a)). 
Affected by gas dynamical friction, accretion torque, and weak gravitational scattering, $v_j$ settles to a stochastic equilibrium between heating and damping (i.e. dashed blue, dashed cyan, purple, pink, and gray lines in panel (d)) and $v_j$ fluctuates stochastically around the equilibrium.
The binary experiences several hard binary-single interactions with a disk BH and star (brown and orange lines in panels (b) and (d)). 
During the interaction, the binary receives recoil kicks due to the binary-single interaction in a random direction (black and blue lines in panel (c) and brown and orange lines in panel (d)). In some of interactions, this binary is chosen as a third object for the binary-single interaction with other binaries. For simplicity, we assume that this third object receives a recoil kick, even though it is a binary (\S \ref{sec:bs}). 
Following a recoil kick, the binary's radial migration is delayed (orange line in panel (c)) since the binary moves out of the AGN disk. 
Then $v_j$ is damped by gas dynamical friction and the accretion torque of the AGN disk (dashed blue and cyan lines infrom  the panel (d)). 
After the binary migrates to $r_j\lesssim 6\times 10^{-2}$ pc, and the separation reaches $s_j \sim 10^{-6}\,\mathrm{pc} \sim 0.2\,\mathrm{AU}$
binary-single interactions become frequent with disk BHs (brown line in panel (b)).
After the binary is hardened to $s_j\sim 1.8\times 10^{-8}$~pc, GW radiation drives it to merge (purple line in panel (b) of Figure~\ref{fig:tot_cont}) at 
4.8 Myr. 
This binary merges outside of the AGN disk since the merger takes place soon after a binary-single interaction. 
During the evolution, $0.8\,\Msun$ and $0.9\,\Msun$ gas mass is accreted onto the primary and secondary BHs, respectively. 
In summary, this binary, formed via the gas-capture mechanism, it is initially mostly hardened by gas dynamical friction, and later by a series of binary-single interactions with other BHs in the AGN disk, and finally by GW radiation.

\begin{figure}
\includegraphics[width=90mm]{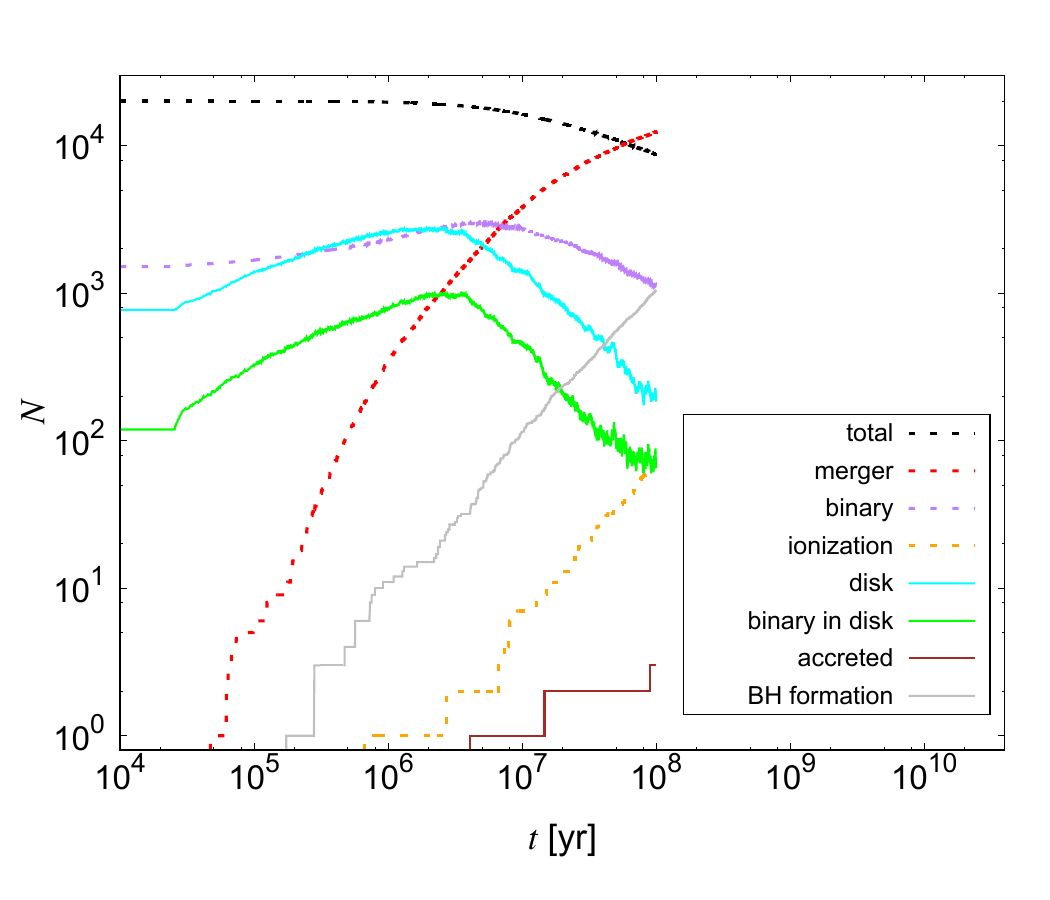}\caption{
The time evolution of the number of various types of objects in
the fiducial model:
the total number of BHs (dotted black), 
BH binaries (dotted purple) 
cumulative number of BH mergers (dotted red), 
the AGN disk's objects shown with solid lines including the BHs (cyan), BH binaries (green), and in-situ formed BHs (gray), 
and BHs that migrated within the inner radius $r_\mathrm{in}=10^{-4}$ pc (brown). 
The number of BHs is significantly reduced by frequent repeated mergers. 
The reduction of binaries due to disruption (orange dashed curve labeled ``ionization'') is relatively less important.
}\label{fig:num_ev}\end{figure}

\begin{figure}
\includegraphics[width=90mm]{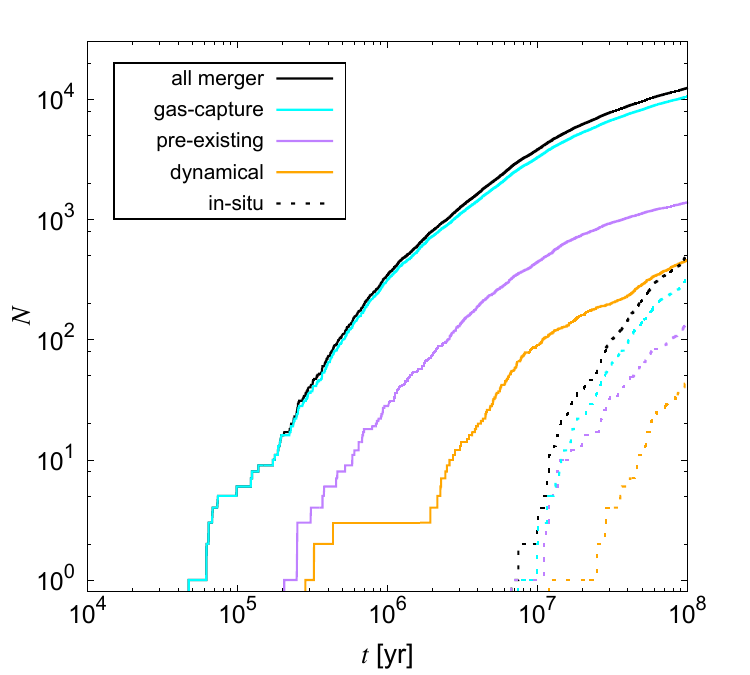}
\caption{
The time evolution of the cumulative number of mergers among binaries formed by different mechanisms in the fiducial model. 
The total number of mergers (black), 
mergers among gas-capture binaries (cyan), 
pre-existing binaries (purple), and 
dynamically formed binaries (orange) are shown by solid lines. 
Dashed lines represent the contribution from in-situ formed BHs. 
Mergers from gas-capture binaries dominate over other binary formation channels. 
}\label{fig:num_mer}\end{figure}

\subsection{Demography of the BH binary population}
\label{sec:number-evolution}

Figure~\ref{fig:num_ev} shows the evolution of the number of BHs (dotted black), mergers (dotted red), all binaries (dotted purple), 
BHs (cyan) and binaries (green) in the AGN disk, BHs that migrated within the inner radius $r_\mathrm{in}=10^{-4}$ pc (brown), and in-situ formed BHs (gray) in the fiducial model. 
Initially the number of binaries is $1.5\times 10^3$, the number of BHs and binaries in the AGN disk are $3.3\times 10^2$ and $35$, respectively. 
The number of BHs and binaries in the AGN disk increase 
for $\sim 3$ Myr (cyan and green lines), 
and then these quantities decrease due to the reduction of the number of BHs due to mergers. 
The reduction due to binary disruption (orange dashed curve labeled ``ionization'') is relatively less important.
The number of mergers and in-situ formed BHs keep increasing for 100 Myr (dashed red and gray lines). 
Up to $100$ Myr, $1.2\times 10^4$ mergers occur and $1.0 \times 10^3$ BHs are formed in-situ. 
This implies that some fraction of BHs merged several times within 100 Myr.

Figure~\ref{fig:num_mer} shows the evolution of mergers among binaries formed by the different mechanisms in the simulation. 
Mergers among gas-capture binaries, pre-existing binaries, and dynamically formed binaries start to take place from $\gtrsim 0.05$, 
$\gtrsim 0.2$, and $\gtrsim 0.3$ Myr (solid cyan, purple, and orange lines),
respectively. 
In the fiducial model, mergers among the binaries formed by the gas-capture mechanism dominate over the other formation channels (cyan line). 
This highlights the importance of gas-capture binary formation when discussing compact object mergers in AGN disks. 
Mergers among in-situ formed BHs contribute only $4.1 \%$ of the total number of mergers at 100 Myr (dashed black line). 
A fraction of 0.85, 0.11, and 
0.04 of mergers at 100 Myr are among gas-capture binaries, pre-existing binaries, and dynamically formed binaries (solid cyan, purple, and orange lines), and 0.64, 0.27, and 0.09
of mergers among in-situ formed BHs at 100 Myr are among gas-capture binaries, pre-existing binaries, and dynamically formed binaries (dashed cyan, purple, and orange lines), respectively. 

\begin{figure}
\includegraphics[width=90mm]{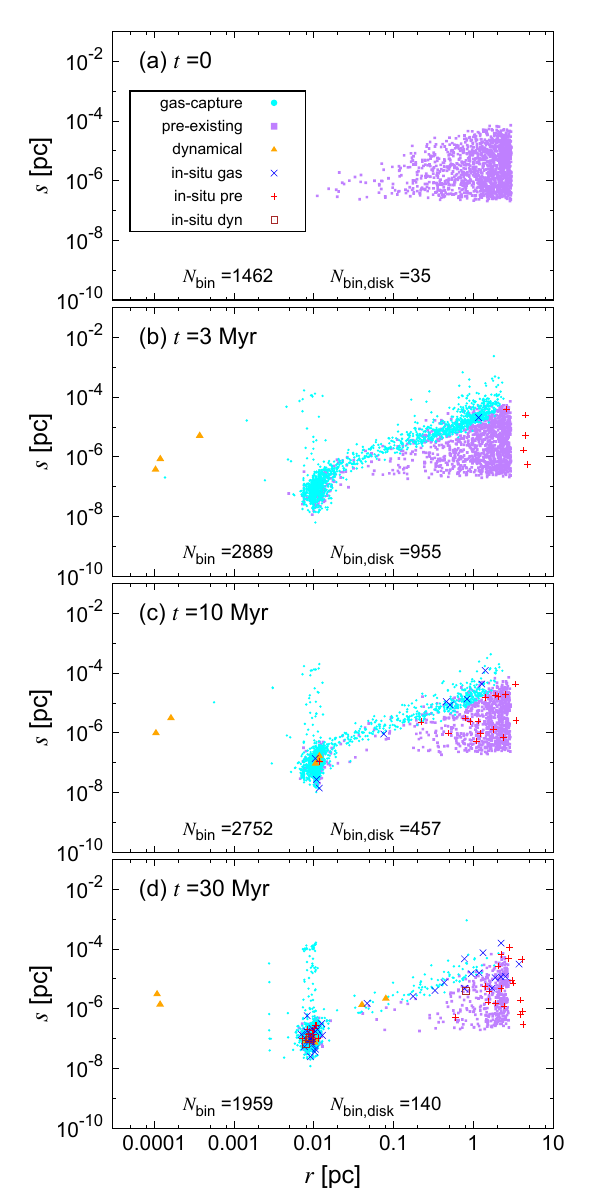}
\caption{
The distribution of binaries in binary separation and distance from the SMBH at four different times. 
Cyan circles, purple squares, and orange triangles represent gas-capture binaries, pre-existing binaries, and dynamically formed binaries, respectively.  
Blue crosses, red pluses, and brown squares represent gas-capture binaries, pre-existing binaries, and dynamically formed binaries for in-situ formed BHs, respectively. 
Panels (a), (b), (c), and (d) show the distribution at $t=0$, 3, 10, and 30 Myr, respectively. 
Binaries formed by different mechanisms 
have distinct spatial- and separation-distributions.
}
\label{fig:binaries}
\end{figure}

\begin{figure}
\includegraphics[width=90mm]{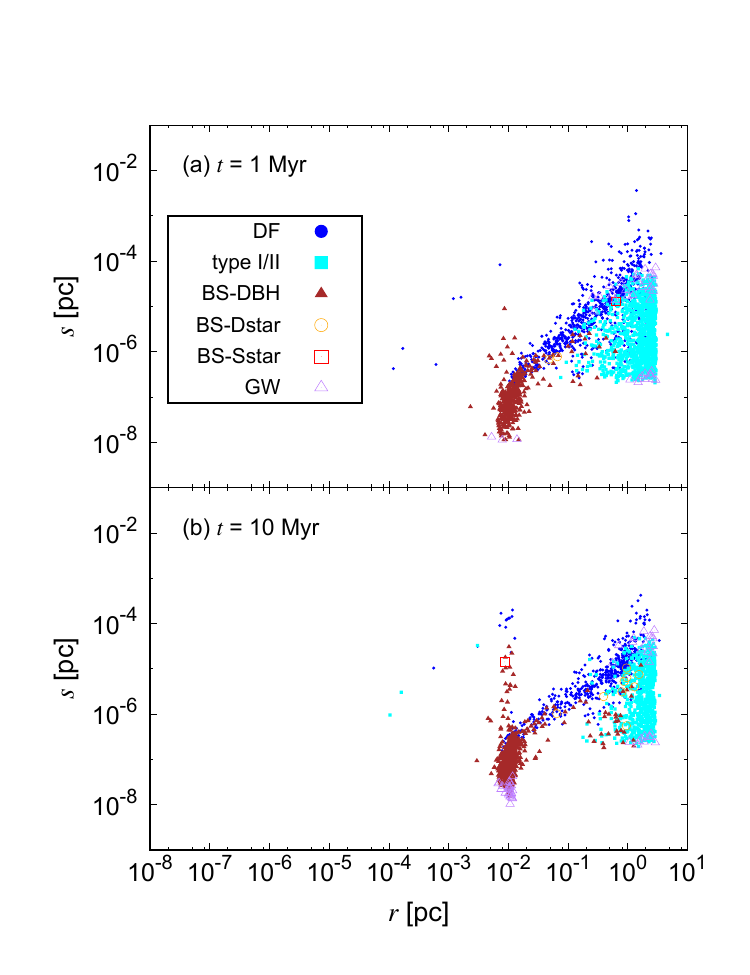}
\caption{
The dominant hardening mechanism is indicated for each binary in the fiducial model at 1 Myr (top panel) and 10 Myr (bottom panel) in the plane of the in-cluster location $r$ vs. the binary separation $s$. 
Different colored points represent binaries hardened mostly by GW (purple triangles), gas dynamical friction (blue circles), type I/II torque of a circumbinary disk (cyan squares), binary-single interaction with the disk BH (brown triangles) and stellar (orange circles) components and the spherical stellar component (red squares). 
Panels (a), and (b) show the distributions at $t=1$, and 10 Myr, respectively.
The dominant hardening mechanism changes with the binary's location. 
}\label{fig:edis}\end{figure}

\begin{figure}
\includegraphics[width=90mm]{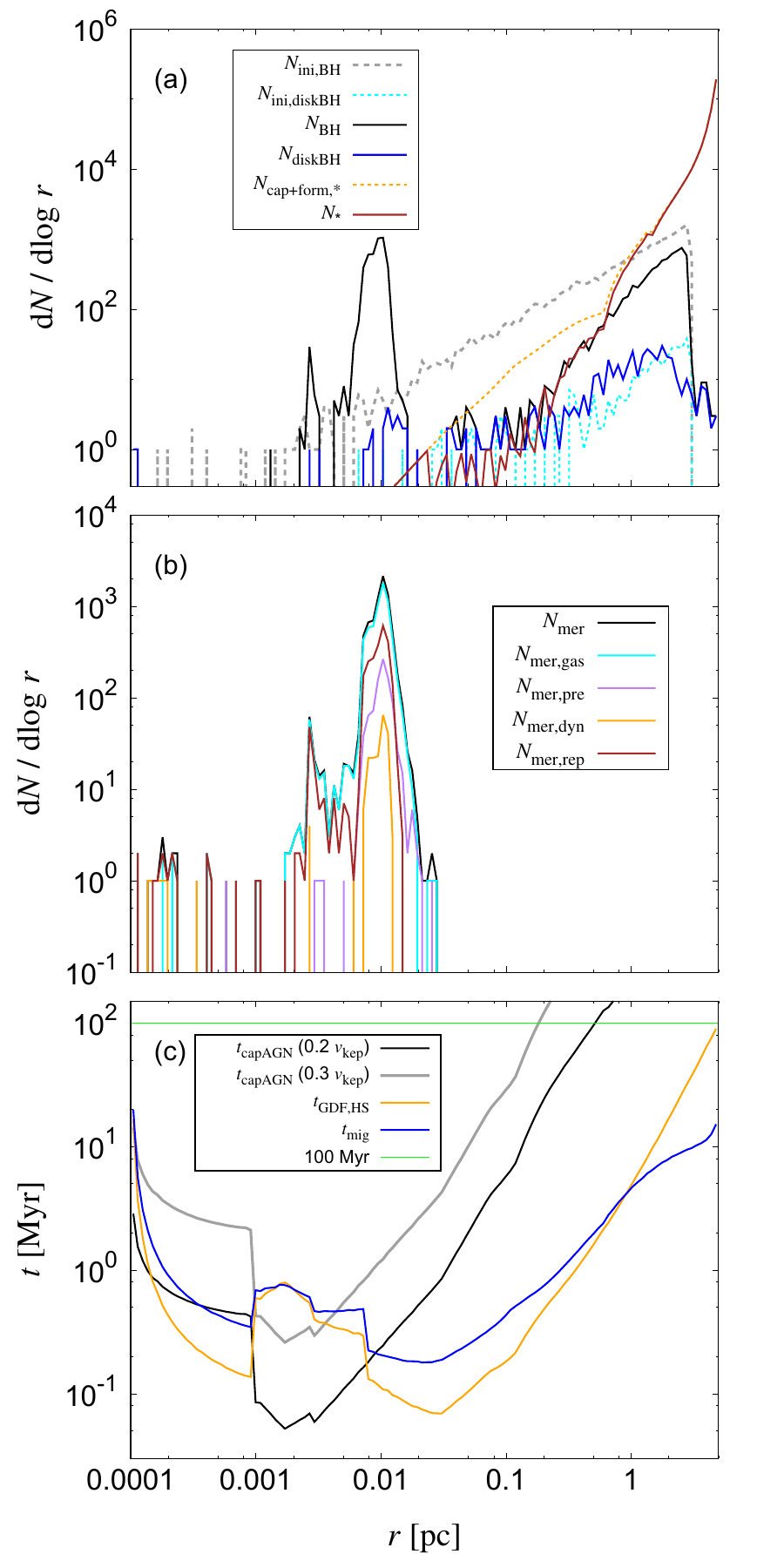}
\caption{
Radial distribution of several quantities measured in different simulation cells. 
(a): The number of BHs at $t$=0 (dashed gray) and 30 Myr (solid black), 
the number of BHs in the AGN disk at $t$=0 (dashed cyan) and 30 Myr (solid blue), 
and the number of stars in the AGN disk (solid brown) and stars which are formed and captured in the AGN disk (dashed orange) at 30 Myr. 
(b): The total number of mergers for all binaries (black), mergers of gas-capture formed binaries (cyan), 
pre-existing binaries (purple), and dynamically formed binaries (orange) at 30 Myr. 
(c): The relevant timescales for an $m_k=10\,\Msun$ binary as a function of radial distance from the SMBH. 
Lines represent the decay timescale of the BH velocity while crossing the disk due to gas dynamical friction ($t_\mathrm{capAGN}$, Eq.~\ref{eq:dt_dec}) for $v_{\mathrm{ini},k}=0.2\,v_\mathrm{Kep}$ (black) and $v_{\mathrm{ini},k}=0.3\,v_\mathrm{Kep}$ (gray), 
the timescale for binary hardening by gas dynamical friction to the hard-soft boundary ($t_\mathrm{GDF,HS}$, Eq.~\ref{eq:df_timescale}, orange), 
the migration timescale (Eq.~\ref{eq:typeI_II}, blue), 
and the typical maximum lifetime of AGN disks (100 Myr, cyan). 
Binaries migrate to $\lesssim$ pc before they become hard binaries. 
}\label{fig:grid_variables}\end{figure}

Figure~\ref{fig:binaries} shows the distribution of binaries in binary separation ($s_j$) and distance from the SMBH ($r_j$). 
At $t=0\,\mathrm{yr}$ (panel (a)), $1462$ pre-existing binaries are distributed according to the initial condition. The upper bound on $s$ arise from the assumption that soft binaries are disrupted prior to the AGN phase, and the lower bound on $s$ come from the assumption that the initial binary separation must be larger than the sum of the radii of the progenitor stars to survive without merging during the main sequence stellar phase.

Since gas-capture binaries form within the AGN disk, they spend a large fraction of their time within the AGN disk. 
Both the radial position ($r_j$) and the binary separation ($s_j$) evolve simultaneously for such binaries (Figure~\ref{fig:tot_cont}).

Figure~\ref{fig:edis} shows the dominant binary hardening mechanisms at two different times.
Gas-capture binaries (cyan circles in Figure~\ref{fig:binaries}) are hardened mostly by gas dynamical friction (blue circles in Figure~\ref{fig:edis}) before binaries migrate to $r \sim 10^{-2}$~pc and their separation reduces to $s \sim 10^{-6}\,\mathrm{pc}\sim 0.2\,\mathrm{AU}$.
On smaller $r$, 
gas-capture binaries are hardened mostly by binary-single interaction with the disk BH component (brown triangles in Figure~\ref{fig:edis}), 
and reach merger (cyan line in panel (b) of Figure~\ref{fig:grid_variables}). 

Most pre-existing binaries (purple squares in Figure~\ref{fig:binaries}) are hardened mostly by type I/II torque of a circumbinary disk (cyan squares in Figure~\ref{fig:edis}). 
Pre-existing binaries also merge after migrating to $\sim 10^{-2}$ pc (purple line in panel (b) of Figure~\ref{fig:grid_variables}).

Panels (a) and (b) of Figure~\ref{fig:grid_variables} show the number of BHs, stars, and mergers as a function of the distance from the SMBH, respectively. Black, brown, and blue lines in panel (a) are all BHs, and stars and BHs in the AGN disk at 30 Myr, respectively. 
At $r \sim 0.01-0.1$ pc, most of BHs are captured in the AGN disk due to the strong gaseous drag from the high-density AGN disk. 

Most mergers occur in $r \lesssim$ 0.01 pc (panel (b) of Figure~\ref{fig:grid_variables}). 
In this region, the surface density of the AGN disk $\Sigma_\mathrm{disk,min}$ is reduced by torques from stellar-mass BHs (Eq. \ref{eq:typeI_II_Sigma}) due to the low scale height for the AGN disk (blue line in Figure~\ref{fig:disk_model}),
and the migration timescale increases by a factor of a few (blue line in panel (c) of Figure~\ref{fig:grid_variables}). 

Panel (c) of Figure~\ref{fig:grid_variables} shows the relevant timescales affecting the evolution of a BH binary with $m_k=10\,\Msun$ in the simulation. 
Black and gray lines show the timescale for the initial supersonic velocity of BHs relative to the local AGN motion $v_{\mathrm{ini},k}$ to decay to due to gas dynamical friction and for the BH binary to be captured in the AGN disk, $t_\mathrm{capAGN}$, for $v_{\mathrm{ini},k}=0.2\,v_\mathrm{Kep}$ and $v_{\mathrm{ini},k}=0.3\,v_\mathrm{Kep}$ (Eq.~\eqref{eq:dt_dec}).

The blue line in panel (c) of Figure~\ref{fig:grid_variables} shows the migration timescale given by Eq.~(\ref{eq:typeI_II}). 
The orange line in panel (c) of Figure~\ref{fig:grid_variables} is the timescale on which the binary is hardened by gas dynamical friction, $t_\mathrm{GDF,HS}$, given by Eq.~\eqref{eq:df_timescale}.
The cyan line marks 100 Myr, which roughly corresponds to the upper limit for the typical lifetime of AGN disks \citep[e.g.][]{Martini04}. 
We do not show the hardening time scale of binaries due to type I/II torque of a circumbinary disk surrounding the binary since it depends on the binary separation \citep{Bartos17}. 
For example, the hardening timescale by type I/II torque of a circumbinary disk around a binary with $m_{j_1}=m_{j_2}=5\,\Msun$ and the Eddington accretion rate is $\sim100$ Myr at $s_j=$1 AU and $\sim200$ Myr at $s_j=$0.1 AU. 
In Figure~\ref{fig:edis} we have shown that binary hardening by type I/II migration dominates over gas dynamical friction for small binary separations at a fixed location in the AGN disk beyond 0.05 pc, but it is subdominant at nearly all separations for $r\lesssim 0.05\,\mathrm{pc}$. 
Note that recently revised type I/II migration timescales \citep[e.g.][]{Duffell14,Kanagawa18} are longer than previously used type II migration timescale in which a clear gap is assumed. 
At around 1 pc, $t_\mathrm{mig}$ and $t_\mathrm{GDF,HS}$ intersect (orange and blue line in panel (c)). 
Most outer binaries in the AGN disk migrate within this radius before they are hardened by gas dynamical friction. 
Also outside of $\sim$1 pc, BHs are not captured by the AGN disk if the initial relative velocity of BHs $v_{\mathrm{ini},k}$ are higher than $0.2-0.3\,v_\mathrm{Kep}$.

\begin{figure}
\includegraphics[width=90mm]{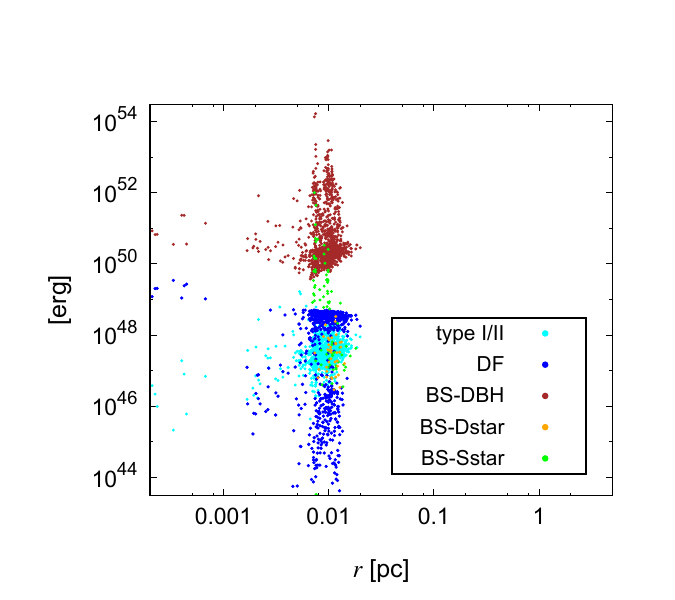}
\caption{The total binding energy lost to different mechanisms as a function of the distance from the central SMBH at merger for all successfully merged binaries over 30 Myr in the fiducial model (Model~1). 
Points represent hardening by type I/II torque of a circumbinary disk (cyan), gas dynamical friction (blue), binary-single interactions with the disk BH (brown) and stellar (orange) and spherical stellar (green) components. 
Binary-single interactions dominate the total energy lost during the binary hardening. 
}
\label{fig:merged_e}
\end{figure}

\subsection{Binary hardening mechanisms}
\label{sec:contribution}

\begin{figure*}\begin{center}
\includegraphics[width=200mm]{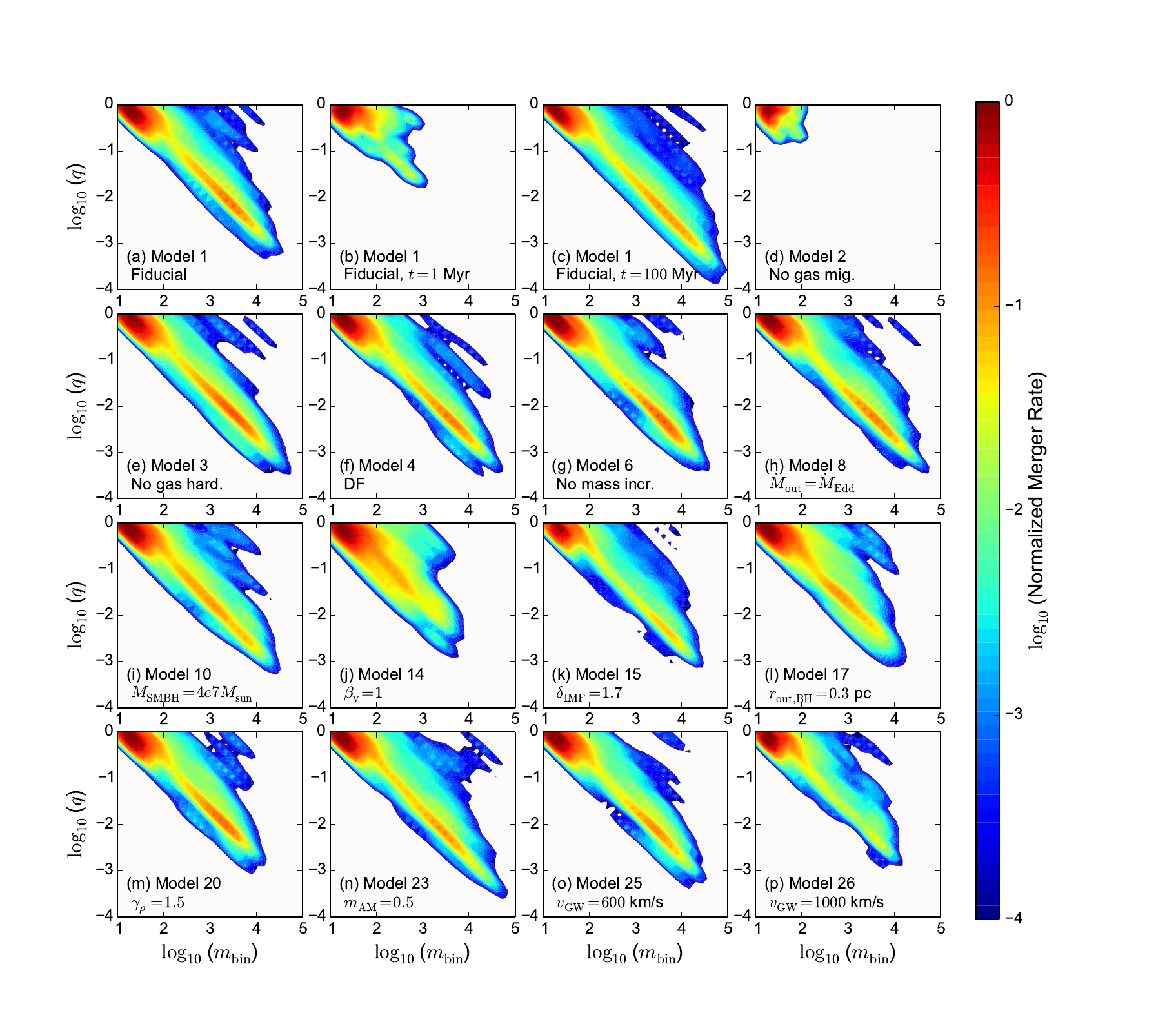}
\caption{
The distribution of mergers in the binary mass vs. binary mass ratio plane are shown for different models and times as labeled.
Panels (a), (d)-(l) shows the merger rate distribution at 10 Myr, and panels (b) and (c) show the distribution at 1 and 100 Myr, respectively. 
The merger rate is normalized by the maximum value in the plane. 
IMBHs form due to repeated mergers in most models. 
}\label{fig:merged_mass}\end{center}\end{figure*}

To clarify the relative importance of different binary hardening mechanisms for the merging binary population in the fiducial model,
Figure~\ref{fig:merged_e} shows the binding energy lost to several mechanisms as a function of the distance from the central SMBH ($r_j$) at merger for all successfully merging binaries (cf. Figure~\ref{fig:edis} showing the instantaneous dominant mechanism for all binaries). 
Most merged binaries are hardened mostly by binary-single interactions 
with the disk BH component (brown dots in Figure~\ref{fig:merged_e}). 
Some fraction of mergers are hardened due to binary-single interaction by $\gtrsim 10^{51}$ erg. 
The masses of these merged binaries are very high $\sim 10^2-10^3\,\Msun$ due to repeated mergers (Figure~\ref{fig:merged_mass} below and panel (b) of Figure~\ref{fig:grid_variables}). 
For comparison note that the separation as a function of binding energy $E$ of a binary with total mass $10 m_{j,10} \Msun$ and symmetric mass ratio $\eta_j=q_j(1+q_j)^{-2}$ is $s_j=10^{-6}\,\mathrm{pc}\, m_{j,10}^2 (4\eta_j) (E/10^{48}\,\mathrm{erg})^{-1}$, where $10^{-6}\,\mathrm{pc}=0.2\,\mathrm{AU}$.

Gas dynamical friction also contributes $\sim 10^{48}-10^{49}$ erg to binary hardening. 
As seen in the Figure~\ref{fig:edis}, gas dynamical friction hardens binaries during the early phases of their evolution. 

For mergers at $r_j \sim0.01$ pc from the SMBH, binary-single interactions with the disk stellar component contribute $10^{46}-10^{48}$ erg to binary hardening (orange dots in Figure~\ref{fig:merged_e}). Since stars in the AGN disk are distributed at $r\gtrsim 0.01$ pc (panel (a) of Figure~\ref{fig:grid_variables}), interactions with the disk stellar component occur before binaries migrate to $\lesssim 0.01$ pc.

\begin{table*}
	\caption{
		Summary of results in different models. 
        The first column indicates the model number. 
        ``DF'', ``type I/II'', ``max'', ``No gas hard.'', and 
        ``Negative type I/II'' 
        in the ``Gas'' column represent models in which binaries are hardened by gas dynamical friction, type I/II torque, 
        maximum of gas dynamical friction and type I/II torque, 
        not hardened by gas interaction, 
        and type I/II torque whose direction is opposite (negative), 
        respectively. 
        ``No gas mig.'', ``No mass incr.'', and 
        ``No acc tor.'' 
        in the ``Gas'' column mean models 
        without the type I/II torque of the AGN disk, 
        without the mass increase by gas accretion, 
        without velocity damping by gas accretion torque,
        respectively. 
        In the ``parameter'' column, 
        we indicate parameters that deviate from the fiducial model, 
        while ``anisotropic BH component'' refers to a model (\#37) in which this component is also taken into account.  
        In the output columns, 
        we summarize the main results in each model:
        $N_\mathrm{mer}$ is the merged number, 
        $N_\mathrm{bin}$ is the binary number, 
        $f_\mathrm{mer,pre}$, $f_\mathrm{mer,gas}$ and $f_\mathrm{mer,dyn}$  
        are, respectively, the fraction of mergers from pre-existing binaries, gas-capture binaries, and dynamically formed binaries compared to total mergers, 
         $f_\mathrm{mer,rep}$ is the fraction of repeated mergers over total mergers, 
         $N_\mathrm{acc}$ is the number of BHs which migrate within $r_\mathrm{in}$, 
        and $N_\mathrm{mer,SF}$ is the merged number from in-situ formed BHs 
        at $t=30$ Myr 
        for the upper 37 rows
        and $t=1$ and $t=100$ Myr for rows 38 and 39, respectively. 
        }
\label{table_results}
\begin{tabular}{c|c|c||c|c|c|c|c|c|c|c}
\hline
\multicolumn{3}{c}{input} \vline& \multicolumn{8}{c}{output}\\\hline
M&Gas&Parameter&
$N_\mathrm{mer}$&$N_\mathrm{bin}$&$f_\mathrm{mer,pre}$&$f_\mathrm{mer,gas}$&$f_\mathrm{mer,dyn}$&$f_\mathrm{mer,rep}$&
$N_\mathrm{acc}$&$N_\mathrm{mer,SF}$
\\\hline

1&max&Fiducial&
$7.7\times 10^3$&$2.0\times 10^3$&0.12&0.85&$0.026$
&0.31&$2$&$93$\\\hline

2&No gas mig.&Fiducial&
$4.0\times 10^2$&$6.6\times 10^3$&0.092&0.91&$2.5\times 10^{-3}$&0.21&
$0$&$0$\\\hline

3&No gas hard.&Fiducial&
$6.5\times 10^3$&$2.9\times 10^3$&0.13&0.83&$0.032$&0.35&
$2$&$49$\\\hline

4&DF&Fiducial&
$7.8\times 10^3$&$2.0\times 10^3$&0.12&0.86&$0.028$&$0.33$&
$1$&$97$
\\\hline

5&type I/II&Fiducial&
$6.9\times 10^3$&$2.6\times 10^3$&0.13&0.84&$0.029$&0.35&
$2$&$48$\\\hline

6&Negative type I/II&Fiducial&
$6.8\times 10^3$&$2.8\times 10^3$&0.14&0.82&$0.035$&0.36&
9&$56$\\\hline

7&No mass incr.&Fiducial&
$7.4\times 10^3$&$2.1\times 10^3$&0.12&0.85&$0.031$&0.31&
2&$78$\\\hline

8&No acc tor.&Fiducial&
$7.8\times 10^3$&$1.9\times 10^3$&0.12&0.86&$0.023$&0.33&
1&$1.0\times 10^2$\\\hline

9&max&${\dot M}_\mathrm{out}=0.4\,{\dot M}_\mathrm{Edd}$&
$8.3\times 10^3$&$2.4\times 10^3$&0.12&0.85&$0.038$&0.33&
$2$&$3.9\times 10^2$\\\hline

10&max&${\dot M}_\mathrm{out}={\dot M}_\mathrm{Edd}$&
$8.6\times 10^3$&$3.8\times 10^3$&0.12&0.82&$0.058$&0.32&
$1$&$6.7\times 10^2$\\\hline

11&max&$M_\mathrm{SMBH}=4\times 10^5\,\Msun$
&$6.3\times 10^3$&$4.0\times 10^3$&0.12&0.85&$0.026$&0.35&
$0$&$0$\\\hline

12&max&$M_\mathrm{SMBH}=4\times 10^7\,\Msun$
&$5.2\times 10^3$&$3.2\times 10^3$&0.078&0.91&$0.011$&0.33&
$3$&$47$\\\hline

13&max&$\beta_\mathrm{v}=0.1$&
$1.0\times 10^4$&$2.5\times 10^3$&0.11&0.86&0.037&0.31&
$2$&$88$\\\hline

14&max&$\beta_\mathrm{v}=0.3$&
$5.4\times 10^3$&$1.7\times 10^3$&0.13&0.85&$0.019$&0.32&
$1$&$91$\\\hline

15&max&$\beta_\mathrm{v}=0.5$&
$3.2\times 10^3$&$1.4\times 10^3$&0.13&0.84&$0.023$&0.36&
$2$&$98$\\\hline

16&max&$\beta_\mathrm{v}=1$&
$1.6\times 10^3$&$1.5\times 10^3$&0.14&0.83&$0.033$&0.45&
$3$&$1.2\times 10^2$\\\hline

17&max&$\delta_\mathrm{IMF}=-1.7$&
$3.8\times 10^4$&$1.2\times 10^4$&0.079&0.86&$0.063$&0.32&
$3$&$5.7\times 10^2$
\\\hline

18&max&$\delta_\mathrm{IMF}=-2$&
$2.2\times 10^4$&$7.7\times 10^3$&0.11&0.86&$0.036$&0.30&
$3$&$3.2\times 10^2$
\\\hline

19&max&$r_\mathrm{out,BH}=0.3$ pc&
$1.5\times 10^3$&$1.9\times 10^2$&0.084&0.89&$0.024$&0.40&
$2$&$1.3\times 10^2$\\\hline

20&max&$r_\mathrm{out,BH}=1$ pc&
$3.5\times 10^3$&$5.6\times 10^2$&0.10&0.87&$0.032$&0.34&
$2$&$86$\\\hline

21&max&$M_\mathrm{star,3pc}=3\times 10^6\,\Msun$&
$2.6\times 10^3$&$8.3\times 10^2$&0.12&0.84&$0.037$&0.32&$5$&$36$
\\\hline

22&max&$\gamma_{\rho}=-0.5$&
$6.8\times 10^3$&$2.0\times 10^3$&0.12&0.85&$0.025$&0.31&
$4$&$87$\\\hline

23&max&$m_\mathrm{AM}=0.1$&
$6.5\times 10^3$&$2.8\times 10^3$&0.11&0.86&$0.026$&0.33&
$1$&$57$
\\\hline

24&max&$m_\mathrm{AM}=0.3$&
$8.9\times 10^3$&$1.2\times 10^3$&0.11&0.86&$0.027$&0.33&
$4$&$39$
\\\hline

25&max&$m_\mathrm{AM}=0.5$&
$8.9\times 10^3$&$1.1\times 10^3$&0.12&0.86&$0.027$&0.33&
$3$&$2$
\\\hline

26&max&$v_\mathrm{GW}=400$ km/s&
$7.9\times 10^3$&$1.9\times 10^3$&0.11&0.86&$0.024$&0.32&
$2$&$90$
\\\hline

27&max&$v_\mathrm{GW}=600$ km/s&
$8.0\times 10^3$&$1.8\times 10^3$&0.12&0.86&$0.024$&0.31&
$2$&$99$
\\\hline

28&max&$v_\mathrm{GW}=1000$ km/s&
$8.5\times 10^3$&$1.6\times 10^3$&0.12&0.86&$0.019$&$0.33$&
$2$&$86$
\\\hline

29&max&$f_\mathrm{pre}=0$&
$7.4\times 10^3$&$1.6\times 10^3$&0&0.97&$0.028$&0.31&
$2$&$98$\\\hline

30&max&$f_\mathrm{pre}=0.3$&
$7.9\times 10^3$&$2.2\times 10^3$&0.21&0.76&$0.027$&0.31&
$2$&$93$\\\hline

31&max&$f_\mathrm{pre}=0.7$&
$8.1\times 10^3$&$3.1\times 10^3$&0.39&0.58&$0.027$&0.32&
$2$&$76$\\\hline

32&max&$\alpha_\mathrm{SS}=0.01$&
$7.9\times 10^3$&$2.0\times 10^3$&0.11&0.86&$0.025$&0.33&
$2$&$1.2\times 10^2$
\\\hline

33&max&$N_\mathrm{cell}=80$&
$7.2\times 10^3$&$2.4\times 10^3$&0.11&0.85&$0.036$&0.33&
$1$&$89$
\\\hline

34&max&$N_\mathrm{cell}=160$&
$7.8\times 10^4$&$1.9\times 10^3$&0.12&0.85&$0.028$&0.31&
$1$&$99$
\\\hline

35&max&$\eta_t=0.4$&
$8.4\times 10^3$&$1.6\times 10^3$&0.12&0.83&$0.056$&0.33&
$1$&$1.2\times 10^2$
\\\hline

36&max&$\eta_t=0.2$&
$7.9\times 10^3$&$1.8\times 10^3$&0.12&0.84&$0.040$&0.31&
$3$&$1.1\times 10^2$
\\\hline

37&Fiducial&anisotropic BH component&
$1.1\times 10^4$&$6.9\times 10^2$&0.11&0.85&$0.042$&0.40&
2&$1.2\times 10^2$\\\hline

1&max&$t=1\,\mathrm{Myr}$&
$3.4\times 10^2$&$2.3\times 10^3$&0.081&0.91&$8.7\times 10^{-3}$
&0.29&$0$&$0$\\\hline

1&max&$t=100\,\mathrm{Myr}$&
$1.2\times 10^4$&$1.2\times 10^3$&0.11&0.85&$0.037$
&0.34&$3$&$5.1\times 10^2$\\\hline

\end{tabular}
\end{table*}

\begin{figure*}\begin{center}
\includegraphics[width=200mm]{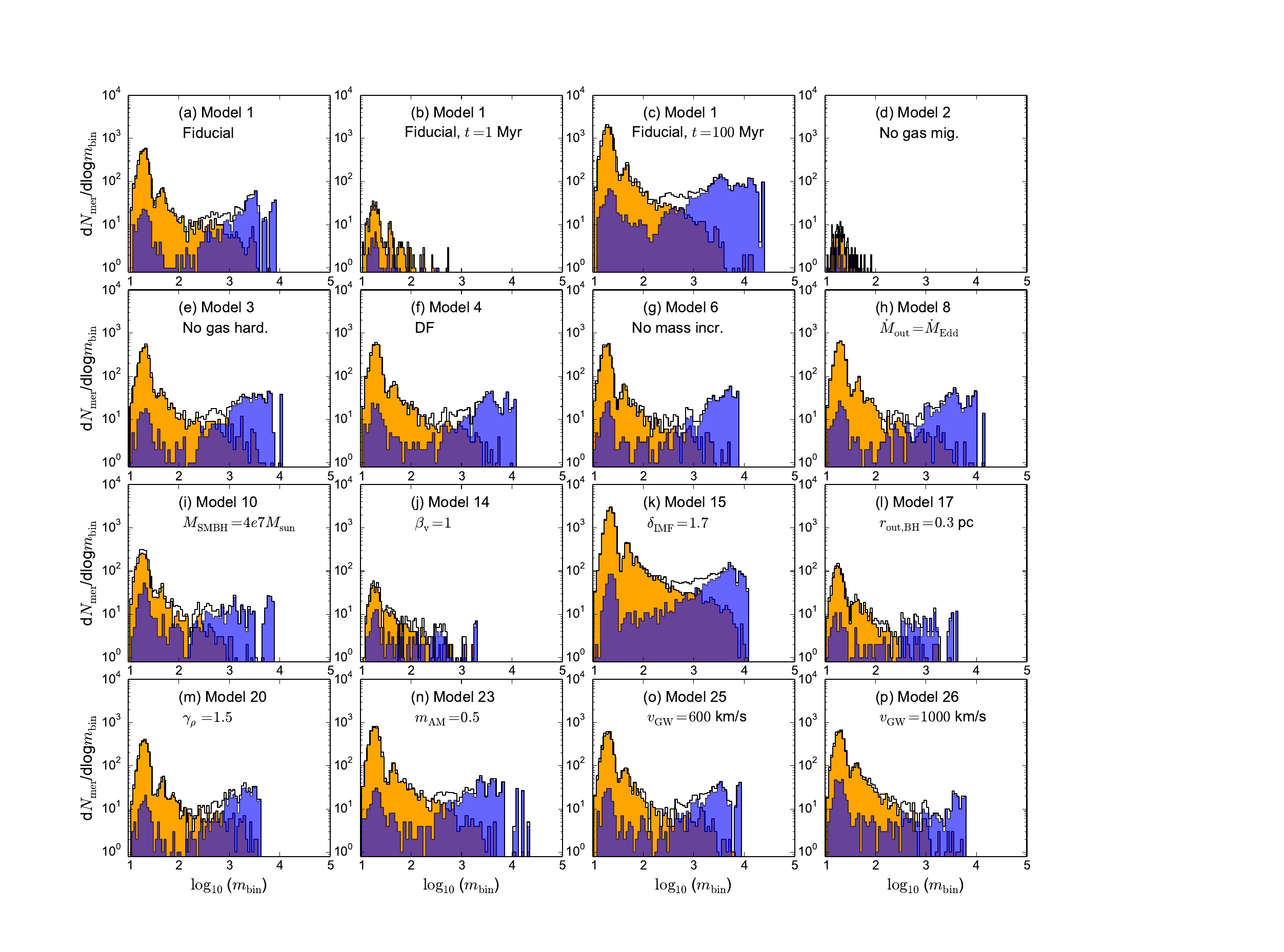}
\caption{
The distribution of binary masses at mergers. 
The layout of models in the different panels is the same as in Figure~\ref{fig:merged_mass}. 
The orange and blue regions present the distribution for mergers outside and inside the AGN disk, respectively. 
}\label{fig:histgram}\end{center}\end{figure*}

\begin{figure*}
\begin{center}
\includegraphics[width=200mm]{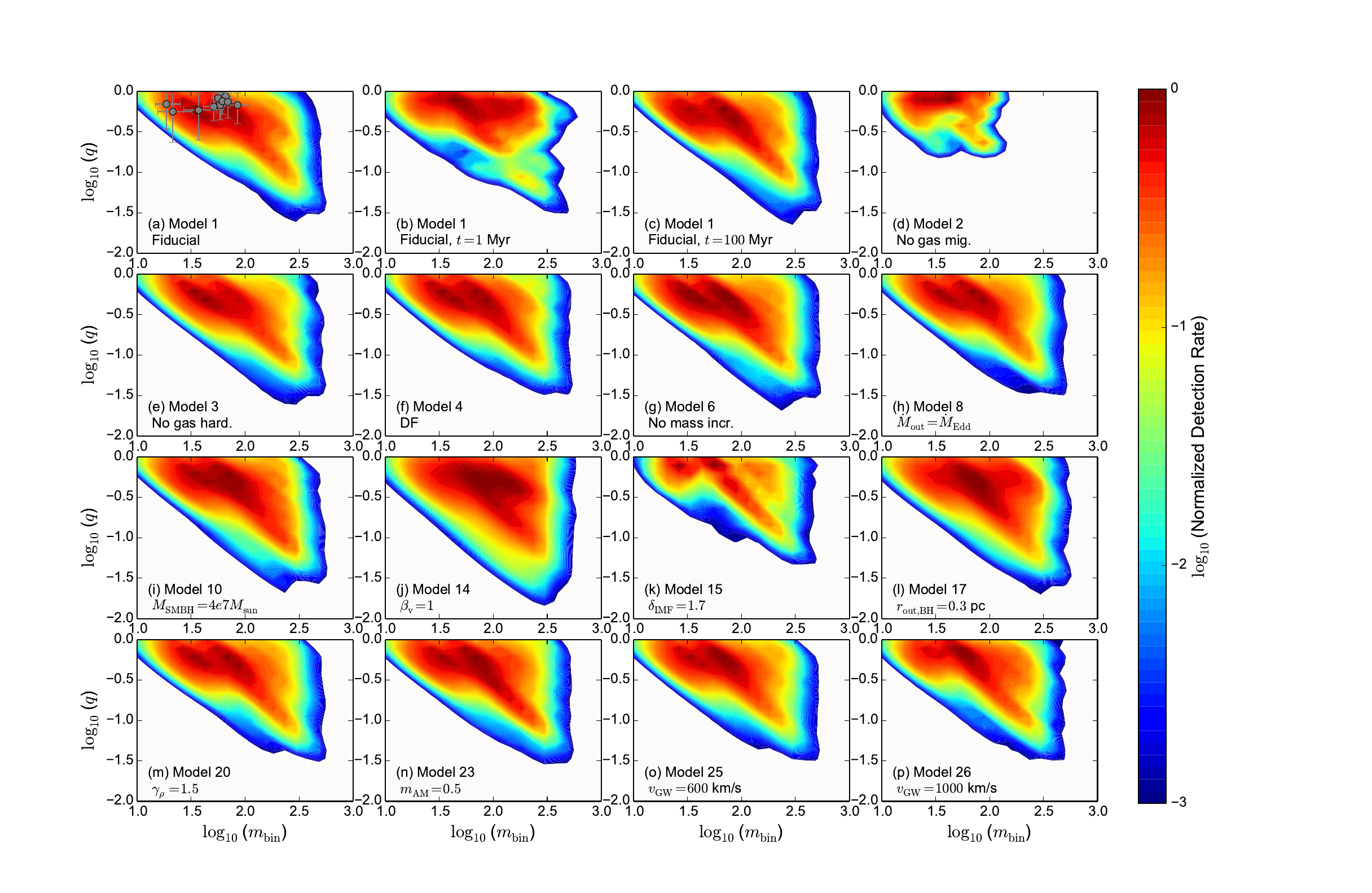}
\caption{
Similar to Figure~\ref{fig:merged_mass}, but showing the normalized detection rate of mergers in the binary mass vs. binary mass ratio plane in several Models. 
We use the noise spectral density of the ER13 (prior to O3) run of LIGO Hanford \citep{Kissel18}. 
The merger rate is normalized by the maximum value in the plane. 
In panel (a), the GW events detected to date are overplotted \citep{TheLIGO18}. 
The mass and mass ratio distribution can well match the observed distribution in some parameter regions. 
Since our simulations do not take into account the exchange of binary components at interactions, this figure presents rough estimates of the mass ratio. 
}\label{fig:detection_mass}\end{center}\end{figure*}

\begin{figure}\begin{center}
\includegraphics[width=90mm]{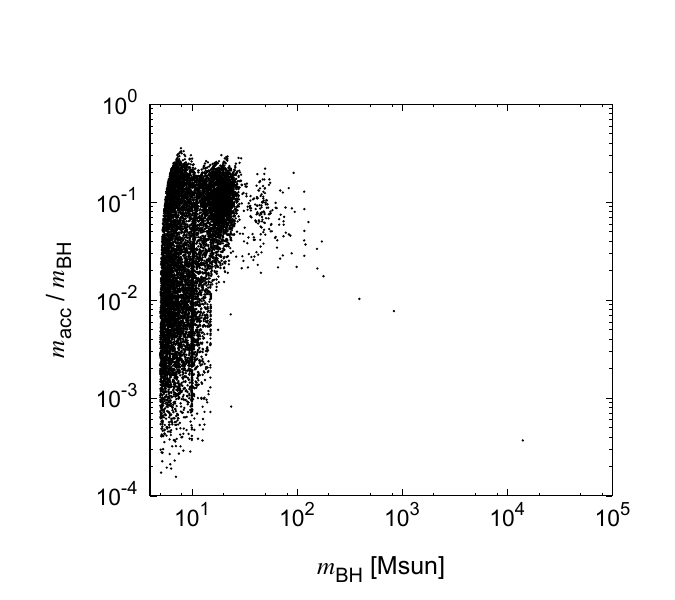}
\caption{
The ratio of the accreted mass over the BH mass as a function of the BH mass at 30 Myr in the fiducial model. BHs typically gain several tens of percent of their final mass via gas accretion. 
}\label{fig:mass_acc}\end{center}\end{figure}

\begin{figure}\begin{center}
\includegraphics[width=90mm]{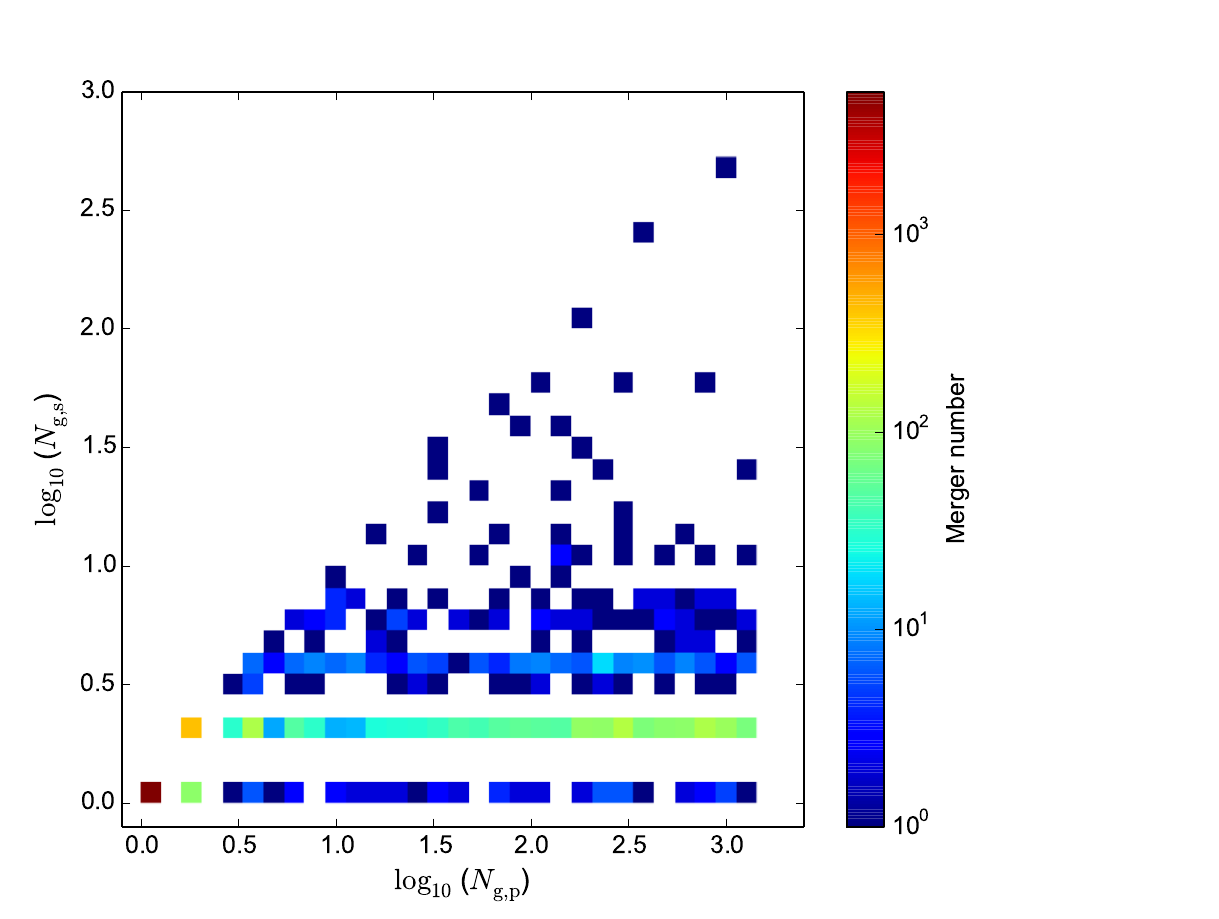}
\caption{
The number of mergers as a function of the generation of the primary ($N_\mathrm{g,p}$) vs. the secondary ($N_\mathrm{g,s}$) BH in the fiducial model up to 30 Myr. The generation of a BH is defined as 
the number of mergers that its progenitor BHs have experienced in the past (plus one). 
Here `primary' and `secondary' refer to the BHs for which $N_\mathrm{g,p}\geq N_\mathrm{g,s}$.
Most  mergers are between 1st generation BHs, but some BHs experience 
over a thousand mergers. 
}\label{fig:generation}\end{center}\end{figure}

\begin{figure}\begin{center}
\includegraphics[width=90mm]{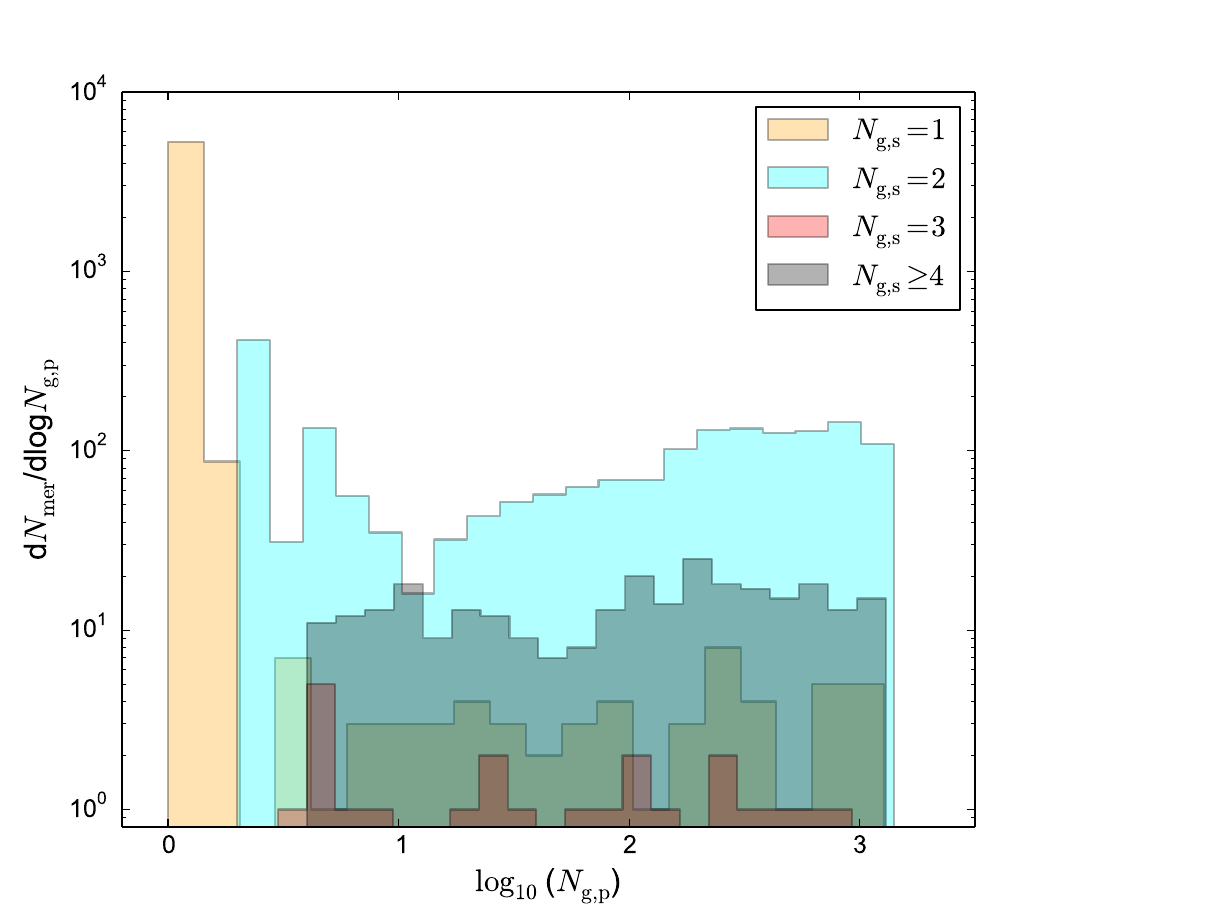}
\caption{
The distribution of the generation of the primary ($N_\mathrm{g,p}$) for several values of the secondary ($N_\mathrm{g,s}$) in the fiducial model up to 30 Myr. 
Orange, blue, red, and gray histograms represents the distribution for $N_\mathrm{g,s}=$1, 2, 3, and $N_\mathrm{g,s}\geq 4$, respectively.
}\label{fig:generation_h}\end{center}\end{figure}

\subsection{Dependence on model parameters}
\label{sec:dependence}

Table~\ref{table_results} shows the results for several different variations with respect to the fiducial model. The input represents the settings for parameters and mechanisms in each model. 
In the output columns, we show the properties of the system at 30 Myr, namely, the number of merged binaries ($N_\mathrm{mer}$), surviving binaries ($N_\mathrm{bin}$), the fraction of mergers among pre-existing binaries ($f_\mathrm{mer,pre}$), gas-capture formed binaries ($f_\mathrm{mer,gas}$), and dynamically formed binaries ($f_\mathrm{mer,dyn}$) compared to the total number of mergers, 
the fraction of repeated mergers over total mergers ($f_\mathrm{mer,rep}$), 
the number of BHs which migrate within the inner boundary $r_\mathrm{in}$ ($N_\mathrm{acc}$), and the number of mergers among BHs formed in-situ due to the fragmentation of the AGN disk ($N_\mathrm{mer,SF}$). 

\subsubsection{Number and fraction of mergers}
The fraction of the merger number ($N_\mathrm{mer}$) over the initial BH number ($N_\mathrm{totBH,ini}$)  ($f_\mathrm{BH,mer}\equiv N_\mathrm{mer}/N_\mathrm{totBH,ini}$) at $t=30$ Myr is $\sim 0.02-0.8$ for Models~1--37, 
where $N_\mathrm{totBH,ini}=2\times 10^4$ in Models~1--16, 22--37 and $N_\mathrm{totBH,ini}=1.0\times 10^5,\,6.2\times 10^4,\,2.0\times 10^3,\,6.6\times 10^3,$ and $6.0\times 10^3$ for Models~17, 18, 19, 20, and 21, respectively. 
We find that the merger fraction is almost unaffected by hardening due to gaseous processes (Models~1, 3--6), 
the recoil kick velocity at mergers (Models~26--28), or the pre-existing binary fraction (Models~29--31). 
On the other hand, 
the merger fraction is lowest when migration does not operate (Model~2), and highest when we only consider the evolution for BHs in the inner regions ($r_\mathrm{out,BH}=0.3\,\mathrm{pc}$, Model~19). 
This is because mergers require BHs to migrate to the inner regions of $\lesssim 0.01$ pc where the hardening by binary-single interaction is efficient. 
In the cases of lower $\beta_\mathrm{v}$, which determines the dispersion of the initial BH velocity distribution, the number of mergers is higher (Models~1, 13--16) since a larger fraction of BHs is captured in the AGN disk where binary formation and hardening are efficient. 
Also the merger fraction is low in the high SMBH mass case (Model~12) because the high Keplerian velocity enhances the decay timescale of the BH velocity $v_k$ (see Eq.~\ref{eq:dt_dec}). 
Note that high-mass SMBHs tends to have larger AGN disks and nuclear star clusters, which also needs to be considered for the estimate of the merger rate (see $\S$~\ref{sec:gw_rate} below). 
If we take into account interactions with BHs outside the AGN disk (Model~37), the merger fraction is enhanced by a factor of 1.4. This is because the high density BHs in the inner regions ($\lesssim 10^{-2}\,\mathrm{pc}$) outside the AGN disk (black line in panel~(a) of Fig.~\ref{fig:grid_variables}) enhance the rate of hard binary-single interactions.

The number of mergers from in-situ formed BHs ($N_\mathrm{mer,SF}$) depends strongly on the accretion rate of the AGN disk ${\dot M_\mathrm{out}}$ (Models~1, 9, 10) and the angular momentum transfer parameter $m_\mathrm{AM}$ (Models~23-25). This is because the star formation rate depends on ${\dot M_\mathrm{out}}$ and $m_\mathrm{AM}$. 
The rate of mergers among in-situ formed BHs is estimated in $\S$ \ref{sec:comparison}.

\subsubsection{Convergence test}
We checked whether the results change due to the time step parameter (Models~1, 35, 36) or the size of cells (Models~1, 33, 34) in which physical quantities for components (Eqs.~\ref{eq:quantity_sph}, \ref{eq:quantity_disk}, \ref{eq:quantity_sdisk}), and the AGN disk (Figure~\ref{fig:disk_model}) are stored. The number of mergers ($N_\mathrm{mer}$) is not significantly affected by these parameters at around the fiducial values (see $N_\mathrm{mer}$ for Models~1 and 34 or Models~1 and 36). Hence the merger fraction, which we are interested in, is not influenced by the numerical resolution. 

\subsubsection{Repeated mergers}
In the fiducial model, the fraction of repeated mergers is as high as $\sim 0.26$ at 30 Myr, allowing massive BHs to form. 
Figure~\ref{fig:merged_mass} shows the distribution of merged BHs in the binary mass vs binary mass ratio plane, and Figure~\ref{fig:histgram} shows the cumulative distribution of the merged mass of binaries. Figure~\ref{fig:merged_mass} and \ref{fig:histgram} show that BHs can grow to $\sim 10^2 - 10^4\,\Msun$ by repeated mergers in most models. 
In the fiducial model at 30 (10) Myr, the maximum BH mass is $1.4\times 10^4\,\Msun$ ($8.6\times 10^3\,\Msun$), 
and 14 (9) and 1 (2) BHs are more massive than $10^2\,\Msun$ and $10^3$ $\Msun$, respectively. 
In Model~17 ($\delta_\mathrm{IMF}=-1.7$) at 30 (10) Myr (panel (k) of Figure~\ref{fig:merged_mass}) in which the initial number of BHs is higher ($N_\mathrm{totBH,ini}=10^{5}$), the maximum BH mass is $2.2\times 10^4\,\Msun$ ($1.2\times 10^4\,\Msun$), and 128 (81) and 6 (4) BHs are more massive than $10^2$ and $10^3$ $\Msun$, respectively. 
Thus we find that a lot of IMBHs of $\gtrsim 10^2\,\Msun$ are reproduced by repeated mergers.

\subsubsection{BH binary parameter distributions}
Figure~\ref{fig:detection_mass} shows the normalized detection rate in the binary mass vs mass ratio plane for several models in Table~\ref{table_results}. 
The normalized detection rate is calculated as the product of the merger rate and the detection volume (e.g. Eq.~6 of \citet{TheLIGO12}).  
We use the noise spectral density from the calibrated sensitivity spectra of aLIGO-Handford on 2018 November 8 \citep{Kissel18}. 
Note that our simulations can only roughly estimate the total binary mass and mass ratio distribution of mergers since we do not take into account the exchange interactions during binary-single and binary-binary interactions.

In Figure~\ref{fig:detection_mass} panel (a), we have overlaid the mass distribution of the observed LIGO/VIRGO sources during the O1 and O2 observing runs (gray points).
Interestingly, despite the fact that the expected maximum BH mass at birth is limited to $\lesssim 15\,\Msun$ due to the solar metallicity environment in galactic nuclei, our results in Figure~\ref{fig:detection_mass} suggest that the total mass of the detectable merging binaries in AGN extends to masses of $250\,\Msun$. This is beyond the mass of the detected sources announced to date 
\citep[but also see][]{Udall19}. 
In the fiducial model, the fraction of mergers with $m_\mathrm{bin}>200\,\Msun$ is $8.1\%$, that with $q<0.1$ is $2.8\%$, and that with $m_\mathrm{bin}>200\,\Msun$ and $q<0.1$ is $2.2\%$. If such mergers will be discovered in the future, it can be a possible signature that mergers are originating in AGN disks.

The hardening of the mass function of mergers in AGN migration traps, if they exist, was previously pointed out by \citet{Yang19a}. Our results confirm the assumption on the existence of a region similar to a migration at $\sim$0.01\,pc. The hardening of the mass function seen in Figure~\ref{fig:merged_mass} is more prominent than previously thought, due to the relatively high likelihood of multiple generations of mergers. 

\subsubsection{Impact of the GW recoil velocity}
In the case that the recoil velocity at mergers due to anisotropic GW radiation is $600$ km/s (Model~27), 19 (19) BHs grow to $\gtrsim 10^2\,\Msun$ by repeated mergers at 30 (10) Myr in a single AGN, which is similar to our fiducial model assuming much lower GW kicks. 
To realize the high kick velocity of $v_\mathrm{GW}\gtrsim 600$ km/s, the BHs need to be rapidly spinning and the directions of BH spins must be misaligned with the angular momentum direction of a binary. 
Hence even if the recoil kick velocity is very high, 
repeated mergers build up binaries with total masses of $\gtrsim 100\,\Msun$, 
which contribute moderately to the detection rate.

Our simulation results confirm previous assumptions on the possibility of mass growth of BHs through mergers in AGN disks \citep{McKernan12,McKernan14,Yang19b}. 
Mergers of massive BHs in the pair-instability supernovae mass gap such as \citet{Zackay19b} can provide a compelling case that mergers are facilitated by AGN disks,
although masses in some cases can be significantly overestimated due to statistical noise fluctuations~\citep{Fishbach19}.

\subsubsection{BH growth by gas accretion}
Figure~\ref{fig:mass_acc} shows the ratio of the accreted mass over the BH mass for the fiducial model. Gas accretion contributes to the BH masses by less than several tens of percent. Thus gas accretion is not a dominant mechanism for the BH growth. 
The contribution of gas accretion decreases as the BH mass increases. This suggests that BHs violently grow through repeated mergers. 
Figure~\ref{fig:generation} shows the merger number as a function of the generation of primary and secondary BHs for the fiducial model. 
The generation of a BH is defined as the number of mergers that the BH has experienced in the past plus one. Some BHs have experienced hundreds mergers in their past, during which the masses of BHs are increased due to repeated mergers also by a factor of about hundreds. 
Hence BHs grow mainly due to repeated mergers. 
Figure~\ref{fig:generation_h} shows the merger number as a function of the generation of primary BHs for the several generation of secondary BHs. Most mergers are between 1st generation BHs (orange histogram), while the growth of massive BHs (large $N_\mathrm{g,p}$) is dominated by mergers with 2nd generation BHs (cyan histogram).  

\section{Discussion}

\subsection{Spin of merging binaries}
\label{sec:spin}

The binary in Figure~\ref{fig:tot_cont} merges outside of the AGN disk ($h_{{z},k}>h_{\mathrm{AGN},l}$).  In the fiducial model, $81\%$, $80\%$, $97\%$ and $19\%$ of all mergers, mergers among gas-capture binaries, pre-existing binaries, and dynamically formed binaries, respectively, occur outside of the AGN disk (evaluated at 30 Myr). 
At $t = 1,$ 10, and 100 Myr, the fractions of mergers outside of the AGN disk are 0.84, 0.81, and 0.78, respectively, and less massive BHs tend to merge outside of the AGN disk  (orange regions in Fig.~\ref{fig:histgram}). 
The direction of the internal orbital angular momentum of binaries orbiting outside of the AGN disk may be misaligned in a random direction following hard binary-single interactions for kicked binaries. Similarly vector resonant relaxation and flyby encounters may reorient the orbital plane pre-existing binaries. 
In these cases, the correlation between the direction of BH spins and the angular momentum of binaries vanishes, 
which can produce mergers with low values of the effective spin parameter. 
GW observations may statistically distinguish a population of higher generation mergers based on the measurement of the effective spin \citep{GerosaBerti2017,Fishbach2017}. 
Note that $91\%$ of mergers among dynamically formed binaries merged in the AGN disk are mergers of $\gtrsim 10^3\,\Msun$ BHs, which are not observed by LIGO.

Furthermore, $19\%$ of mergers occur within the AGN disk in the fiducial model. In these mergers, the binary's orbital angular momentum is expected to be aligned or antialigned with the angular momentum direction of the AGN disk \citep{McKernan17,Secunda19,Yang19b,McKernan19}. 
Also previous studies suggest that a few to ten percent mass increase by gas accretion might be sufficient to 
align the spin directions of BHs in a binary with the angular momentum direction of a circumbinary disk \citep[e.g.][]{Scheuer96,Natarajan98,Ogilvie99,Hughes03,King05,Volonteri07,Bogdanovic07,Lodato13}, 
and the angular momentum direction of a circumbinary disk is suggested to be the same as for the AGN disk \citep[e.g.][]{Lubow99}. 
GW sources with a moderate to high effective spin 
such as GW151216, GW170403 and GW170817A \citep{Zackay19,Venumadhav19,Zackay19b,LIGO20_GW190412} may represent mergers with multiple generation mergers in AGN disks \citep[see also][]{McKernan17,Gayathri19,Tagawa20}.

\subsection{Doppler acceleration and gravitational lensing}
\label{sec:doppler}

Figure~\ref{fig:merged_e} and panel (b) of Figure~\ref{fig:grid_variables} shows that most mergers occur between 0.001--0.02 pc from the SMBH. 
Due to the high acceleration in the gravitational field of the SMBH in these regions, the effect of the Doppler acceleration may be detectable in a multiyear detection campaign.
The resulting frequency drift (or, change in apparent redshift)
allows the binary's distance from the SMBH to be estimated from the LISA GW waveforms \citep{Meiron17,Inayoshi17b,Wong19}. 
\citet{Inayoshi17b} found that when the projected separation $r \sin I$ of a 10-10 $\Msun$ BH binary from a $4\times 10^6$ SMBH  is smaller than $\sim 0.2$ pc  (where $I$ is the inclination of the orbital plane of the center of mass for a binary), 
the strain perturbation by Doppler acceleration is detectable. 
\citet{Wong19} estimated that when $r \sin I$ is smaller than $\sim 0.01-0.03$ pc for such binaries, the orbital period and velocity around the SMBH are determined with $\lesssim10\%$ uncertainties. In the fiducial model of a single AGN until 30 Myr, the expected number of mergers within 0.001, 0.003, 0.01, and 0.03 pc from the SMBH is $0.25\%$, $1.4\%$, $45\%$, and $100\%$, respectively. 
Even in the case that migration does not operate (Model~2), the fraction of all mergers within 0.003, 0.01, 0.03, 0.1, and 0.3 pc from the SMBH  at 30 Myr is $0.50\%$, $5.0\%$, $25\%$, $78\%$, and $99\%$, respectively.

Furthermore, for configurations in which the binary's orbit around the SMBH is not too far from edge-on, the GW emission from the binary can be lensed by the SMBH~\citep{Kocsis2013}.
\citet{DOrazio19} found that GW wave signals for a tight orbit, repeatedly lensed by a SMBH, are observable with the probability of more than a few percent by LISA if the orbital period around a SMBH is less than $\sim1$ yr. 
In our fiducial model, a fraction of $0.25\%$ of all mergers satisfies this condition.

Thus perturbations in the GW waveforms due to Doppler acceleration and gravitational lensing are expected to be observable by future GW instruments, provided that the rates in this channel are sufficiently high for detections.

\begin{figure}
\includegraphics[width=90mm]{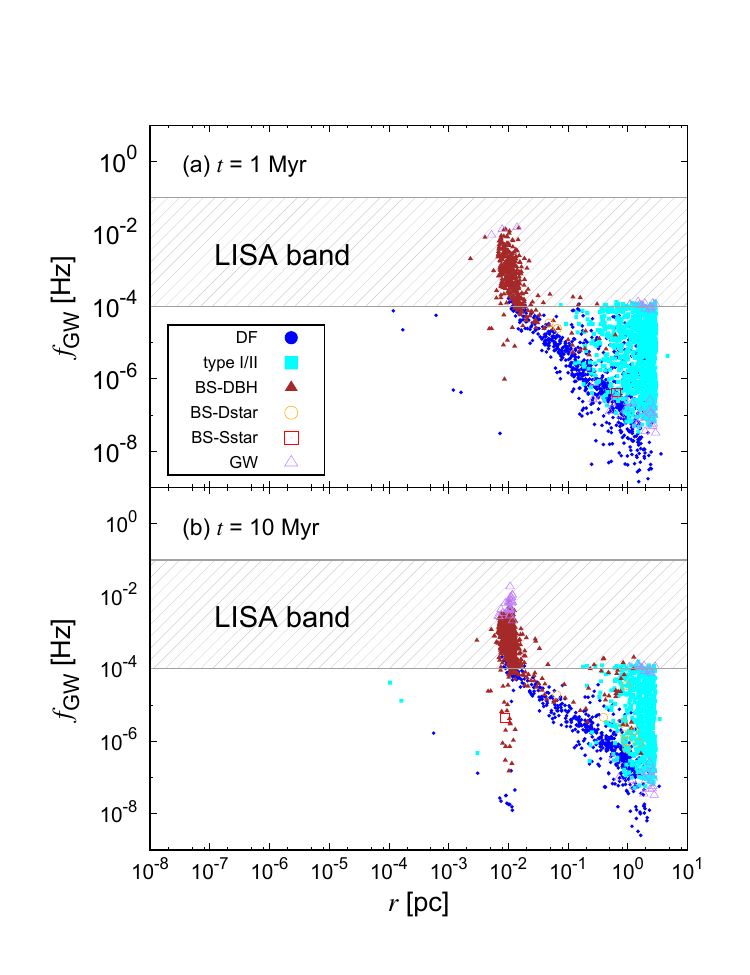}
\caption{ 
Same as Fig.~\ref{fig:edis}, but the vertical-axis represents the peak GW frequency $f_\mathrm{GW}$. To calculate $f_\mathrm{GW}$, we assume the eccentricity of the binary to be 0.7, which is the median value of the thermal distribution. 
The hatched gray regions enclose 
$10^{-1}$ and $10^{-4}$ 
Hz, in which LISA is sensitive to GWs \citep{LISA}. 
}\label{fig:fgw_edis}\end{figure}

\subsection{Eccentricity of merging binaries}
\label{sec:ecc}

When mergers are facilitated by binary-single interactions,  
mergers may have some residual binary eccentricities \citep[e.g.][]{Hills75,Heggie75,Trani18}, which is expected to be observed by future low-frequency GW instruments \citep{Brown10,Nishizawa16,Gondan18b,Gondan18c,Hinder18,Huerta18,Klein18,Lower18,Romero-Shaw19,Abbott19_Ecc}. 
Also, when mergers are initially driven by type I/II torque, a significant binary eccentricity is expected to remain at mergers \citep[e.g.][]{Artymowicz91,Armitage2005,Cuadra09,Rodig11,Fleming17,Munoz19}. 

The binary eccentricity can be enhanced to typically around 0.7 during binary-single interactions if the eccentricity distribution approaches a thermal shape \citep[e.g.][]{Geller19}.\footnote{For an isotropic thermal distribution, $e^2$ is uniformly distributed. The median eccentricity is $1/\sqrt{2}\sim 0.7$, and $19\%$ of sources have $e>0.9$.}
Once the binaries are driven by GWs, the eccentricity decreases \citep{Peters64}. To highlight an example, consider a $5+5\,\Msun$ BH binary, whose initial binary separation is $3\times 10^{10}$ cm, and the initial binary eccentricity is 0.7. 
This setting corresponds to the typical binding energy ($\sim2\times 10^{50}$ erg) down to which first-generation binaries in an AGN disk can be hardened by binary-single interactions. The orbital frequency of such a binary is $1.1\times 10^{-3}\,\mathrm{Hz}$ and the peak GW frequency is $f_\mathrm{GW}=1.2\times 10^{-2}$\,Hz \citep{Wen2003}.
Since LISA will be able to detect the eccentricity of $\gtrsim 10^{-3}$ at $f_\mathrm{GW}\sim 0.01\,\mathrm{Hz}$ \citep{Nishizawa16}, 
a nonzero eccentricity is expected to be measurable if BH mergers originate from AGN disks. Following \citet{Peters64}, 
the binary eccentricity subsequently evolves by GW radiation 
to $\sim 3\times 10^{-4}$ at $f_\mathrm{GW}=10$ Hz. The eccentricity is slightly higher than typical values for in-cluster mergers in globular clusters 
and somewhat smaller than for GW capture binaries in galactic nuclei \citep[e.g.][]{Rodriguez18b,Gondan18a,Zevin19}. 
Thus, the value of the eccentricity along with masses and spins may provide distinctive footprints to identify BH binaries in AGN.

\subsection{GW phase shift and population distribution}\label{sec:GWperturbation}
Figure~\ref{fig:fgw_edis} shows the dominant hardening mechanism as a function of the distance from the SMBH and the peak GW frequency $f_\mathrm{GW}$ \citep{Wen2003} at two snapshots, 1 Myr (top panel) and 10 Myr (bottom panel) for the fiducial model assuming an eccentricity of 0.7. Most binaries are hardened predominantly by binary-single interactions in the LISA band at 
$\sim 10^{-2}$--$10^{-4}\,\mathrm{Hz}$ 
peak frequencies before GWs drive the binaries to merger.\footnote{Note however, that we use the $e=0$ approximation to simulate the evolution of binaries. Binaries may be hardened more efficiently by GWs at at a somewhat lower frequency for higher eccentricity than shown in Figure~\ref{fig:fgw_edis}.
Thus, LISA may detect the GWs of stellar mass BH and IMBH binaries embedded in AGN where astrophysical environmental effects are still significant \citep[see also][]{Kocsis11,Barausse2015}. For these binaries, the mean inspiral rate ($d f_{\rm GW}/dt$, the ``chirp'') is higher than that of an isolated binary in vacuum, which leads to an astrophysical GW phase shift for non-stationary sources \citep[see also][]{Yunes2011,Kocsis11,Derdzinski18}. For nearly stationary GW sources (with respect to the observation time), the $f_{\rm GW}$-distribution of binaries may be used to measure the residence time that the binaries spend at a particular frequency to infer the underlying astrophysical mechanism driving the evolution of the binary separation \citep{KocsisSesana2011}. }

\subsection{GW event rates}
\label{sec:gw_rate}

\subsubsection{Stellar-mass BH-BH mergers in AGN}

Here we estimate the merger rate among stellar-mass BHs in AGN disks. We calculate the merger rate density $\mathcal{R}_\mathrm{sBH}$ for stellar mass BH mergers as 
\begin{align}\label{eq:merger_rate_int}
\mathcal{R}_\mathrm{sBH} = & \int_{M_\mathrm{SMBH,min}}^{M_\mathrm{SMBH,max}}\frac{\mathrm{d} n_\mathrm{AGN}}{\mathrm{d} M_\mathrm{SMBH}}\frac{f_\mathrm{BH,mer} N_\mathrm{BH,cross}}{t_\mathrm{AGN}} \mathrm{d} M_\mathrm{SMBH} 
\end{align}
where $N_\mathrm{BH,cross}$ is the number of BHs crossing AGN disks, $t_\mathrm{AGN}$ is the average life time of AGN disks,  $n_\mathrm{AGN}$ is the average number density of AGNs in the Universe, $f_\mathrm{BH,mer}$ is the merger fraction per BH in AGN given by Table~\ref{table_results}, and $M_\mathrm{SMBH,min}$ and $M_\mathrm{SMBH,max}$ are the minimum and maximum SMBH masses we consider here.

To derive $N_\mathrm{BH,cross}$, 
we assume that the power law exponent of the probability distribution function of the BH orbital 
radii around the SMBH (Eqs.~\ref{eq:bh_density} and \ref{eq:bh_density2}) is $\gamma_\rho=0$ on large scales (100 pc).  
Within pc scales, this value is motivated by theoretical studies of a relaxed mass-segregated cluster \citep[e.g.][]{Hopman06,Antonini14}. 
Outside of the radius of a nuclear cluster, the slope of the one-dimensional stellar density is observed to be $\gamma_\rho\sim 0$ up to $\sim 100$ pc \citep{Feldmeier14,Schodel14}, although some fluctuation exists at each $r$. 
In the case of $\gamma_\rho=0$, the number of BHs which exist within $r$ is given by $N_\mathrm{BH,NSC}\times r/r_\mathrm{eff,NSC}$, where $N_\mathrm{BH,NSC}$ is the number of BHs in a nuclear star cluster, and $r_\mathrm{eff,NSC}$ is the effective (half-mass) radius of a nuclear star cluster. 
Also in this distribution ($\gamma_\rho=0$), the number of BHs crossing the AGN disk for a thermal eccentricity distribution is enhanced by a factor of $\sim$ten compared to the number of BHs on strictly circular orbits
\citep[see Eq.~7 in][]{Bartos17}. 
Conservatively neglecting this enhancement due to eccentricity, we assume that the number of BHs crossing the AGN disk is 
\begin{equation}\label{eq:NBHcross}
N_\mathrm{BH,cross}= N_\mathrm{BH,NSC}  \frac{r_\mathrm{AGN}}{r_\mathrm{eff,NSC}}\,,    
\end{equation}
where $r_\mathrm{AGN}$ is the typical size of the AGN disk. 
The sizes of the AGN disks are highly uncertain. Radiation hydrodynamical simulations suggest that thin dense gas disks extend to pc scales, which are beyond the dust sublimation radius \citep[e.g.][]{Namekata16,Wada16,Williamson19}. 
Mid-infrared observations show that the sizes of AGN disks are 
\begin{align}
\label{eq:r_agn}
r_\mathrm{AGN} \sim \mathrm{pc}\left(\frac{L_\mathrm{bol}}{10^{45}\,\mathrm{erg}}\right)^{1/2}
\sim 0.1\,\mathrm{pc}M_\mathrm{SgrA}^{1/2}\left(\frac{f_\mathrm{Edd}}{0.03}\right)^{1/2} 
\end{align}
(see Figure~36 in \citealt{Burtscher13}), where $L_\mathrm{bol}$ is the bolometric luminosity of an AGN disk. 
We set $r_\mathrm{AGN}=r_\mathrm{AGN,MW}(M_\mathrm{SMBH}/4\times 10^6\,\Msun)^{1/2}$ assuming a fixed Eddington accretion rate as in Eq.~\eqref{eq:r_agn}, 
where $r_\mathrm{AGN,MW}$ is the size of the AGN disk for $M_\mathrm{SMBH}=4\times 10^6\,\Msun$.

We substitute the $M_{\mathrm{SMBH}}$ dependence of $r_\mathrm{eff,NSC}$ and $N_{\mathrm{BH,NSC}}$ in Eq.~\eqref{eq:NBHcross} using empirical correlations as follows. The number of BHs may be expressed with the stellar mass of the nuclear star cluster as 
\begin{equation}\label{eq:etaBH}
    N_\mathrm{BH,NSC}=\eta_\mathrm{n,BH} M_\mathrm{NSC},
\end{equation}
where the parameter $\eta_\mathrm{n,BH} \sim 0.002\,\Msun^{-1}$
represents the number of BHs per unit stellar mass for a Salpeter IMF (but see discussion below for BHs in NSCs).
The mass of the nuclear star cluster follows
\begin{equation}\label{eq:Mnsc-sigma}
    M_\mathrm{NSC}=4.3\times10^6\,\Msun\left(\frac{\sigma_\mathrm{Bulge}}{54\,\mathrm{km/s}}\right)^{2.11}\,
\end{equation}
\citep{Scott13}, where the velocitiy dispersion of the bulge is given by \citep{Kormendy13}
\begin{equation}\label{eq:Msmbh-sigma}
\sigma_\mathrm{Bulge}
  =200\,\mathrm{km/s}\left(\frac{M_\mathrm{SMBH}}{3.1\times 10^8\,\Msun}\right)^{0.228}\,.
\end{equation}.
The radius of the nuclear star clusters in late-type galaxies is expressed as \citep{Georgiev16}
\begin{equation}\label{eq:Mnsc-r}
    r_\mathrm{eff,NSC}=3.23\,\mathrm{pc}\left(\frac{M_\mathrm{NSC}}{3.6\times 10^6\,\Msun} \right)^{0.321}\,.
\end{equation} 
Thus, $r_\mathrm{AGN}/r_\mathrm{eff,NSC}$ in Eq.~\eqref{eq:NBHcross} increases from 0.01 to 0.1 as $M_\mathrm{SMBH}$ increases from $10^5$ to $10^9\,\Msun$.

Following \citet{Bartos17}, we use the log-normal fit to the observed AGN mass function in the local universe \citep{Greene07,Greene09} in Eq.~\eqref{eq:merger_rate_int}
\begin{align}
\label{eq:dnagn_dmsmbh}
\frac{\mathrm{d} n_\mathrm{AGN}}{\mathrm{d} M_\mathrm{SMBH}}=&\frac{3.4\times 10^{-5}\,\mathrm{Mpc^{-3}}}{M_\mathrm{SMBH}}\nonumber\\
&\times 10^{-[\mathrm{log}(M_\mathrm{SMBH}/\Msun)-6.7]^2/1.22}. 
\end{align}
This mass function includes low-luminosity AGN 
with Eddington ratios of $f_{\rm Edd}\equiv L/L_{\rm Edd}=0.01$.
We include mergers in such low luminosity AGN 
since the rate of mergers from pre-existing BHs is not very sensitive to the accretion rate onto the SMBH (Models~1, 9, 10). 
Below 
$f_{\rm Edd}\approx 0.01$
geometrically thin AGN disks transition to geometrically thick advection-dominated accretion flows \citep{Narayan08}. In such low-density flows, it is not obvious if mergers proceed as we modeled in this paper. 
Here, to be conservative, we consider the rate of mergers only in AGN disks
with $f_{\rm Edd}\gtrsim 0.01$.

By integrating Eq.~(\ref{eq:merger_rate_int}) between $M_\mathrm{SMBH,min}=10^5\,\Msun$ and $M_\mathrm{SMBH,max}=10^9\,\Msun$ using the assumptions above, we find
\begin{align}\label{eq:merger_rate}
\mathcal{R}_\mathrm{sBH} \sim & 3\,\mathrm{Gpc^{-3}yr^{-1}}\left(\frac{f_\mathrm{BH,mer}}{0.5}\right)\left(\frac{t_\mathrm{AGN}}{30\,\mathrm{Myr}}\right)^{-1}\nonumber\\
&\times\left(\frac{r_\mathrm{AGN,MW}}{0.1\,\mathrm{pc}}\right) \left(\frac{\eta_\mathrm{n,BH}}{0.005\,\Msun^{-1}}\right)\,.
\end{align}
The relative contribution of the mass ranges 
$M_\mathrm{SMBH}=10^{5-6}$, $10^{6-7}$, $10^{7-8}$, $10^{8-9}\,\Msun$ are 
$0.96\,\%$, $34\,\%$, $59\,\%$ and $6.1\,\%$, respectively. 
The rate is dominated by $M_\mathrm{SMBH}\approx 10^{7-8}\,\Msun$; this is because the peak of the AGN mass function is at $10^{6.7}\,\Msun$ \citep{Greene07} and the number of BHs crossing the AGN disk increases with $M_\mathrm{SMBH}^{0.83}$.

\subsubsection{Uncertainties in the merger rate estimate}
The merger rate estimate in AGN is parameterized in Eq.~\eqref{eq:merger_rate} with the merger fraction per AGN lifetime $f_\mathrm{BH,mer}/t_{\rm AGN}$, the radius of the AGN for a MW-sized galaxy, $r_\mathrm{AGN,MW}$, and the number of BHs per unit mass in the nuclear star cluster. The uncertainties in these parameters may be estimated as follows. 

The radius of the AGN disk based on  mid-infrared observations is given by Eq.~\eqref{eq:r_agn}.
We assume that the allowed range of $r_\mathrm{AGN,MW}=0.06-0.2$ pc for $M_\mathrm{SMBH}=4\times 10^6\,\Msun$, considering that the merger rate is dominated by low-luminosity AGNs 
with $f_{\rm Edd}= 0.01-0.1$ \citep{Kelly13}. 

Table~\ref{table_results} shows the value of $f_\mathrm{BH,mer}$ for the different models.
For a stationary accretion disk with a fixed accretion rate and Eddington rate, the merger fraction per unit time ($f_\mathrm{BH,mer}/t$) is 
$1.7\times 10^{-8}$, $2.0\times 10^{-8}$, $1.9\times 10^{-8}$, 
$1.3\times 10^{-8}$, and $6.2\times 10^{-9}\,\mathrm{yr}^{-1}$ at 1, 3, 10, 30, and 100 Myr, respectively
in the fiducial model. 
Compared to its value at 30 Myr, $f_\mathrm{BH,mer}/t$ is higher by a factor of $\sim 1.6$ at 3 Myr and lower by a factor of $\sim 2.1$ at 100 Myr. 
Thus the merger fraction is correlated with the lifetime of the AGN disk (Figure~\ref{fig:num_mer}), 
which reduces the uncertainties of the merger rate caused by the uncertain AGN disk lifetime. 
From Table~\ref{table_results}, the maximum and minimum $f_\mathrm{BH,mer}/t$ at 30 Myr are $\sim 2.5\times 10^{-8}$ (Model~19) 
and $\sim 6.7\times 10^{-10}\,\mathrm{yr}^{-1}$ (Model~2), respectively. 
Based on these results, we consider a reasonable range of uncertainty to be $f_\mathrm{BH,mer}/t_\mathrm{AGN}\sim 3\times 10^{-10}\, \mathrm{yr}^{-1}$ -- $4\times10^{-8}\, \mathrm{yr}^{-1}$, so that $0.018\lesssim (f_\mathrm{BH,mer}/0.5)(t_\mathrm{AGN}/30\mathrm{Myr})^{-1}\lesssim 2.4$ in Eq.~\eqref{eq:merger_rate}.

In comparison, $f_\mathrm{BH,mer}/t_\mathrm{AGN}$ has been poorly constrained in previous studies, with the allowed range as wide as $\sim 5\times 10^{-13} - 3 \times 10^{-7}$ \citep{McKernan17}. 
The uncertainty in $f_\mathrm{BH,mer}$ comes from the uncertainties of the binary fraction, the capture fraction of BHs by AGN disks, and the merger fraction of binaries. 
The typical lifetime of AGN disks ($t_\mathrm{AGN}$) also has large uncertainties of $1-100$ Myr \citep[e.g.][]{Martini01,Haiman01,Marconi04,Martini04}. 
In the present study, we find that binaries are efficiently formed by gas-capture and dynamical mechanisms, 
and as a result, the merger fraction is relatively insensitive to the highly uncertain pre-existing binary fraction (Models~1, 29-31). 
The number of mergers is a factor 5 smaller than in the fiducial model if the initial BH population is isotropic (Model~16 $\beta_v=1$, cf. Model~1 with $\beta_v=0.2$.).

For a Salpeter initial mass function with $\delta_\mathrm{IMF}=-2.35$, the number of BHs per unit mass $\eta_\mathrm{n,BH}$ is $\sim 0.002\,\Msun^{-1}$ if we assume that the mass range of stars is $0.1-140\,\Msun$ and $20-140\,\Msun$ stars form BHs. 
On the other hand, the initial mass function in galactic centers is suggested to be top-heavy referring to observational ($\delta_\mathrm{IMF}=-1.7\pm0.2$ by \citealt{Lu13}) and theoretical studies \citep{Nayakshin07}. 
For example, $\eta_\mathrm{n,BH}$ is $\sim 0.01\,\Msun^{-1}$ for $\delta_\mathrm{IMF}=-1.7$. 
Furthermore, within parsec regions, numerical studies \citep{MiraldaEscude00,Freitag06,Hopman06,Antonini14} show that the number of BHs per unit mass $\eta_\mathrm{n,BH}$ is enhanced by a factor of $\sim 2$ by mass segregation. 
We assume that the range of $\eta_\mathrm{n,BH}$ is $\sim 0.002-0.02\,\Msun^{-1}$ allowing for the possibility of both a top-heavy initial mass function and mass segregation.

In summary, with the uncertainties adopted above,  
$f_\mathrm{BH,mer}/t_\mathrm{AGN}=(3-400)\times 10^{-10}\,\mathrm{yr}^{-1}$, 
$r_\mathrm{AGN,MW}=0.06-0.2$ pc,
and $\eta_\mathrm{n,BH}=0.002-0.02\,\Msun^{-1}$, 
the allowed range of merger rates is estimated to be
\begin{equation}
    0.02\,\mathrm{Gpc^{-3}yr^{-1}}\lesssim \mathcal{R}_\mathrm{sBH}\lesssim 60\,\mathrm{Gpc^{-3}yr^{-1}}
\end{equation}

In comparison, the current measurement of the merger rate using the LIGO--VIRGO observations is in the range
9.7-101
$\,\mathrm{Gpc^{-3}yr^{-1}}$ \citep{TheLIGO18}. 
Thus mergers in AGN disks are plausible candidates for the observed GW events.

\subsubsection{Extreme-mass-ratio inspirals}
\label{sec:emri}

We next predict the rate of extreme-mass-ratio inspirals (EMRIs). 
EMRIs into SMBHs of $\sim 10^4-10^7\,\Msun$ are promising targets for future low-frequency GW observations by LISA \citep{AmaroSeoane07}. 
In our simulation, the inner boundary is at $10^{-4}$ pc, which corresponds to $\sim 530\,R_g$ for a $4\times 10^6\,\Msun$ SMBH. 
The distance from which a $5\,\Msun$ BHs on a circular orbit can migrate to the SMBHs within a Hubble time via GW radiation is $\sim1000\,R_g$. 
However, if BHs accumulate at migration traps ($\lesssim 490\,R_g$) and repeatedly merge with one another \citep{Bellovary16,Secunda19,Yang19a,Yang19b,Gayathri19}, the number of BHs which can spiral into SMBHs may decrease significantly. 
Thus not all BHs which migrate within our inner boundary may spiral into the SMBH, 
and so $N_\mathrm{acc}$ is an upper limit for the number of EMRIs.
According to the results of our simulations (Table~\ref{table_results}), $N_\mathrm{acc}\sim 1-5$ at 30 Myr. 
If we assume that $N_\mathrm{acc}$ is independent of ${\dot M}_\mathrm{SMBH}$ 
and most AGNs are low-luminosity AGN with $\sim0.01\,f_\mathrm{Edd}$ \citep{Kelly13}, and 
$N_\mathrm{acc}$ is roughly proportional to time (brown line in Figure~\ref{fig:num_ev}),
the number of BHs accreted onto the SMBH per accreted gas mass can be approximated as $N_\mathrm{acc}/({\dot M}_\mathrm{out} t_\mathrm{AGN}) \sim (4-20) \times 10^{-5} \,\Msun^{-1}$. 
Then we can roughly estimate 
the EMRI event rate density as 
\begin{align}
\label{eq:rate_emri}
R_\mathrm{EMRI}&\sim
\frac{N_\mathrm{acc}{\dot \rho}_\mathrm{SMBH}}{t_\mathrm{AGN} {\dot M_\mathrm{SMBH}}}\nonumber\\
&\sim(0.1-0.6)\,\mathrm{Gpc^{-3}yr^{-1}}
\left[\frac{N_\mathrm{acc}/({\dot M}_\mathrm{SMBH}t_\mathrm{AGN})}{(4-20) \times 10^{-5} \,\Msun^{-1}}\right]\nonumber\\
&\qquad \left(\frac{{\dot \rho}_\mathrm{SMBH}}{3\times 10^3\,\mathrm{Gpc}^{-3}\Msun \mathrm{yr}^{-1}}\right).
\end{align}
where ${\dot \rho}_\mathrm{SMBH}$ is the total mass accretion rate onto all SMBHs in the local Universe, 
and we adopt ${\dot \rho}_\mathrm{SMBH}=3\times 10^3\,\Msun \mathrm{Gpc}^{-3}\mathrm{yr}^{-1}$ \citep{Marconi04}. 
If the EMRI rate by migration in AGN disks is $\sim 0.1-0.6\,\mathrm{Gpc^{-3}yr^{-1}}$, 
this channel may largely contribute to the EMRI rate, 
since the EMRI rate by stellar relaxation processes is predicted to be comparable,
$\sim 0.02-2\,\mathrm{Gpc^{-3}yr^{-1}}$
\citep[e.g.][]{Miller05,Hopman06,Freitag06,AmaroSeoane07,AmaroSeoane11,Aharon16,BarOr16,Babak17}. 
Even if repeated mergers take place at migration traps and reduce the number of EMRIs significantly, 
at least one massive BH can migrate and merge with the central SMBH during each AGN phase of every galaxy, 
which can be observed as intermediate-mass-ratio inspirals \citep[IMRIs, e.g.][]{Derdzinski18}. 
Note that the LISA detection rate is enhanced by the increased detection volume that corresponds to the increased migrator mass following repeated mergers. 

In our simulation, all BHs that migrate within the inner boundary $R_\mathrm{min}$ are single (not binary). This is because binaries either merge or are disrupted by binary-single interactions before they migrate to $R_\mathrm{min}$.

\subsection{Comparison with previous works}
\label{sec:comparison}

In this section, we compare our models and results with those in previous works on BH mergers in AGN disks by \citet{Bartos17} and \citet{Stone17}. 
A major difference between the present study and esrlier works is that we here model the time-evolving system explicitly, 
enabling us to follow the formation and destruction of binaries, together with the evolution of their separation, their center-of-mass velocity, and their radial distance from the SMBH consistently. 

\citet{Bartos17} considered the capture of binaries due to linear momentum exchange with an AGN disk within 0.01 pc from the SMBH, and the hardening of binaries by gas dynamical friction of the AGN disk and type I/II torques of a circumbinary disk. 
\citet{Bartos17} found the merger rate to be $\sim 1.2\,\mathrm{Gpc^{-3}yr^{-1}}$. 
Our study is a more detailed and extended version of \citet{Bartos17}. 
One difference between \citet{Bartos17} and this work is the binary fraction. 
\citet{Bartos17} considered the evolution of pre-existing binaries, whose fraction is assumed to be 0.3, 
while our study finds that a large fraction of single BHs captured within AGN disks later form binaries by gas-capture and dynamical mechanisms. 
A second difference is the assumption of the BH distribution. 
\citet{Bartos17} assumed an isotropic velocity distribution, in which the merger rate is lower by a factor of $\sim$ 6 than with the anisotropic velocity distributions adopted here (see Models~1 and 13-16 in Table~\ref{table_results}). 
Such anisotropic velocity distribution is predicted by theoretical \citep{Kocsis11b,Szolgyen18} and observational studies \citep{Trippe08,Yelda14,Feldmeier14,Feldmeier15}. 
Also \citet{Bartos17} assumed the strongly mass-segregated number density of $\gamma_\rho=0.5$, while our study assumes $\gamma_\rho=-0.5-0$ (Models~1, 22). 
A third difference is the assumed size of the AGN disks. 
\citet{Bartos17} consider a size of 0.01 pc, compared to the $\sim 0.03-0.1$ pc $(M_\mathrm{SMBH}/4\times10^6\,\Msun)^{1/2}$ adopted here.
The larger disk sizes are motivated by observations of AGN disks \citep{Burtscher13}. 

\citet{Stone17} considered the evolution of binaries formed in-situ in AGN disks at $\sim$pc from the SMBH. 
Both \citet{Stone17} and our study use the disk model proposed by \citet{Thompson05}, and consider the hardening by type I/II torque and binary-single interactions. 
The differences are that we treat binary-single interactions considering the evolution of the distribution of BHs, and we include a new mechanism of gas-capture to form new binaries. 
\citet{Stone17} estimated the merger rate from in-situ formed binaries to be $\sim 3\,\mathrm{Gpc}^{-3}\mathrm{yr}^{-1}$, in which the binary fraction (the binary number over the BH number) and the merger fraction are assumed to be 0.56 and 1, respectively. 
In our simulation, the fraction of the number of mergers from in-situ formed BHs ($N_\mathrm{mer,SF}$) over the number of in-situ formed BHs ($N_\mathrm{SF}$) is $\sim 0.14-0.67$ in Models~1, 9, 10, and 23-25. 
Note that these values are upper limits, since we assume that BHs form immediately at star formation. 
Following the estimate in \citet{Stone17}, 
we find the merger rate density for in-situ formed BHs to be
\begin{eqnarray}
    R_\mathrm{IS} &\sim& f_\mathrm{SF/AGN} (N_\mathrm{mer,SF}/N_\mathrm{SF}) f_\mathrm{BH} {\dot \rho}_\mathrm{SMBH}/{\bar m_\mathrm{BH}} \nonumber\\
    &\sim& 0.7-22 \,\mathrm{Gpc^{-3}yr^{-1}},
\end{eqnarray}
where $\bar m_\mathrm{BH}$ is the average BH mass, 
$f_\mathrm{BH}$ is the mass fraction of BHs over stars, 
${\dot \rho}_\mathrm{SMBH}$ is the total mass accretion rate onto all SMBHs in the local Universe, 
and $f_\mathrm{SF/AGN}$ is the star formation rate within the AGN disk over the accretion rate onto the SMBH. 
In the estimate above, we use $f_\mathrm{SF/AGN}=1$ and 
${\dot \rho}=3\times 10^3\,\Msun \mathrm{Gpc}^{-3}$ \citep{Marconi04} as adopted in \citet{Stone17}. 
For a top-heavy initial mass function with $\delta_\mathrm{IMF}=1.7-2.35$ \citep{Lu13}, we find $f_\mathrm{BH}=0.016-0.092$ and ${\bar m_\mathrm{BH}}=8.4-9.2\,\Msun$. 
Since these differences are relatively small, our estimated rate is consistent with that by \citet{Stone17}. 
We find that mergers 
between pre-existing BHs captured by AGN disks
($\sim 0.02-60\,\mathrm{Gpc^{-3}yr{-1}}$) 
are roughly comparable to 
the mergers from BHs formed in-situ in AGN disks ($\sim 0.7-22\,\mathrm{Gpc^{-3}yr^{-1}}$).

\begin{figure}
\includegraphics[width=90mm]{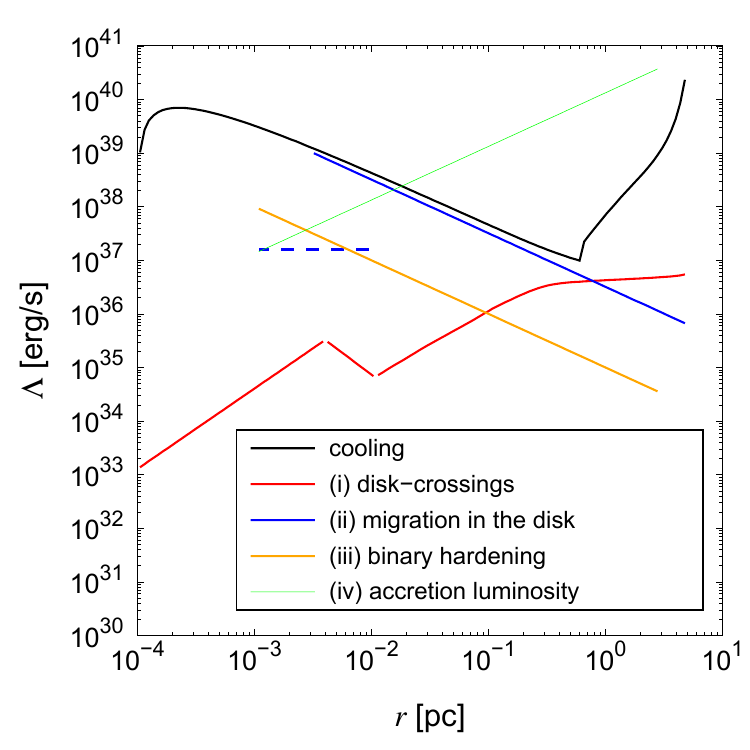}
\caption{
The cooling and heating rates in each cell $l$ as a function of the distance from the SMBH for the fiducial model. 
Solid and dashed black lines corresponds to the optically thick and thin cooling rates. 
Red, blue, orange, and green lines correspond to heating via the processes 
(i) disk-crossings
(ii) migration in the disk
(iii) binary hardening
and (iv) accretion luminosity, respectively.
Solid and dashed blue lines present the heating rate due to migration torque and damping of velocity by gas dynamical friction, respectively. 
}
\label{fig:cooling_rate}
\end{figure}

\subsection{Neglected effects}

In this section, we discuss several effects which have been neglected in our model. 

\subsubsection{Heating of AGN disks}
\label{sec:heating_disk}

We first discuss several heating 
processes which we have not included. 
All heating rates must be compared to the cooling rate of the disk,  which we show for our fiducial disk model in Fig.~\ref{fig:cooling_rate} as the black curve. 
The effective volumetric cooling rate is given by
\begin{align}
\Lambda_\mathrm{cool}\sim \frac{\sigma_\mathrm{SB}T_\mathrm{eff}^4}{h_\mathrm{disk}}V
\end{align}
where 
$V=4 \pi r_l \Delta r_l h_\mathrm{disk}$ is the disk's volume in the cell $r_l$, $T_\mathrm{eff}$ is the effective surface temperature, and $\sigma_\mathrm{SB}$ is the Stefan-Boltzmann constant. 
In the outer regions ($r\gtrsim 0.3\,\mathrm{pc}$), gas is optically thin, and is stabilized by star formation feedback.

In addition to the $\alpha$-viscosity and star formation feedback, which are implicitly part of our disk model, there are several processes which heat the AGN disk that we have not taken into account: 
\begin{enumerate}
 \item\label{i:heating:vdamp} damping of the orbital velocity of stars and compact objects in the spherical cluster when they cross the AGN disk,
\item \label{i:heating:torque} torques operating on single objects and (the center-of-mass of) binaries embedded in the AGN disk,  
\item \label{i:heating:binaries} hardening of binaries, as well as 
\item \label{i:heating:accretion} radiation feedback from accretion onto stellar-mass BHs. 
\end{enumerate}
We estimate the order of magnitude of these effects in turn below.

First, we consider the total heating rate due to the damping of the velocities of a spherical cluster of stars during disk crossings (process~\ref{i:heating:vdamp}). 
During one orbital period around the SMBH, the orbital velocity of an object is damped by 
$\Delta v \sim v_\mathrm{orb} \Delta M_\mathrm{cross}/m_i$ \citep[e.g.][]{Bartos17}, 
where $\Delta M_\mathrm{cross}=\pi [\mathrm{max}(r_\mathrm{BHL},R_{\mathrm{star},i})]^2 h_\mathrm{cross} \rho_\mathrm{gas}$ is the gas mass that passes within the Bondi-Hoyle-Lyttleton radius of the object when crossing the disk (twice per orbit),
$h_\mathrm{cross}\sim 2h_{\rm disk}/\sin\iota$ is the distance in which the object crosses the  disk where $\iota$ is the inclination of the orbit and recall that $h_{\rm disk}$ is the disk half-thickness, and $v_\mathrm{orb}$ is the orbital velocity of the object.  
In practice $r_{\rm BHL}$ is smaller than the Hill radius (Eq.~\ref{eq:rHill}) if the orbital inclination satisfies $\iota \gtrsim = 3^{1/6}(m_i/M_{\rm SMBH})^{1/3} = 5\times 10^{-3} (m_i/0.36\Msun)^{1/3}$, so the accretion rate during disk-crossing is typically  not limited by the Hill radius.\footnote{For $\iota \lesssim 5\times 10^{-3}$, Bondi accretion must be limited to the Hill sphere, which reduces the accretion rate by $(r_{\rm Hill}/r_{\rm BHL})^2$.} 
If this condition holds and if
$\sin\iota \geq h_{\rm disk}/r$, then for circular orbits $r_{\rm BHL}$ is determined by the $\iota$ inclination of the orbit as\footnote{recall that $v_i$ is the relative velocity with respect to the gas that follows the equatorial Keplerian orbit}
\begin{equation}
    r_{\rm BHL} = \frac{G m_i}{c_s^2 + v_i^2}
    \approx \frac{G m_i}{2(1-\cos\iota) v_{\rm orb}^2}
    =
    \frac{\cos^2\frac{\iota}2}{\sin^2\iota}\frac{m_i}{ M_{\rm SMBH}}r
    \,.
\end{equation}
Since $v_\mathrm{orb} \gg \Delta v$, the kinetic energy loss rate averaged over an orbit is
$\sim m_{i} \Delta v v_\mathrm{orb} / t_\mathrm{orb}$. The total heating rate by damping of the orbital velocity of all stars in cell $l$ is estimated as 
\begin{align}
\label{eq:heat_i}
\Lambda_\mathrm{i,damp}\sim & \frac{N_{\mathrm{ob},l} m_i \Delta v v_\mathrm{orb}}{ t_\mathrm{orb}}
= N_{\mathrm{ob},l} \frac{\cos^4 \frac{\iota}{2}}{\sin^5\iota}\frac{G^{3/2} m_i^2 \rho_{\rm gas}h_{\rm disk}}{M_{\rm SMBH}^{1/2}r^{1/2}}
\nonumber\\
\sim &2\times 10^{34}\,\mathrm{erg~s^{-1}}
\left(\frac{N_{\mathrm{ob},l}}{2\times 10^4}\right)
\frac{m_i^2}{(0.36\,\Msun)^2}\nonumber\\
&\times\left(\frac{M_\mathrm{SMBH}}{4\times 10^6\,\Msun}\right)^{-1/2}
\left(\frac{r_l}{0.1\,\mathrm{pc}}\right)^{-2.5}\nonumber\\
&\times\left(\frac{\rho_\mathrm{gas}r_l^3}{7\times 10^5\,\mathrm{\Msun}}\right)
\left(\frac{h_{\rm disk}/r_l}{10^{-3}}\right)\left(\frac{\sin\iota}{0.3}\right)^{-5}\nonumber\\
&\qquad\qquad\qquad\mathrm{for}~R_{\mathrm{star},i}<r_\mathrm{BHL}\,,
\end{align}
and 
\begin{align}
\label{eq:heat_i_r}
\Lambda_\mathrm{i,damp}
=& N_{\mathrm{ob},l}\frac{G^{3/2}M_{\rm SMBH}^{3/2}R_{\rm star}^2\rho_{\rm gas}h_{\rm disk}}{\sin\iota \,r^{5/2}}\nonumber\\
\sim &4\times 10^{32}\,\mathrm{erg~s^{-1}}
\left(\frac{N_{\mathrm{ob},l}}{2\times 10^4}\right)
\left(\frac{m_i}{0.36\,\Msun}\right)\nonumber\\
&\times\left(\frac{M_\mathrm{SMBH}}{4\times 10^6\,\Msun}\right)^{3/2}
\left(\frac{r_l}{0.1\,\mathrm{pc}}\right)^{-4.5}\nonumber\\
&\times\left(\frac{\rho_\mathrm{gas}r_l^3}{7\times 10^5\,\mathrm{\Msun}}\right)\left(\frac{h_{\rm disk} /r_l}{10^{-3}}\right)
\left(\frac{\sin\iota}{0.3}\right)^{-1}\nonumber\\
&\qquad\qquad\qquad\mathrm{for}~R_{\mathrm{star},i}>r_\mathrm{BHL}\,,
\end{align}
where $N_{\mathrm{ob},l}$ is the number of objects\footnote{Since we have $N_{\rm bin}=120$ bins with a log-uniform distribution in radius between $r_{\max}=5\mathrm{pc}$ and $r_{\min}=10^{-4}\mathrm{pc}$, $\Delta r/r=(\ln \Lambda)/N_{\rm bin}$ where $\ln \Lambda=\ln (r_{\rm max}/r_{\rm min})$ and the number density of objects in the spherical component is $n(r)$ is given by Eq.~\eqref{eq:stellar_density}, the number of objects in a cell at $r$ is $N_{\mathrm{ob},l}(r)= 4\pi r^2 n(r) \Delta r$ which scales as $\propto r^{2.5}$ for $r\ll 0.3\,\mathrm{pc}$ and $\propto r^{1.2}$ for $r\gg 0.3\,\mathrm{pc}$.}
in cell-$l$ and $R_{\mathrm{star},i}$ is given by Eq.~\eqref{eq:Rstar}.
Additionally, the total heat of process~\ref{i:heating:vdamp} is limited by the initial kinetic energy of objects in the Keplerian corotating frame with the gas.\footnote{assuming that the gravitational potential is dominated by the SMBH} If we average the total heat over $t_\mathrm{AGN}$, the upper limit for the time-averaged heating rate by disk-crossing is given by  
\begin{align}
\label{eq:heat_i_limit}
\Lambda_\mathrm{i,lim}\sim & \frac{N_{\mathrm{ob},l} m_i (2v_\mathrm{orb})^2}{ 2t_\mathrm{AGN}}\nonumber\\
\sim & 5\times 10^{37}\,\mathrm{erg~s^{-1}}
\left(\frac{N_{\mathrm{ob},l}}{2\times 10^4}\right)
\left(\frac{m_i}{0.36\,\Msun}\right)\nonumber\\
&\left(\frac{M_\mathrm{SMBH}}{4\times 10^6\,\Msun}\right)
\left(\frac{r_l}{0.1\,\mathrm{pc}}\right)^{-1}
\left(\frac{t_\mathrm{AGN}}{30\,\mathrm{Myr}}\right)^{-1}
\end{align}
The red curve in Fig.~\ref{fig:cooling_rate} shows $\mathrm{min}(\Lambda_\mathrm{i,damp},\Lambda_\mathrm{i,lim})$. 
For $R_{\mathrm{star},i}>r_\mathrm{BHL}$ ($r\lesssim 0.2\,\mathrm{pc} \times\sin^2\iota/\cos^2(\iota/2)$, Eq.~\ref{eq:heat_i_r}), 
we assume $1/\sin\iota=1.5$ as a typical inclination for isotropic orbits. 

On the other hand, since the heating rate for $R_{\mathrm{star},i}<r_\mathrm{BHL}$
(Eq.~\ref{eq:heat_i}) 
is sensitive to $\sin\iota$, the allowed values for $\sin\iota$ need to be examined carefully. 
Even if the heating rate is much higher than the AGN cooling rate, this may still be a small effect if the heating timescale is very short. Although stars with small $\sin\iota$ strongly heat gas as $\Lambda_\mathrm{i,damp} \propto \sin\iota^{-5}$, 
the probability (timescale) for a star to have $\iota$ also scales with $\sin\iota^{5}$ while $r_\mathrm{BHL}<r_\mathrm{Hill}$. Thus, the product of the heating rate at $\iota$ and the timescale for a star to have $\iota$ is roughly constant. On the other hand, the maximum dissipated energy during the AGN lifetime differs for different $\iota$. We assume that the upper limit for the heating rate averaged on the AGN lifetime 
is given by setting all stars to have the inclination $\iota_\mathrm{cap}$ at which the relative velocities of stars are damped on the AGN lifetime. 
Such inclination is derived to satisfy $v_\mathrm{rel}/t_\mathrm{AGN}=\Delta v/t_\mathrm{orb}$. 
Assuming $v_\mathrm{rel}\sim v_\mathrm{orb}\mathrm{sin}\iota$, 
\begin{align}
(\mathrm{sin}\iota_\mathrm{cap})
\sim &
\left(\frac{t_\mathrm{AGN}}{t_\mathrm{orb}}\frac{4 \pi m_i \rho_\mathrm{gas}h_\mathrm{disk}r^2}{ M_\mathrm{SMBH}^2}\right)^{1/6}\nonumber\\
\sim & 0.13 
\times\left(\frac{\rho_\mathrm{gas}r_l^3}{7\times 10^5\,\mathrm{\Msun}}\right)^{1/6}
\left(\frac{h_{\rm disk} /r_l}{10^{-3}}\right)^{1/6}\nonumber\\
&\left(\frac{m_i}{0.36\,\Msun}\right)^{1/6}
\left(\frac{M_\mathrm{SMBH}}{4\times 10^6\,\Msun}\right)^{-1/4}\nonumber\\
&\left(\frac{r_l}{0.1\,\mathrm{pc}}\right)^{-1/4}
\left(\frac{t_\mathrm{AGN}}{30\,\mathrm{Myr}}\right)^{1/6}. 
\end{align}
We use $i_\mathrm{cap}$ to estimate the heating due to disk crossing for $R_{\mathrm{star},i}<r_\mathrm{BHL}$ ($r\gtrsim 10^{-2}\,\mathrm{pc}$) in Fig.~\ref{fig:cooling_rate}. Since the fraction of stars having $v_\mathrm{rel} \lesssim \iota_\mathrm{cap} v_\mathrm{orb}$ over all stars is $\lesssim \iota_\mathrm{cap}$, and the average heating rate for stars with $\iota>\iota_\mathrm{cap}$ rapidly decreases, the average heating rate is overestimated by $\sim 1/\iota_\mathrm{cap}$.

On the other hand, 
for $R_{\mathrm{star},i}<r_\mathrm{BHL}$, 
the heating rate using $m_i=0.36\Msun$ in Eq.~\eqref{eq:heat_i} underestimates the true heating rate since $\Lambda_{i,\rm damp}\propto m_i^2$ and $\langle m^2\rangle$ may be much larger than $\langle m\rangle^2$, and so massive stars dominate the heating due to the strong dependence on the mass. 
If we assume pre-existing stars are mostly old ($\gtrsim$ a few 100 Myr), the maximum mass of massive stars is $\sim 3\,\Msun$. Here we consider the extreme case that all stars are as massive as $3\,\Msun$. As the mass of a stellar cluster is fixed, $N_{\mathrm{ob},l}$ for $m_i=3\,\Msun$ is lowered by a factor of $\sim 10$ compared to that for $m_i=0.36\,\Msun$. 
Then, $\Lambda_\mathrm{i,damp}\propto N_{\mathrm{ob},l} \langle m^2\rangle$ is enhanced by a factor of $\sim 10$ compared with Eq.~\eqref{eq:heat_i} for $m_i=0.36\Msun$. 

In conclusion, the dynamical heating rate by disk crossing of stars  (process \ref{i:heating:vdamp}) may exceed the cooling rate at $r\sim 0.5\,\mathrm{pc}$. If the number of BHs compared to the number of stars is between $10^{-4}$--$10^{-3}$, and the typical BH mass is $10$--$100$ times larger than the typical stellar mass, the heating due to disk crossing $\Lambda_\mathrm{i,damp} \propto N_{{\rm ob},l} m_i^2$ (Eq.\ref{eq:heat_i}) may be comparable to that of stars within a factor 10.

Next, we consider heating corresponding to torques operating on objects embedded in the disk (process \ref{i:heating:torque}). 
Since these objects are assumed to migrate towards the SMBH via gas torques,  the disk can be heated by the corresponding increase in these objects' binding energy. 
If we assume that roughly half of all BHs 
migrate to $\sim 10^{-3}-10^{-2}\,\mathrm{pc}$ in our models (see black line in panel (a) of Fig.~\ref{fig:grid_variables}), the total dissipation rate  in cell~$l$, averaged over $t_\mathrm{AGN}$, is 
\begin{align}
\label{eq:heat_ii}
\Lambda_\mathrm{ii,mig}\sim & 
\frac{N_\mathrm{mig} G m_i M_\mathrm{SMBH}}{t_\mathrm{AGN}}\left(\frac{1}{r_l-\Delta r_l}-\frac{1}{r_l}\right)\nonumber\\
\sim & 3\times 10^{36}~\mathrm{erg}~\mathrm{s}^{-1}
\left(\frac{N_\mathrm{mig}}{10^4}\right)
\left(\frac{m_i}{10\,\Msun}\right)\nonumber\\
&\left(\frac{M_\mathrm{SMBH}}{4\times 10^6\,\Msun}\right)
\left(\frac{r_l}{1\,\mathrm{pc}}\right)^{-1}
\left(\frac{\Delta r_l/r_l}{0.08}\right)
\left(\frac{t_\mathrm{AGN}}{30\,\mathrm{Myr}}\right),
\end{align}
where $N_\mathrm{mig}$ is the number of objects that migrated to $r\lesssim 10^{-2}\,\mathrm{pc}$. $\Delta r_l/r_l=0.08$ is the value adopted in the fiducial model ($N_\mathrm{cell}=120$). 
The solid blue line in Fig.~\ref{fig:cooling_rate} presents $\Lambda_\mathrm{ii,mig}$, with parameter values as in Eq.~\eqref{eq:heat_ii}. 
We can see that this heating rate roughly matches the cooling rate in the optically thick region ($r\lesssim 0.3\,\mathrm{pc}$), but is significantly below it in the optically thin region ($r\gtrsim 0.3\,\mathrm{pc}$).
As the number of objects in the AGN disk varies with time (cyan line in Fig.~\ref{fig:num_ev}), 
the heating rate can be momentarily be even higher than the average value in Eq.~\eqref{eq:heat_ii}, which may require revisions of the AGN disk model.

Eq.~\eqref{eq:heat_ii} neglects the contribution from stars for simplicity.  In our simulation, the total mass of stars captured by the AGN disk over 30 Myr is $\sim 10^5\,\Msun$ (brown line in panel~(a) of Fig.~\ref{fig:grid_variables}). If all these stars migrate to $\lesssim 10^{-2}\,\mathrm{pc}$, the heating rate is enhanced by a factor of 2 compared to Eq.~\eqref{eq:heat_ii}. 
However, the migration timescale is inversely proportional to the mass of migrating objects as long as long as a gap does not form, which suggests that typical-mass stars ($\sim 0.36\,\Msun$) do not migrate inward within 30 Myr if they begin at $r\gtrsim 0.2\,\mathrm{pc}$ (blue line in panel~(c) of Fig.~\ref{fig:grid_variables}). 
This suggests a minor contribution from stars to heating compared with BHs.

After binary BHs migrate within $\sim 10^{-2}\,\mathrm{pc}$ of the SMBH, 
they receive kicks due to binary-single interactions, which dominates the heating of the orbital velocities of binaries (e.g. panel (c) of Fig.~\ref{fig:tot_cont}). 
Following each kick, the center-of-mass velocity of the binary is damped by gas dynamical friction, heating the disk. 
If we assume that each binary experiences $N_\mathrm{damp}\sim 10$ binary-single interactions, 
the total energy transferred to the disk can be estimated as
\begin{align}
    E_\mathrm{ii,damp}&\sim \frac{1}{2}m_\mathrm{tot,cap}v_\mathrm{typ}^2 N_\mathrm{damp} \nonumber\\
    &\sim 4\times 10^{53}\,\mathrm{erg}\frac{m_\mathrm{tot,cap}}{ 10^5\,\Msun}\left(\frac{v_\mathrm{typ}}{200\,\mathrm{km/s}}\right)^2 \frac{N_\mathrm{damp}}{10},
\end{align}
where $v_\mathrm{typ}$ is the typical 
excess velocity
damped by gas dynamical friction, 
$m_\mathrm{tot,cap}$ is the total mass of binaries captured by the AGN disk, 
and $N_\mathrm{damp}$ is the typical number of velocity damping episodes per binary. 
If we spread this total energy over the total simulation time of 30 Myr, the corresponding average heating rate is 
\begin{align}
\Lambda_\mathrm{ii,damp}&\sim E_\mathrm{ii,damp}/t_\mathrm{AGN}\nonumber\\
    &\sim 4\times 10^{38}\,\mathrm{erg/s}\left(\frac{E_\mathrm{ii,damp}}{10^{53}\,\mathrm{erg}}\right)\left(\frac{t_\mathrm{AGN}}{30\,\mathrm{Myr}}\right)^{-1}. 
\end{align}
We further assume that binary-single interactions occur uniformly between $10^{-2}-10^{-3}\,\mathrm{pc}$, where mergers typically occur (panel~(b) of Fig.~\ref{fig:grid_variables}). The heating rate under these assumptions is shown by the dashed blue line in Fig.~\ref{fig:cooling_rate}. We can see that this process represents a relatively minor contribution compared with the cooling or the other heating processes. 
We speculate that $\Lambda_\mathrm{ii,damp}$ is not strongly time-dependent, as the number of merged binaries is roughly proportional to the elapsed time (Fig.~\ref{fig:num_mer}).

Interactions with the gas disk also 
harden BH binaries (process~\ref{i:heating:binaries}), 
further heating the ambient gas. According to Fig.~\ref{fig:edis}, binaries are hardened by gas dynamical friction and Type I/II torques to $s\sim 10^{-6}\,\mathrm{pc}$ at $r\sim 10^{-2}\,\mathrm{pc}$ and to $s\sim 10^{-4}\,\mathrm{pc}$ at $r\sim 1\,\mathrm{pc}$ in the AGN disk.  For simplicity, we assume that binaries harden to $s\sim 10^{-4} r$. Then, the corresponding heating rate as a function of distance from the SMBH is approximately
\begin{align}
    \Lambda_\mathrm{iii,hard}\sim &\frac{N_\mathrm{hard,bin}G m_1 m_2}{  t_\mathrm{AGN}}\left(\frac{1}{s-\Delta s}-\frac{1}{s}\right)
    \nonumber\\
    \sim &10^{36}\,\mathrm{erg/s}
    \left(\frac{N_\mathrm{hard,bin}}{ 10^4}\right)
    \left(\frac{m_1}{10\,\Msun}\right)
    \left(\frac{m_2}{10\,\Msun}\right)\nonumber\\
     &\left(\frac{r_l}{1\,\mathrm{pc}}\right)^{-1}
     \left(\frac{\Delta r_l/ r_l}{0.08}\right)
    \left(\frac{t_\mathrm{AGN}}{30\,\mathrm{Myr}}\right)^{-1},
\end{align}
which is shown by the orange line in Fig.~\ref{fig:cooling_rate}. 
Although we assume that all binaries are hardened as they migrate, 
if all binaries form in some cell and they are hardened to $s\sim 10^{-4}r$, $\Lambda_\mathrm{iii,hard}$ is enhanced by a factor of 10, which is slightly lower than the cooling rate (black lines). 
Hence, we estimate that the heating by process (iii) does not affect the disk properties.

To estimate the radiation luminosity from accretion onto stellar-mass BHs (process \ref{i:heating:accretion}), 
we assume that the number of BHs in the  disk at cell $l$ follows $N_{\mathrm{disk},l}\sim 40 (r_l/3\,\mathrm{pc})$ for $r_l<3\,\mathrm{pc}$. This is motivated by the initial BH distribution, which remains roughly
in place
at $t=30\,\mathrm{Myr}$ (cyan and blue lines in panel (a) of Fig.~\ref{fig:grid_variables}). 
The luminosity of the population of stellar-mass BHs can then be estimated by 
\begin{align}
    L_\mathrm{vi,rad}&
    \sim N_{\mathrm{disk},l}L_\mathrm{Edd}\Gamma_\mathrm{Edd,cir}\nonumber\\
    &\sim 4\times 10^{40}\,\mathrm{erg/s}
    \left(\frac{r_l}{3\,\mathrm{pc}}\right)
     \left(\frac{m_\mathrm{BH}}{10\,\Msun}\right)
    \left(\frac{\Gamma_\mathrm{Edd,cir}}{1}\right),
\end{align}
which is shown by the green line in Fig.~\ref{fig:cooling_rate}. 
This luminosity exceeds the cooling rate at $r\gtrsim 10^{-2}\,\mathrm{pc}$, which is consistent with the estimate by \citet{Levin03}. 
However, the majority of this radiation might escape in directions perpendicular to the AGN disk due to predicted anisotropic radiation \citep[e.g.][]{Sugimura18} and the small disk height compared with the Bondi-Hoyle-Lyttleton radii and the Hill radii. 
Also, the HII regions are mostly confined inside the Bondi-Hoyle-Lyttleton radii in simulations. In this case, the gas on larger scales is not affected by radiation feedback \citep[e.g.][]{Toyouchi20}. 
Here, the size of HII regions for stellar-mass BHs at $r\sim \mathrm{pc}$ in the AGN disk is 
\begin{align}
    R_\mathrm{HII}
  \sim  & 0.01~\mathrm{pc} 
    \left(1+710\frac{Z}{Z_\odot} \right)^{-1/3}
    \left(\frac{m_i}{10\,\Msun}\right)^{1/3}\nonumber\\
    & \left(\frac{\rho_\mathrm{gas}}{10^6\,\mathrm{\Msun~pc^{-3}}}\right)^{-2/3}
    \left(\frac{\sqrt{c_s^2+v_i^2}}{1\,\mathrm{km/s}}\right)^{-4/3}
\end{align}
\citep[][]{Toyouchi20}, 
where $Z$ is the metallicity of gas, and $Z_\odot$ is the solar value, 
while the Bondi-Hoyle-Lyttleton radius is 
\begin{align}
    r_\mathrm{BHL}
  \sim  & 0.04~\mathrm{pc} 
  \left(\frac{m_i}{10\,\Msun}\right)
     \left(\frac{\sqrt{c_s^2+v_i^2}}{1\,\mathrm{km/s}}\right)^{-2}.
\end{align}
Due to the small filling factor of the HII regions, we expect that process \ref{i:heating:accretion} does not significantly affect gas properties on large scales in the inner regions ($r\lesssim$ pc) of the AGN disk. 
On the other hand, in the outer regions ($r\gtrsim$ a few pc), since the HII regions become larger than the Bondi-Hoyle-Lyttleton radii, radiation from stellar-mass BHs can significantly heat the AGN disk. 

Let us now consider the luminosity from stars embedded in the disk. If we assume the Salpeter initial mass function with mass ranges from $0.1$ to $140\,\Msun$ and the luminosity by Eq.~\eqref{eq:mass_luminosity}, the average luminosity per stellar mass over 100 Myr is $\sim 30\,\Lsun$. 
If we roughly assume that the number of stars captured by the AGN disk is two orders of magnitude larger than that of BHs with $10\,\Msun$ (see cyan and orange lines in panel~(a) of Fig.~\ref{fig:grid_variables}), the total luminosity from stars is lower than that from BHs by a factor of $\sim 300$. Hence, we expect that the luminosity from stars are negligible compared to that from BHs.

Overall, we conclude that migration torques (process \ref{i:heating:torque}) may heat and thicken the AGN disk relative to the model we adopted. 
Furthermore, radiation from accreting stellar-mass BHs (process \ref{i:heating:accretion}) may also significantly heat the outer regions of the AGN disk. 
Additionally, more localized structures in the disk (e.g. the widths of the annular gaps around compact objects, local inhomogeneities) may be affected by mechanical feedback from both BHs and stars. Such effects need to be  investigated in the future. 

\subsubsection{Kozai-Lidov mechanism}

There are several possibilities for the Kozai-Lidov (KL) mechanism to affect the dynamical evolution of compact objects in AGN disks. 
The KL mechanism operates on triple systems when the motion of an inner binary orbit is strongly misaligned with the outer orbit, 
and the timescale of the KL mechanism is given as 
\begin{equation}
    t_\mathrm{KL}=\frac{2}{3\pi}\frac{P_\mathrm{out}^2}{P_\mathrm{in}}(1-e_\mathrm{out}^2)^{3/2}\frac{m_1+m_2+m_3}{m_3}
\end{equation}
\citep{Kiseleva98}, where $P_\mathrm{out}$ and $P_\mathrm{in}$ are the orbital periods of the outer and inner orbits, respectively, $e_\mathrm{out}$ is the eccentricity of the outer orbit, $m_1$ and $m_2$ are the masses of inner binary components, and $m_3$ is the mass of the third body.
Here, we consider three triple-system configurations, composed respectively of 
\begin{enumerate}
\item \label{num_smbh_bhbh} a compact object binary and the central SMBH 
\item \label{num_bhbhbh} three compact objects 
\item \label{num_disk_smbh_bh} the central SMBH, a compact object, and the AGN disk. 
\end{enumerate}

For \ref{num_smbh_bhbh} and \ref{num_bhbhbh}, the angle between the inner and outer orbits is damped by gas dynamical friction in the AGN disk, which weakens the effect of the KL mechanism. 
However, if the outer orbit is eccentric, the coplanar configuration may also lead to very high eccentricities and an orbital flip typically on a timescale between 1 to 10$t_{\rm KL}$ \citep{Li_Naoz_Kocsis2014}.
Additionally, after binary-single interaction and before the binary is captured by the AGN disk, the angle is randomized. During this period, the KL mechanism may efficiently induce mergers. 

For \ref{num_smbh_bhbh}, the KL timescale is estimated as 
\begin{align}
    t_\mathrm{KL,i}\sim 2\,\mathrm{Myr}& \left(\frac{a_\mathrm{out}}{0.1\,\mathrm{pc}}\right)^3
    \left(\frac{a_\mathrm{in}}{1\,\mathrm{AU}}\right)^{-3/2}\nonumber\\
    &\left(\frac{m_\mathrm{bin}}{20\,\Msun}\right)^{1/2}
    \left(\frac{M_\mathrm{SMBH}}{4\times 10^6\,\Msun}\right)^{-1}
\end{align}
assuming $e_\mathrm{out}\sim 0$. The KL timescale is shorter than the capture timescale by the AGN disk due to gas dynamical friction (gray and black lines in panel~(c) of Fig.~\ref{fig:grid_variables}) in the inner regions ($a_\mathrm{out}\lesssim 0.1\,\mathrm{pc}$). 
On the other hand, if the timescale of apsidal precession for either the inner or the outer orbit due to additional mass or general relativity is shorter than the KL timescale, the KL mechanism is typically suppressed \citep[e.g.][]{Fabrycky07,Naoz2013}. 
Since the timescale for precession due to a circumbinary disk is 
\begin{align}
    t_\mathrm{pre}&\sim P_\mathrm{in}\frac{m_1+m_2}{M_\mathrm{cir}}\nonumber\\
    &\sim 40\,\mathrm{yr}\left(\frac{a_\mathrm{in}}{1\,\mathrm{AU}}\right)^{3/2}
    \left(\frac{m_1+m_2}{20\,\Msun}\right)^{1/2}
    \left(\frac{M_\mathrm{cir}}{0.1\,\Msun}\right)^{-1}
\end{align}
\citep[e.g.][]{Chang09}, where $M_\mathrm{cir}$ is the mass of a circumbinary disk, 
the KL mechanism \ref{num_smbh_bhbh} is suppressed at $r\gtrsim 10^{-3}\,\mathrm{pc}$ where binary-single interactions occur efficiently.

Similarly the KL timescale for \ref{num_bhbhbh} is 
\begin{align}
    t_\mathrm{KL,ii}\sim 3\,\mathrm{yr} &\left(\frac{a_\mathrm{out}}{1\,\mathrm{AU}}\right)^3
    \left(\frac{a_\mathrm{in}}{0.1\,\mathrm{AU}}\right)^{-3/2}\nonumber\\
    &\left(\frac{m_\mathrm{bin}}{20\,\Msun}\right)^{1/2}
    \left(\frac{M_\mathrm{3}}{10\,\Msun}\right)^{-1}
\end{align}
assuming $e_\mathrm{out}\sim 0$. 
This timescale is much shorter than the timescale of capture by the AGN disk and shorter than the precession timescale. 
In AGN disks, such three-body systems presumably efficiently form due to the gas-capture mechanism. 
If the angle between the inner and outer orbits becomes $\sim 90^\circ$ after inner binary-single interactions and a third body is stably bound after the interactions \citep{Mardling01}, mergers can be driven by the KL mechanism. 
Octupole order corrections may lead to close encounters and mergers even in cases where the inclination angle is far from $90^\circ$ \citep{Naoz2013a,Hoang2018}. 
Due to the KL mechanism, mergers in AGN disks may be further accelerated. 
Such systems would be worth investigating in the future.

For \ref{num_disk_smbh_bh}, during the damping of the velocity of a compact object due to gas dynamical friction when crossing the AGN disk, the eccentricity and the inclination of the orbital motion of the compact object can be exchanged with each other due to the perturbation by the AGN disk as investigated by \citet{Vokrouhlicky98} and \citet{Subr05}. 
\citet{Chang09} estimated that for this KL oscillation to operate, the mass of stars and BHs within a radius $r$ needs to be smaller than the mass of the gas disk within the same radius. 
In our fiducial model, the mass of stars and BHs is always higher than the gas disk's mass, making the KL mechanism inefficient.

In summary, the KL mechanism for systems composed of three-compact objects \ref{num_bhbhbh} formed by gas-capture mechanism may further facilitate mergers.
This can be the case if the angular momentum directions between the inner and outer orbits become close to orthogonal after a binary-single interaction, or if the inner orbit is highly eccentric.

\subsubsection{Interaction with a SMBH-IMBH binary}

A hard IMBH-SMBH binary can form whenever IMBHs are present in the AGN disk. 
Here we consider the interaction of compact objects with such an IMBH-SMBH binary. 
If compact objects migrate inward in the AGN disk, they can gradually approach the IMBH.
In this case, after entering the Hill radius of the IMBH, the third body is captured by the IMBH due to the high gas density (Eq.~\ref{eq:gas_capture_time_ratio}).

On the other hand, if a third body orbiting outside of the AGN disk comes close to an IMBH, it can suffer a strong kick via the slingshot mechanism. 
For high-mass ratio binaries, it takes a long time for a third body to receive a strong kick; presumably due to the small Hill radius of an IMBH (Fig. 2 in \citealt{Bonetti20}). 
Furthermore, stellar-mass binaries can be disrupted by an IMBH as they push the binary close to the SMBH where the tidal forces disrupt the binary \citep{Deme2019}. 
Treating the dynamical evolution of stars and BHs interacting with a SMBH-IMBH binary is beyond the scope of our present study, but should be studied in the future via direct $N$-body simulation.

\begin{figure}\includegraphics[width=90mm]{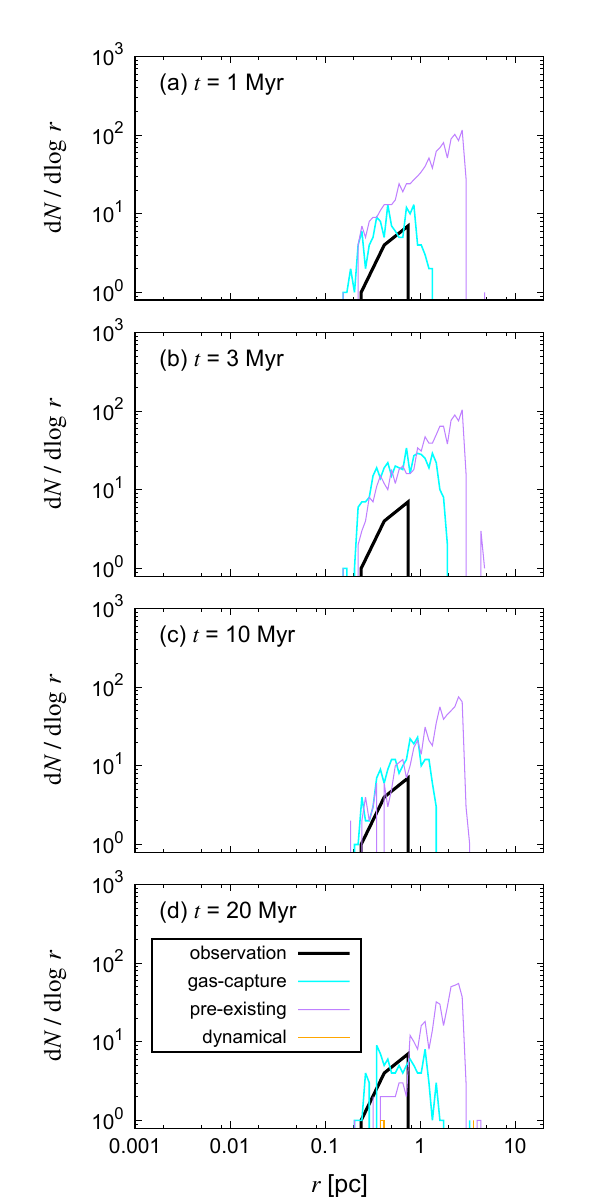}
\caption{
The final binary distribution as a function of the distance from the SMBH $r$ for the fiducial model at $t=$1, 3, 10, and 20 Myr for the fiducial model. 
Cyan, purple, and orange lines are the distribution for gas-capture formed binaries, pre-existing binaries, and dynamically formed binaries, respectively. 
Thick black line shows the observed distribution of X-ray binary candidates \citep{Hailey18} as a function of the projected distance from the Galactic center. 
These binaries satisfy the following conditions: 
(i) The timescale of merger by GW radiation is longer than  $10$ Gyr. 
(ii) The binary is hard compared to the spherical stellar component ($Gm_{j_1}m_{j_2}/(2s_j)>(1/2){\bar m}_\mathrm{star}v_\mathrm{Kep}^2$). 
(iii) The timescale of binary-single interaction with the spherical stellar component at the separation $s_\mathrm{GW}$ is longer than $10$ Gyr. 
Gas capture binaries can reproduce the distribution of the observed X-ray binaries. 
}
\label{fig:rad_dist}\end{figure}

\subsection{Spatial distribution of surviving binaries}

\label{sec:surviving_binary}

The observed low-mass X-ray binaries found by \citet{Hailey18} provide useful constraints for theories of binary formation and evolution in the Galactic center. 
Interestingly, \citet{Hailey18} have shown that the LMXB candidates are found only within $\sim 1$~pc despite the fact that other X-ray sources have been observed out to $\sim$~4~pc~\citep{Muno09}. 
The possible reason for the cutoff at $\sim1$ pc in the LMXB distribution has not been previously explained.

Figure~\ref{fig:rad_dist} shows the spatial distribution of hard binaries at 1, 3, 10, and 20 Myr in the fiducial model. 
Different colored lines show the distribution of binaries formed by different mechanisms. 
To select the binaries that would survive until today, 
we impose the following conditions: 
\begin{enumerate}
 \item\label{i:tGW} The timescale of merger by GW radiation is longer than 10 Gyr. 
 \item\label{i:hard} The binary is hard compared to the spherical stellar component \citep{Gould91,Binney08}, 
   \begin{equation}
   \label{eq:shs}
   s_j<\frac{Gm_{j_1}m_{j_2}}{{\bar m}_\mathrm{star}v_\mathrm{Kep}^2},
  \end{equation}
  where ${\bar m}_\mathrm{star}$ is defined below Eq.~\eqref{eq:IMF}.
 \item\label{i:binarysingle} The timescale of binary-single interactions with the spherical stellar component at the separation $s_\mathrm{GW}$ is longer than 10 Gyr, where $s_\mathrm{GW}$ is the separation from which the binary merges in a Hubble time for the mass of the binary and assuming zero eccentricity \citep{Peters64}. 
\end{enumerate}
Condition~\ref{i:tGW} removes binaries whose separation is smaller than 
\begin{align}
\label{eq:sgw}
s_\mathrm{GW}=0.053\,\mathrm{AU} \left(\frac{m_{j_1}}{5\,\Msun}\right)^{1/4}
\left(\frac{m_{j_2}}{5\,\Msun}\right)^{1/4}\left(\frac{m_{j}}{10\,\Msun}\right)^{1/4}, 
\end{align}
which corresponds to lower regions ($s\lesssim 10^{-7}$ pc) in Figure~\ref{fig:binaries}. 
Condition~\ref{i:hard} 
accounts for the lack of the purple points in the upper left region in panel (a)). 
Condition~\ref{i:binarysingle} produces the inner cut-off at $\sim 0.1$ pc. 
Conditions \ref{i:tGW}, \ref{i:tGW}+\ref{i:hard}, and  \ref{i:tGW}+\ref{i:hard}+\ref{i:binarysingle}  respectively exclude 1622, 1709, and 1880 binaries from the set of 2750 binaries at 10 Myr.

The outer cutoff seen in gas-capture binaries (cyan lines in Figure~\ref{fig:rad_dist}) can be explained 
by the competition between binary hardening due to gas dynamical friction and  type I/II migration in the AGN disk, whose timescales are shown by 
the blue and orange lines in panel (c) of Figure~\ref{fig:grid_variables}, respectively. 
In Model~1, these timescales become equal at around $\sim$ pc. 
At $\gtrsim$~pc, migration is faster than 
hardening. 
Thus, before gas capture binaries become hard, they migrate in to $\lesssim$ pc. 
The exact position of the outer cutoff for gas-capture binaries is influenced by the scale height of the AGN disk (black line in Figure \ref{fig:disk_model}) and the binary mass, which affect the ratio of the migration and hardening rates  (Eqs.~\ref{eq:typeI} and~\ref{eq:df_timescale}). As the scale height decreases or the binary mass decreases, the cutoff moves inner regions. 

For pre-existing binaries, the outer cut-off is at $\sim 3$ pc (purple lines in Figure~\ref{fig:rad_dist}), reflecting our assumed initial BH distribution. 
Since the BH distribution extends to larger radii  \citep[e.g.][]{Freitag06,Antonini14,Generozov18,Schodel14,Feldmeier14}, pre-existing binaries may not have the outer cut-off at $\sim1$ pc as in the LMXB observations. 
However, the fraction of the pre-existing binaries among all binaries in the Galactic center is poorly constrained, it may be subdominant. Due to rapid hardening expected during the common envelope phase, the formation of LMXBs is puzzling
\citep[][]{Podsiadlowski03,Wiktorowicz14,Li15,Wang16}. This is especially the case in galactic nuclei where only hard binaries are long-lived, so that surviving the common envelope phase without merging is even more difficult \citep{Stephan19}. 

These considerations suggest gas-capture binaries as a possible origin of LMXBs. Therefore, we compare the number of LMXBs between observations and our models. Including the undetected population of faint sources, the total number of LMXBs in the Galactic Center is estimated to be hundreds to a thousand \citep{Hailey18}. 
In our simulation, several hundreds to thousands of binaries survive until the dissipation of the AGN disk (Table~\ref{table_results}). 
Also the number of gas-capture binaries, which would survive until today, at 1, 3, 10, and 20 Myr in Figure~\ref{fig:rad_dist} are 134, 441, 205, and 96 respectively. 
Note that we only consider the evolution of BH-BH binaries, leaving the corresponding number of LMXBs uncertain. 
Since stars outnumber BHs in the disk at $\gtrsim 0.1$ pc (Figure~\ref{fig:grid_variables}), stars are expected to reside frequently in binaries. 
However, to estimate the number of LMXBs, 
we need to consider the fact that 
massive objects more commonly reside in binaries after binary-single interactions \citep[e.g.][]{Heggie96}. 
Also after the dissipation of the AGN disk, the distribution of BHs may evolve due to two-body relaxation whose timescale is uncertain and comparable to the Hubble time \citep[e.g.][]{Merritt10,Emami19}. 
To understand the relation between the binary evolution in the AGN disk and the observed LMXBs in detail, requires one to account for the effect of the stellar relaxation processes. 
Also, vector resonant relaxation redistributes the orbital angular momentum vector directions of LMXBs around the SMBH~\citep{Kocsis11b,Szolgyen18}, which is roughly consistent with the 2 dimensional distribution of LMXBs \citep{Hailey18}. 

Finally, the apparent paucity of high-mass X-ray binaries (HMXBs) in the Galactic Center is an interesting additional piece of information for binary formation and evolution. 
The lack of HMXBs in our scenario may be explained by the short lifetime of massive stars if these HMXBs formed during long-past AGN phases, although there are suggestions that the last AGN phase occurred as recently as several Myrs ago in the Galaxy \citep{Nayakshin07,Wardle08,Nayakshin07,Su10,Bland-Hawthorn13}.

\section{Conclusions}

In this paper, we investigated mergers of BHs in accretion disks during the active phase of galactic nuclei. 
We performed 
one-dimensional $N$-body simulations combined with a semi-analytical model. 
While simplified, this model allows us to incorporate the formation and disruption of binaries, 
binary-single interactions, weak gravitational scattering with ambient stars, 
gas dynamical friction in the AGN disk, 
binary hardening due to type I/II torques from circumbinary disks and gas dynamical friction, migration in the AGN disk, star formation, 
and gravitational wave radiation. 
Our main results can be summarized as follows:

\renewcommand{\labelenumi}{\arabic{enumi}. }
\begin{enumerate}
\item Gas-capture binaries, which have been neglected in previous studies, contribute $\sim 58-97\,\%$ of all BH-BH mergers in the AGN disk, with dynamically formed binaries contributing up to $\sim 6\,\%$ (Table~\ref{table_results}). 
\item After their formation, binaries in the AGN disk are hardened by gas dynamical friction, 
binary-single interactions with disk stellar and BH components, 
and finally by gravitational radiation (Figure~\ref{fig:tot_cont}). 
Since binaries efficiently migrate to the inner regions of the disk ($\lesssim 0.01$ pc) before they merge, Doppler acceleration due to their center-of-mass motion around the SMBH is expected to be observable in many cases by the future gravitational wave detector LISA (\S \ref{sec:doppler}). 
\item Due to the recoil kick at frequent binary-single interactions, $\sim 80\%$ of mergers occur
outside the AGN disk. 
In this case, the binaries' orbital planes at merger are randomly oriented.  
\item On the other hand, $\sim 20\%$ of mergers occur inside the AGN disk. 
In this case, due to accretion from a circumbinary disk, some degree of (anti-) alignment of the BH spins with the binary's internal orbital angular momentum is expected, as suggested in the high-effective spin merger candidates, GW151216, GW170403 and GW170817A (\S \ref{sec:spin}). 
\item Due to frequent binary-single interactions during the binary inspiral, 
the binary eccentricity is thermalized in the LISA band. 
LISA will detect such highly eccentric binaries (\S \ref{sec:ecc}). 
\item The binary separation is driven predominantly by binary-single scattering interactions in the evolutionary phase preceding the GW-driven merger. The GW frequency is in the LISA band already in the binary-single interaction driven phase which may be identified by a significant GW phase shift for individual sources and/or the frequency distribution of a population of sources (\S \ref{sec:GWperturbation}). 
\item
We explicitly compute the binary fraction, the capture fraction to AGN disks, the merger fraction of binaries, and the dependence of the merger fraction on the lifetime of the AGN disk.   Accounting for uncertainties in these quantities, we find a volumetric BH merger rate of  $\sim 0.02-60\,\mathrm{Gpc^{-3}yr^{-1}}$, whose uncertainties
are reduced by several orders of magnitude compared to prior works~\citep{McKernan17}.
\item Due to repeated mergers, this pathway naturally explains ``heavy'' BHs detected in existing GW observations, even if BHs are born with much smaller ($\lesssim 15\Msun$) masses. Our model also predicts that mergers of yet more massive BHs ($\gtrsim 10^2\,\Msun$) will be detected by LIGO in the future (Figure~\ref{fig:detection_mass}). IMBHs formed during repeated mergers in AGN in most of our models (one exception is Model 2 where gas-driven migration was turned off).
\item The maximum rate of extreme-mass-ratio inspirals (EMRIs) involving stellar-mass BHs is roughly estimated to be $\sim 0.1-0.6\,\mathrm{Gpc^{-3}yr^{-1}}$, 
which could largely contribute to the total EMRI rate, 
as well as possibly the total intermediate-mass-ratio inspiral (IMRI) rate (\S \ref{sec:emri}). 
\item The distribution of surviving binaries formed by the gas-capture mechanism can reproduce the spatial distribution of LMXBs observed in the Galactic center,  including their outer cutoff at $\sim1$ pc. This cut-off arises from the competition between binary hardening by gas dynamical friction and Type I/II migration in the disk (\S \ref{sec:surviving_binary}). 
Binaries migrate inside $\lesssim1$ pc before they are hardened. 
\end{enumerate}

In this paper, we employed simplified prescriptions, and ignored the exchange of binary components during binary-single interactions and the evolution in the binary eccentricities, as well as other possibly important processes (see \S~\ref{sec:overview}).  These issues will be further investigated in the future.

\acknowledgments

We thank Imre Bartos, Barry McKernan, Yuri Levin, Brian Metzger, Hidekazu Tanaka, Alexander Rasskazov, Massimo Ricotti and Alessandro Trani for useful discussions.
This project was supported by funds from the European Research Council (ERC) under the European Union's Horizon 2020 research and innovation programme under grant agreement No 638435 (GalNUC) and by the Hungarian National Research, Development, and Innovation Office grant NKFIH KH-125675. This research was supported in part by the National Science Foundation under Grant No. NSF PHY-1748958.  ZH acknowledges support from NASA grant NNX15AB19G and NSF grant 1715661.
Simulations and analyses were carried out on Cray XC50 and other computers at the Center for Computational Astrophysics, National Astronomical Observatory of Japan.

\bibliographystyle{yahapj.bst}
\bibliography{agn_bhm}

\begin{thebibliography}{}
\providecommand\natexlab[1]{#1}
\providecommand\JournalTitle[1]{#1}

\bibitem[{{Aarseth} \& {Heggie}(1976)}]{Aarseth76}
{Aarseth}, S.~J., \& {Heggie}, D.~C. 1976, \JournalTitle{\aap}, 53, 259

\bibitem[{{Abbott} {et~al.}(2016){Abbott}, {Abbott}, {Abbott}, {Abernathy},
  {Acernese}, {Ackley}, {Adams}, {Adams}, {Addesso}, {Adhikari}, {Adya},
  {Affeldt}, {Agathos}, {Agatsuma}, {Aggarwal}, {Aguiar}, {Aiello}, {Ain}, \&
  {Ajith}}]{Abbott16b}
{Abbott}, B.~P., {Abbott}, R., {Abbott}, T.~D., {et~al.} 2016,
  \href{http://dx.doi.org/10.1103/PhysRevLett.116.241103}{\JournalTitle{\prl},
  116, 241103}

\bibitem[{Abbott {et~al.}(2016)Abbott, Abbott, Abbott, Abernathy, Acernese,
  Ackley, Adams, Adams, Addesso, Adhikari, Adya, \& Affeldt}]{Abbott16a}
Abbott, B.~P., Abbott, R., Abbott, T.~D., {et~al.} 2016,
  \href{http://dx.doi.org/10.1103/PhysRevLett.116.061102}{\JournalTitle{Phys.
  Rev. Lett.}, 116, 061102}

\bibitem[{{Abbott} {et~al.}(2017{\natexlab{a}}){Abbott}, {Abbott}, {Abbott},
  {Acernese}, {Ackley}, {Adams}, {Adams}, {Addesso}, {Adhikari}, {Adya},
  {Affeldt}, {Afrough}, {Agarwal}, {Agathos}, {Agatsuma}, {Aggarwal}, {Aguiar},
  {LIGO Scientific}, \& {Virgo Collaboration}}]{Abbott17}
{Abbott}, B.~P., {Abbott}, R., {Abbott}, T.~D., {et~al.} 2017{\natexlab{a}},
  \href{http://dx.doi.org/10.1103/PhysRevLett.118.221101}{\JournalTitle{\prl},
  118, 221101}

\bibitem[{{Abbott} {et~al.}(2017{\natexlab{b}}){Abbott}, {Abbott}, {Abbott},
  {Acernese}, {Ackley}, {Adams}, {Adams}, {Addesso}, {Adhikari}, {Adya},
  {Affeldt}, {Afrough}, {Agarwal}, {Agathos}, {Agatsuma}, {Aggarwal}, {Aguiar},
  {Aiello}, {Ain}, {Ajith}, {Allen}, {Allen}, {Allocca}, {Altin}, {Amato},
  {Ananyeva}, {LIGO Scientific Collaboration}, \& {Virgo
  Collaboration}}]{TheLIGO17a}
---. 2017{\natexlab{b}},
  \href{http://dx.doi.org/10.1103/PhysRevLett.119.141101}{\JournalTitle{\prl},
  119, 141101}

\bibitem[{Abbott {et~al.}(2017)Abbott, Abbott, Abbott, Acernese, Ackley, Adams,
  Adams, Addesso, Adhikari, Adya, Affeldt, Afrough, Agarwal, Agathos, Agatsuma,
  Aggarwal, \& Aguiar}]{TheLIGO17b}
Abbott, B.~P., Abbott, R., Abbott, T.~D., {et~al.} 2017,
  \href{http://dx.doi.org/10.1103/PhysRevLett.119.161101}{\JournalTitle{Phys.
  Rev. Lett.}, 119, 161101}

\bibitem[{{Abbott} {et~al.}(2019{\natexlab{a}}){Abbott}, {Abbott}, {Abbott},
  {Abraham}, {Acernese}, {Ackley}, {Adams}, {Adhikari}, {Adya}, {Affeldt},
  {Agathos}, {Agatsuma}, {Aggarwal}, {Aguiar}, {Aiello}, {Ain}, {Ajith},
  {Allen}, {Allocca}, {LIGO Scientific Collaboration}, \& {Virgo
  Collaboration}}]{TheLIGO18}
{Abbott}, B.~P., {Abbott}, R., {Abbott}, T.~D., {et~al.} 2019{\natexlab{a}},
  \href{http://dx.doi.org/10.1103/PhysRevX.9.031040}{\JournalTitle{Physical
  Review X}, 9, 031040}

\bibitem[{{Abbott} {et~al.}(2019{\natexlab{b}}){Abbott}, {Abbott}, {Abbott},
  {Abraham}, {Acernese}, {Ackley}, {Adams}, {Adhikari}, {Adya}, {Affeldt},
  {Agathos}, {Agatsuma}, {Aggarwal}, \& {Aguiar}}]{Abbott19_Ecc}
---. 2019{\natexlab{b}},
  \href{http://dx.doi.org/10.3847/1538-4357/ab3c2d}{\JournalTitle{\apj}, 883,
  149}

\bibitem[{{Abt}(1983)}]{Abt83}
{Abt}, H.~A. 1983,
  \href{http://adsabs.harvard.edu/abs/1983ARA%26A..21..343A}{\JournalTitle{ARA\&A},
  21, 343}

\bibitem[{{Aharon} \& {Perets}(2016)}]{Aharon16}
{Aharon}, D., \& {Perets}, H.~B. 2016,
  \href{http://adsabs.harvard.edu/abs/2016ApJ...830L...1A}{\JournalTitle{ApJL},
  830, L1}

\bibitem[{{Alexander} \& {Hopman}(2009)}]{AlexanderHopman2009}
{Alexander}, T., \& {Hopman}, C. 2009,
  \href{http://dx.doi.org/10.1088/0004-637X/697/2/1861}{\JournalTitle{\apj},
  697, 1861}

\bibitem[{{Amaro-Seoane} {et~al.}(2007){Amaro-Seoane}, {Gair}, {Freitag},
  {Miller}, {Mandel}, {Cutler}, \& {Babak}}]{AmaroSeoane07}
{Amaro-Seoane}, P., {Gair}, J.~R., {Freitag}, M., {et~al.} 2007,
  \href{http://adsabs.harvard.edu/abs/2007CQGra..24R.113A}{\JournalTitle{CQGra},
  24, R113}

\bibitem[{{Amaro-Seoane} \& {Preto}(2011)}]{AmaroSeoane11}
{Amaro-Seoane}, P., \& {Preto}, M. 2011,
  \href{http://adsabs.harvard.edu/abs/2011CQGra..28i4017A}{\JournalTitle{CQGra},
  28, 094017}

\bibitem[{{Amaro-Seoane} {et~al.}(2017){Amaro-Seoane}, {Audley}, {Babak},
  {Baker}, {Barausse}, {Bender}, {Berti}, {Binetruy}, {Born}, {Bortoluzzi},
  {Camp}, {Caprini}, {Cardoso}, {Colpi}, {Conklin}, {Cornish}, {Cutler},
  {Danzmann}, {Dolesi}, {Ferraioli}, {Ferroni}, {Fitzsimons}, {Gair}, {Gesa
  Bote}, {Giardini}, {Gibert}, {Grimani}, {Halloin}, {Heinzel}, {Hertog},
  {Hewitson}, {Holley-Bockelmann}, {Hollington}, {Hueller}, {Inchauspe},
  {Jetzer}, {Karnesis}, {Killow}, {Klein}, {Klipstein}, {Korsakova}, {Larson},
  {Livas}, {Lloro}, {Man}, {Mance}, {Martino}, {Mateos}, {McKenzie},
  {McWilliams}, {Miller}, {Mueller}, {Nardini}, {Nelemans}, {Nofrarias},
  {Petiteau}, {Pivato}, {Plagnol}, {Porter}, {Reiche}, {Robertson},
  {Robertson}, {Rossi}, {Russano}, {Schutz}, {Sesana}, {Shoemaker}, {Slutsky},
  {Sopuerta}, {Sumner}, {Tamanini}, {Thorpe}, {Troebs}, {Vallisneri},
  {Vecchio}, {Vetrugno}, {Vitale}, {Volonteri}, {Wanner}, {Ward}, {Wass},
  {Weber}, {Ziemer}, \& {Zweifel}}]{LISA}
{Amaro-Seoane}, P., {Audley}, H., {Babak}, S., {et~al.} 2017,
  \JournalTitle{arXiv e-prints}, arXiv:1702.00786

\bibitem[{{Antonini}(2014)}]{Antonini14}
{Antonini}, F. 2014,
  \href{http://adsabs.harvard.edu/abs/2014ApJ...794..106A}{\JournalTitle{\apj},
  794, 106}

\bibitem[{{Antonini} {et~al.}(2017){Antonini}, {Toonen}, \&
  {Hamers}}]{Antonini17}
{Antonini}, F., {Toonen}, S., \& {Hamers}, A.~S. 2017,
  \href{http://dx.doi.org/10.3847/1538-4357/aa6f5e}{\JournalTitle{\apj}, 841,
  77}

\bibitem[{{Arca-Sedda} {et~al.}(2018){Arca-Sedda}, {Li}, \&
  {Kocsis}}]{ArcaSedda18}
{Arca-Sedda}, M., {Li}, G., \& {Kocsis}, B. 2018, \JournalTitle{arXiv
  e-prints}, arXiv:1805.06458

\bibitem[{{Armitage} \& {Natarajan}(2005)}]{Armitage2005}
{Armitage}, P.~J., \& {Natarajan}, P. 2005,
  \href{http://dx.doi.org/10.1086/497108}{\JournalTitle{\apj}, 634, 921}

\bibitem[{{Artymowicz} {et~al.}(1991){Artymowicz}, {Clarke}, {Lubow}, \&
  {Pringle}}]{Artymowicz91}
{Artymowicz}, P., {Clarke}, C.~J., {Lubow}, S.~H., \& {Pringle}, J.~E. 1991,
  \href{http://adsabs.harvard.edu/abs/1991ApJ...370L..35A}{\JournalTitle{\apj},
  370, L35}

\bibitem[{{Artymowicz} {et~al.}(1993){Artymowicz}, {Lin}, \&
  {Wampler}}]{Artymowicz93}
{Artymowicz}, P., {Lin}, D.~N.~C., \& {Wampler}, E.~J. 1993,
  \href{http://dx.doi.org/10.1086/172690}{\JournalTitle{\apj}, 409, 592}

\bibitem[{{Askar} {et~al.}(2017){Askar}, {Szkudlarek}, {Gondek-Rosińska},
  {Giersz}, \& {Bulik}}]{Askar17}
{Askar}, A., {Szkudlarek}, M., {Gondek-Rosińska}, D., {Giersz}, M., \&
  {Bulik}, T. 2017,
  \href{http://adsabs.harvard.edu/abs/2017MNRAS.464L..36A}{\JournalTitle{MNRASL},
  464, L36}

\bibitem[{{Babak} {et~al.}(2017){Babak}, {Gair}, {Sesana}, {Barausse}, {Berry},
  {Berti}, {Amaro-Seoane}, {Petiteau}, \& {Klein}}]{Babak17}
{Babak}, S., {Gair}, J., {Sesana}, A., {et~al.} 2017,
  \href{http://adsabs.harvard.edu/abs/2017PhRvD..95j3012B}{\JournalTitle{Phys.
  Rev. D.}, 95, 103012}

\bibitem[{{Bai} \& {Stone}(2013)}]{Bai13}
{Bai}, X.-N., \& {Stone}, J.~M. 2013,
  \href{http://dx.doi.org/10.1088/0004-637X/767/1/30}{\JournalTitle{\apj}, 767,
  30}

\bibitem[{{Baker} {et~al.}(2007){Baker}, {Boggs}, {Centrella}, {Kelly},
  {McWilliams}, {Miller}, \& {van Meter}}]{Baker07}
{Baker}, J.~G., {Boggs}, W.~D., {Centrella}, J., {et~al.} 2007,
  \href{http://dx.doi.org/10.1086/521330}{\JournalTitle{\apj}, 668, 1140}

\bibitem[{{Banerjee}(2017)}]{Banerjee17}
{Banerjee}, S. 2017,
  \href{http://adsabs.harvard.edu/abs/2017MNRAS.467..524B}{\JournalTitle{\mnras},
  467, 524}

\bibitem[{{Banerjee}(2018{\natexlab{a}})}]{Banerjee18a}
---. 2018{\natexlab{a}},
  \href{http://adsabs.harvard.edu/abs/2018MNRAS.473..909B}{\JournalTitle{\mnras},
  473, 909}

\bibitem[{{Banerjee}(2018{\natexlab{b}})}]{Banerjee18b}
---. 2018{\natexlab{b}},
  \href{http://adsabs.harvard.edu/abs/2018MNRAS.481.5123B}{\JournalTitle{\mnras},
  481, 5123}

\bibitem[{{Bar-Or} \& {Alexander}(2016)}]{BarOr16}
{Bar-Or}, B., \& {Alexander}, T. 2016,
  \href{http://adsabs.harvard.edu/abs/2016ApJ...820..129B}{\JournalTitle{\apj},
  820, 129}

\bibitem[{{Barausse} {et~al.}(2015){Barausse}, {Cardoso}, \&
  {Pani}}]{Barausse2015}
{Barausse}, E., {Cardoso}, V., \& {Pani}, P. 2015,
  \href{http://dx.doi.org/10.1088/1742-6596/610/1/012044}{in Journal of Physics
  Conference Series, Vol. 610, Journal of Physics Conference Series}, 012044

\bibitem[{{Bartko} {et~al.}(2010){Bartko}, {Martins}, {Trippe}, {Fritz},
  {Genzel}, {et~al.}}]{Bartko10}
{Bartko}, H., {Martins}, F., {Trippe}, S., {et~al.} 2010,
  \href{http://adsabs.harvard.edu/abs/2010ApJ...708..834B}{\JournalTitle{\apj},
  708, 834}

\bibitem[{{Bartos} {et~al.}(2017){Bartos}, {Kocsis}, {Haiman}, \&
  {M{\'a}rka}}]{Bartos17}
{Bartos}, I., {Kocsis}, B., {Haiman}, Z., \& {M{\'a}rka}, S. 2017,
  \href{http://adsabs.harvard.edu/abs/2017ApJ...835..165B}{\JournalTitle{\apj},
  835, 165}

\bibitem[{{Baruteau} {et~al.}(2011){Baruteau}, {Cuadra}, \& {Lin}}]{Baruteau11}
{Baruteau}, C., {Cuadra}, J., \& {Lin}, D.~N.~C. 2011,
  \href{http://adsabs.harvard.edu/abs/2011ApJ...726...28B}{\JournalTitle{\apj},
  726, 28}

\bibitem[{{Baruteau} \& {Lin}(2010)}]{Baruteau10}
{Baruteau}, C., \& {Lin}, D.~N.~C. 2010,
  \href{http://dx.doi.org/10.1088/0004-637X/709/2/759}{\JournalTitle{\apj},
  709, 759}

\bibitem[{{Bavera} {et~al.}(2019){Bavera}, {Fragos}, {Qin}, {Zapartas},
  {Neijssel}, {Mandel}, {Batta}, {Gaebel}, {Kimball}, \&
  {Stevenson}}]{Bavera19}
{Bavera}, S.~S., {Fragos}, T., {Qin}, Y., {et~al.} 2019, \JournalTitle{arXiv
  e-prints}, arXiv:1906.12257

\bibitem[{{Belczynski} {et~al.}(2010){Belczynski}, {Bulik}, {Fryer}, {Ruiter},
  {Valsecchi}, {Vink}, \& {Hurley}}]{Belczynski10}
{Belczynski}, K., {Bulik}, T., {Fryer}, C., {et~al.} 2010,
  \href{http://adsabs.harvard.edu/abs/2010ApJ...714.1217B}{\JournalTitle{\apj},
  714, 1217}

\bibitem[{{Belczynski} {et~al.}(2016){Belczynski}, {Daniel}, {Bulik}, \&
  {O'Shaughnessy}}]{Belczynski16}
{Belczynski}, K., {Daniel}, E.~H., {Bulik}, T., \& {O'Shaughnessy}, R. 2016,
  \href{http://adsabs.harvard.edu/abs/2016Natur.534..512B}{\JournalTitle{\nat},
  534, 512}

\bibitem[{{Belczynski} {et~al.}(2008){Belczynski}, {Kalogera}, {Rasio}, {Taam},
  {Zezas}, {Bulik}, {Maccarone}, \& {Ivanova}}]{Belczynski08}
{Belczynski}, K., {Kalogera}, V., {Rasio}, F.~A., {et~al.} 2008,
  \href{http://adsabs.harvard.edu/abs/2008ApJS..174..223B}{\JournalTitle{ApJS},
  174, 223}

\bibitem[{{Belczynski} {et~al.}(2004){Belczynski}, {Sadowski}, \&
  {Rasio}}]{Belczynski04}
{Belczynski}, K., {Sadowski}, A., \& {Rasio}, F.~A. 2004,
  \href{http://adsabs.harvard.edu/abs/2004ApJ...611.1068B}{\JournalTitle{\apj},
  611, 1068}

\bibitem[{{Belczynski} {et~al.}(2017){Belczynski}, {Askar}, {Arca-Sedda},
  {Chruslinska}, {Donnari}, {Giersz}, {Benacquista}, {Spurzem}, {Jin},
  {Wiktorowicz}, \& {Belloni}}]{Belczynski17}
{Belczynski}, K., {Askar}, A., {Arca-Sedda}, M., {et~al.} 2017,
  \href{http://adsabs.harvard.edu/abs/2018A%26A...615A..91B}{\JournalTitle{A\&A},
  615, A91}

\bibitem[{{Bell} \& {Lin}(1994)}]{Bell94}
{Bell}, K.~R., \& {Lin}, D.~N.~C. 1994,
  \href{http://adsabs.harvard.edu/abs/1994ApJ...427..987B}{\JournalTitle{\apj},
  427, 987}

\bibitem[{{Bellovary} {et~al.}(2016){Bellovary}, {Mac Low}, {McKernan}, \&
  {Ford}}]{Bellovary16}
{Bellovary}, J.~M., {Mac Low}, M.~M., {McKernan}, B., \& {Ford}, K.~E.~S. 2016,
  \href{http://adsabs.harvard.edu/abs/2016ApJ...819L..17B}{\JournalTitle{\apj},
  819, L17}

\bibitem[{{Binney} \& {Tremaine}(2008)}]{Binney08}
{Binney}, J., \& {Tremaine}, S. 2008,
  \href{http://adsabs.harvard.edu/abs/2008gady.book.....B}{\JournalTitle{Galactic
  Dynamics (2nd ed.; Princeton, NJ; Princeton Univ. Press)}}

\bibitem[{{Bland-Hawthorn} {et~al.}(2013){Bland-Hawthorn}, {Maloney},
  {Sutherland }, \& {Madsen}}]{Bland-Hawthorn13}
{Bland-Hawthorn}, J., {Maloney}, P.~R., {Sutherland }, R.~S., \& {Madsen},
  G.~J. 2013,
  \href{http://dx.doi.org/10.1088/0004-637X/778/1/58}{\JournalTitle{\apj}, 778,
  58}

\bibitem[{{Bogdanovic} {et~al.}(2007){Bogdanovic}, {Reynolds}, \&
  {Miller}}]{Bogdanovic07}
{Bogdanovic}, T., {Reynolds}, C.~S., \& {Miller}, M.~C. 2007,
  \href{http://adsabs.harvard.edu/abs/2007ApJ...661L.147B}{\JournalTitle{\apj},
  661, L147}

\bibitem[{{Bonetti} {et~al.}(2020){Bonetti}, {Rasskazov}, {Sesana}, {Dotti},
  {Haardt}, {Leigh}, {Arca Sedda}, {Fragione}, \& {Rossi}}]{Bonetti20}
{Bonetti}, M., {Rasskazov}, A., {Sesana}, A., {et~al.} 2020,
  \href{http://dx.doi.org/10.1093/mnrasl/slaa018}{\JournalTitle{\mnras}, 493,
  L114}

\bibitem[{{Bouffanais} {et~al.}(2019){Bouffanais}, {Mapelli}, {Gerosa}, {Di
  Carlo}, {Giacobbo}, {Berti}, \& {Baibhav}}]{Bouffanais19}
{Bouffanais}, Y., {Mapelli}, M., {Gerosa}, D., {et~al.} 2019,
  \href{https://ui.adsabs.harvard.edu/abs/2019arXiv190511054B/abstract}{\JournalTitle{arXiv
  e-prints}}, \href{http://arxiv.org/abs/1905.11054}{{\sffamily
  arXiv:1905.11054}}

\bibitem[{{Breivik} {et~al.}(2016){Breivik}, {Rodriguez}, {Larson}, {Kalogera},
  \& {Rasio}}]{Breivik16}
{Breivik}, K., {Rodriguez}, C.~L., {Larson}, S.~L., {Kalogera}, V., \& {Rasio},
  F.~A. 2016,
  \href{http://dx.doi.org/10.3847/2041-8205/830/1/L18}{\JournalTitle{\apjl},
  830, L18}

\bibitem[{{Brown} \& {Zimmerman}(2010)}]{Brown10}
{Brown}, D.~A., \& {Zimmerman}, P.~J. 2010,
  \href{http://dx.doi.org/10.1103/PhysRevD.81.024007}{\JournalTitle{Physical
  Review D}, 81, 024007}

\bibitem[{{Broz} {et~al.}(2018){Broz}, {Chrenko}, {Nesvornu}, \&
  {Lambrechts}}]{Broz18}
{Broz}, M., {Chrenko}, O., {Nesvornu}, D., \& {Lambrechts}, M. 2018,
  \href{http://adsabs.harvard.edu/abs/2018A%26A...620A.157B}{\JournalTitle{A\&A},
  620, A157}

\bibitem[{{Burtscher} {et~al.}(2013){Burtscher}, {Meisenheimer}, {Tristram},
  {et~al.}}]{Burtscher13}
{Burtscher}, L., {Meisenheimer}, K., {Tristram}, K.~R.~W., {et~al.} 2013,
  \href{http://adsabs.harvard.edu/abs/2013A%26A...558A.149B}{\JournalTitle{A\&A},
  558, 149}

\bibitem[{{Campanelli} {et~al.}(2007){Campanelli}, {Lousto}, {Zlochower}, \&
  {Merritt}}]{Campanelli07}
{Campanelli}, M., {Lousto}, C., {Zlochower}, Y., \& {Merritt}, D. 2007,
  \href{http://dx.doi.org/10.1086/516712}{\JournalTitle{\apjl}, 659, L5}

\bibitem[{{Chang}(2009)}]{Chang09}
{Chang}, P. 2009,
  \href{http://dx.doi.org/10.1111/j.1365-2966.2008.14202.x}{\JournalTitle{\mnras},
  393, 224}

\bibitem[{{Chapon} {et~al.}(2013){Chapon}, {Mayer}, \& R.}]{Chapon13}
{Chapon}, D., {Mayer}, L., \& R., T. 2013,
  \href{http://adsabs.harvard.edu/abs/2013MNRAS.429.3114C}{\JournalTitle{\mnras},
  429, 3114}

\bibitem[{{Crida} {et~al.}(2006){Crida}, {Morbidelli}, \& {Masset}}]{Crida06}
{Crida}, A., {Morbidelli}, A., \& {Masset}, F. 2006,
  \href{http://adsabs.harvard.edu/abs/2006Icar..181..587C}{\JournalTitle{Icarus},
  181, 587}

\bibitem[{{Cuadra} {et~al.}(2009){Cuadra}, {Armitage}, {Alexander}, \&
  {Begelman}}]{Cuadra09}
{Cuadra}, J., {Armitage}, P.~J., {Alexander}, R.~D., \& {Begelman}, M.~C. 2009,
  \href{http://adsabs.harvard.edu/abs/2009MNRAS.393.1423C}{\JournalTitle{\mnras},
  393, 1423}

\bibitem[{{de Mink} \& {Mandel}(2016)}]{deMink16}
{de Mink}, S.~E., \& {Mandel}, I. 2016,
  \href{http://adsabs.harvard.edu/abs/2016MNRAS.460.3545D}{\JournalTitle{\mnras},
  460, 3545}

\bibitem[{{del Valle} \& {Volonteri}(2018)}]{delValle18}
{del Valle}, L., \& {Volonteri}, M. 2018,
  \href{http://adsabs.harvard.edu/abs/2018MNRAS.480..439D}{\JournalTitle{\mnras},
  480, 439}

\bibitem[{{Deme} {et~al.}(2019){Deme}, {Meiron}, \& {Kocsis}}]{Deme2019}
{Deme}, B., {Meiron}, Y., \& {Kocsis}, B. 2019, \JournalTitle{arXiv e-prints},
  arXiv:1909.04678

\bibitem[{{Derdzinski} {et~al.}(2019){Derdzinski}, {D'Orazio}, {Duffell},
  {Haiman}, \& {MacFadyen}}]{Derdzinski18}
{Derdzinski}, A.~M., {D'Orazio}, D., {Duffell}, P., {Haiman}, Z., \&
  {MacFadyen}, A. 2019,
  \href{http://dx.doi.org/10.1093/mnras/stz1026}{\JournalTitle{\mnras}, 486,
  2754}

\bibitem[{{Di Carlo} {et~al.}(2019){Di Carlo}, {Giacobbo}, {Mapelli},
  {Pasquato}, {Spera}, {Wang}, \& {Haardt}}]{DiCarlo19}
{Di Carlo}, U.~N., {Giacobbo}, N., {Mapelli}, M., {et~al.} 2019,
  \href{http://dx.doi.org/10.1093/mnras/stz1453}{\JournalTitle{\mnras}, 487,
  2947}

\bibitem[{{Dominik} {et~al.}(2012){Dominik}, {Belczynski}, {Fryer}, {Holz},
  {Berti}, {Bulik}, {Mandel}, \& {O'Shughnessy}}]{Dominik12}
{Dominik}, M., {Belczynski}, K., {Fryer}, C., {et~al.} 2012,
  \href{http://adsabs.harvard.edu/abs/2012ApJ...759...52D}{\JournalTitle{\apj},
  759, 52}

\bibitem[{{D'Orazio} \& {Loeb}(2019)}]{DOrazio19}
{D'Orazio}, D.~J., \& {Loeb}, A. 2019, \JournalTitle{arXiv e-prints},
  arXiv:1910.02966

\bibitem[{{Duffell} {et~al.}(2019){Duffell}, {D'Orazio}, {Derdzinski},
  {Haiman}, {MacFadyen}, {Rosen}, \& {Zrake}}]{Duffell19}
{Duffell}, P.~C., {D'Orazio}, D., {Derdzinski}, A., {et~al.} 2019,
  \JournalTitle{\apj, submitted; e-print arXiv:1911.05506}

\bibitem[{{Duffell} {et~al.}(2014){Duffell}, {Haiman}, {MacFadyen}, {D'Orazio},
  \& {Farris}}]{Duffell14}
{Duffell}, P.~C., {Haiman}, Z., {MacFadyen}, A.~I., {D'Orazio}, D.~J., \&
  {Farris}, B.~D. 2014,
  \href{http://adsabs.harvard.edu/abs/2014ApJ...792L..10D}{\JournalTitle{ApJL},
  792, L10}

\bibitem[{{Durmann} \& {Kley}(2015)}]{Durmann15}
{Durmann}, C., \& {Kley}, W. 2015,
  \href{http://adsabs.harvard.edu/abs/2015A%26A...574A..52D}{\JournalTitle{A\&A},
  574, 52}

\bibitem[{{Edgar}(2007)}]{Edgar07}
{Edgar}, R.~G. 2007,
  \href{http://adsabs.harvard.edu/abs/2007ApJ...663.1325E}{\JournalTitle{\apj},
  663, 1325}

\bibitem[{{Emami} \& {Loeb}(2019)}]{Emami19}
{Emami}, R., \& {Loeb}, A. 2019,
  \href{http://adsabs.harvard.edu/abs/2019arXiv190302578E}{\JournalTitle{arXiv
  e-prints}}, \href{http://arxiv.org/abs/1903.02578}{{\sffamily
  arXiv:1903.02578}}

\bibitem[{{Escala} {et~al.}(2004){Escala}, {Larson}, {Coppi}, \&
  {Mardones}}]{Escala04}
{Escala}, A., {Larson}, R.~B., {Coppi}, P.~S., \& {Mardones}, D. 2004,
  \href{http://adsabs.harvard.edu/abs/2004ApJ...607..765E}{\JournalTitle{\apj},
  607, 765}

\bibitem[{{Fabrycky} \& {Tremaine}(2007)}]{Fabrycky07}
{Fabrycky}, D., \& {Tremaine}, S. 2007,
  \href{http://dx.doi.org/10.1086/521702}{\JournalTitle{\apj}, 669, 1298}

\bibitem[{{Farris} {et~al.}(2014){Farris}, {Duffell}, {MacFadyen}, \&
  {Haiman}}]{Farris14}
{Farris}, B.~D., {Duffell}, P., {MacFadyen}, A.~I., \& {Haiman}, Z. 2014,
  \href{http://dx.doi.org/10.1088/0004-637X/783/2/134}{\JournalTitle{\apj},
  783, 134}

\bibitem[{{Feldmeier} {et~al.}(2014){Feldmeier}, {Neumayer}, {Seth}, {Sch{\"
  o}del}, {Lutzgendorf}, {de Zeeuw}, {Kissler-Patig}, {Nishiyama}, \&
  {Walcher}}]{Feldmeier14}
{Feldmeier}, A., {Neumayer}, N., {Seth}, A., {et~al.} 2014,
  \href{http://adsabs.harvard.edu/abs/2014A%26A...570A...2F}{\JournalTitle{A\&A},
  570, A2}

\bibitem[{{Feldmeier} {et~al.}(2015){Feldmeier}, {Neumayer}, {Sch{\" o}del},
  {Seth}, {Hilker}, {de Zeeuw}, {Kuntschner}, {Walcher}, {Lutzgendorf}, \&
  {Kissler-Patig}}]{Feldmeier15}
{Feldmeier}, A., {Neumayer}, N., {Sch{\" o}del}, R., {et~al.} 2015,
  \href{http://adsabs.harvard.edu/abs/2015A%26A...584A...2F}{\JournalTitle{A\&A},
  584, A2}

\bibitem[{{Fishbach} {et~al.}(2019){Fishbach}, {Farr}, \& {Holz}}]{Fishbach19}
{Fishbach}, M., {Farr}, W.~M., \& {Holz}, D.~E. 2019, \JournalTitle{arXiv
  e-prints}, arXiv:1911.05882

\bibitem[{{Fishbach} {et~al.}(2017){Fishbach}, {Holz}, \&
  {Farr}}]{Fishbach2017}
{Fishbach}, M., {Holz}, D.~E., \& {Farr}, B. 2017,
  \href{http://dx.doi.org/10.3847/2041-8213/aa7045}{\JournalTitle{\apjl}, 840,
  L24}

\bibitem[{{Fleming} \& {Quinn}(2017)}]{Fleming17}
{Fleming}, D.~P., \& {Quinn}, T.~R. 2017,
  \href{http://adsabs.harvard.edu/abs/2017MNRAS.464.3343F}{\JournalTitle{\mnras},
  464, 3343}

\bibitem[{{Fragione} {et~al.}(2019){Fragione}, {Grishin}, {Leigh}, {Perets}, \&
  {Perna}}]{Fragione19b}
{Fragione}, G., {Grishin}, E., {Leigh}, N. W.~C., {Perets}, H.~B., \& {Perna},
  R. 2019,
  \href{http://dx.doi.org/10.1093/mnras/stz1651}{\JournalTitle{\mnras}, 488,
  47}

\bibitem[{{Fragione} \& {Kocsis}(2019)}]{Fragione19}
{Fragione}, G., \& {Kocsis}, B. 2019,
  \href{http://dx.doi.org/10.1093/mnras/stz1175}{\JournalTitle{\mnras}, 486,
  4781}

\bibitem[{{Freitag} {et~al.}(2006){Freitag}, {Amro-Seoane}, \&
  {Kalogera}}]{Freitag06}
{Freitag}, M., {Amro-Seoane}, P., \& {Kalogera}, V. 2006,
  \href{http://adsabs.harvard.edu/abs/2006ApJ...649...91F}{\JournalTitle{\apj},
  649, 91}

\bibitem[{{Fujii} {et~al.}(2017){Fujii}, {Tanikawa}, \& {Makino}}]{Fujii17}
{Fujii}, M.~S., {Tanikawa}, A., \& {Makino}, J. 2017,
  \href{http://adsabs.harvard.edu/abs/2017PASJ...69...94F}{\JournalTitle{PASJ},
  69, 94}

\bibitem[{{Fung} {et~al.}(2014){Fung}, {Shi}, \& {Chiang}}]{Fung14}
{Fung}, J., {Shi}, J.~M., \& {Chiang}, E. 2014,
  \href{http://adsabs.harvard.edu/abs/2014ApJ...782...88F}{\JournalTitle{\apj},
  782, 88}

\bibitem[{{Gallego-Cano} {et~al.}(2018){Gallego-Cano}, {Sch{\" o}del}, {Dong},
  {Nogueras-Lara}, {Gallego-Calvente}, {Amaro-Seoane}, \&
  {Baumgardt}}]{GallegoCano18}
{Gallego-Cano}, E., {Sch{\" o}del}, R., {Dong}, H., {et~al.} 2018,
  \href{http://adsabs.harvard.edu/abs/2018A%26A...609A..26G}{\JournalTitle{A\&A},
  609, 26}

\bibitem[{{Gayathri} {et~al.}(2019){Gayathri}, {Bartos}, {Haiman}, {Klimenko},
  {Kocsis}, {Marka}, \& {Yang}}]{Gayathri19}
{Gayathri}, V., {Bartos}, I., {Haiman}, Z., {et~al.} 2019, \JournalTitle{arXiv
  e-prints}, arXiv:1911.11142

\bibitem[{{Geller} {et~al.}(2019){Geller}, {Leigh}, {Giersz}, {Kremer}, \&
  {Rasio}}]{Geller19}
{Geller}, A.~M., {Leigh}, N.~W.~C., {Giersz}, M., {Kremer}, K., \& {Rasio},
  F.~A. 2019,
  \href{http://adsabs.harvard.edu/abs/2019ApJ...872..165G}{\JournalTitle{\apj},
  872, 165}

\bibitem[{{Generozov} {et~al.}(2018){Generozov}, {Stone}, {Metzger}, \&
  {Ostriker}}]{Generozov18}
{Generozov}, A., {Stone}, N.~C., {Metzger}, B.~D., \& {Ostriker}, J.~P. 2018,
  \href{http://adsabs.harvard.edu/abs/2018MNRAS.478.4030G}{\JournalTitle{\mnras},
  478, 4030}

\bibitem[{{Georgiev} {et~al.}(2016){Georgiev}, {Boker}, {Leigh}, {Lutzgendorf},
  \& {Neumayer}}]{Georgiev16}
{Georgiev}, I.~Y., {Boker}, T., {Leigh}, N., {Lutzgendorf}, N., \& {Neumayer},
  N. 2016,
  \href{http://adsabs.harvard.edu/abs/2016MNRAS.457.2122G}{\JournalTitle{\mnras},
  457, 2122}

\bibitem[{{Gerosa} \& {Berti}(2017)}]{GerosaBerti2017}
{Gerosa}, D., \& {Berti}, E. 2017,
  \href{http://dx.doi.org/10.1103/PhysRevD.95.124046}{\JournalTitle{\prd}, 95,
  124046}

\bibitem[{{Giacobbo} {et~al.}(2018){Giacobbo}, {Mapelli}, \&
  {Spera}}]{Giacobbo18}
{Giacobbo}, N., {Mapelli}, M., \& {Spera}, M. 2018,
  \href{http://adsabs.harvard.edu/abs/2018MNRAS.474.2959G}{\JournalTitle{\mnras},
  474, 2959}

\bibitem[{{Goldreich} {et~al.}(2002){Goldreich}, {Lithwick}, \&
  {Sari}}]{Goldreich02}
{Goldreich}, P., {Lithwick}, Y., \& {Sari}, R. 2002,
  \href{http://adsabs.harvard.edu/abs/2002Natur.420..643G}{\JournalTitle{\nat},
  420, 643}

\bibitem[{{Gondan} \& {Kocsis}(2019)}]{Gondan18c}
{Gondan}, L., \& {Kocsis}, B. 2019,
  \href{http://adsabs.harvard.edu/abs/2019ApJ...871..178G}{\JournalTitle{\apj},
  871, 178}

\bibitem[{{Gondan} {et~al.}(2018){Gondan}, {Kocsis}, {Raffai}, \&
  {Frei}}]{Gondan18b}
{Gondan}, L., {Kocsis}, B., {Raffai}, P., \& {Frei}, Z. 2018,
  \href{http://adsabs.harvard.edu/abs/2018ApJ...855...34G}{\JournalTitle{\apj},
  855, 34}

\bibitem[{{Gond{\'a}n} {et~al.}(2018){Gond{\'a}n}, {Kocsis}, {Raffai}, \&
  {Frei}}]{Gondan18a}
{Gond{\'a}n}, L., {Kocsis}, B., {Raffai}, P., \& {Frei}, Z. 2018,
  \href{http://dx.doi.org/10.3847/1538-4357/aabfee}{\JournalTitle{\apj}, 860,
  5}

\bibitem[{{Goodman} \& {Tan}(2004)}]{Goodman04}
{Goodman}, J., \& {Tan}, J.~C. 2004,
  \href{http://adsabs.harvard.edu/abs/2004ApJ...608..108G}{\JournalTitle{\apj},
  608, 108}

\bibitem[{{Gould}(1991)}]{Gould91}
{Gould}, A. 1991, \href{http://dx.doi.org/10.1086/170502}{\JournalTitle{\apj},
  379, 280}

\bibitem[{{Graham} {et~al.}(2017){Graham}, {Djorgovski}, {Drake}, {Stern},
  {Mahabal}, {Glikman}, {Larson}, \& {Christensen}}]{Graham17}
{Graham}, M.~J., {Djorgovski}, S.~G., {Drake}, A.~J., {et~al.} 2017,
  \href{http://dx.doi.org/10.1093/mnras/stx1456}{\JournalTitle{\mnras}, 470,
  4112}

\bibitem[{{Greene} \& {Ho}(2007)}]{Greene07}
{Greene}, J.~E., \& {Ho}, L.~C. 2007,
  \href{http://adsabs.harvard.edu/abs/2007ApJ...667..131G}{\JournalTitle{\apj},
  667, 131}

\bibitem[{{Greene} \& {Ho}(2009)}]{Greene09}
---. 2009,
  \href{http://dx.doi.org/10.1088/0004-637X/704/2/1743}{\JournalTitle{\apj},
  704, 1743}

\bibitem[{{Gruzinov} {et~al.}(2019){Gruzinov}, {Levin}, \&
  {Matzner}}]{Gruzinov19}
{Gruzinov}, A., {Levin}, Y., \& {Matzner}, C.~D. 2019, \JournalTitle{arXiv
  e-prints}, arXiv:1906.01186

\bibitem[{{Hailey} {et~al.}(2018){Hailey}, {Mori}, {Bauer}, {Berkowitz},
  {Hong}, \& {Hord}}]{Hailey18}
{Hailey}, C.~J., {Mori}, K., {Bauer}, F.~E., {et~al.} 2018,
  \href{http://adsabs.harvard.edu/abs/2018Natur.556...70H}{\JournalTitle{Nature
  Letter}, 556, 70}

\bibitem[{{Haiman} \& {Hui}(2001)}]{Haiman01}
{Haiman}, Z., \& {Hui}, L. 2001,
  \href{http://adsabs.harvard.edu/abs/2001ApJ...547...27H}{\JournalTitle{\apj},
  547, 27}

\bibitem[{{Haiman} {et~al.}(2009){Haiman}, {Kocsis}, \& {Menou}}]{Haiman09}
{Haiman}, Z., {Kocsis}, B., \& {Menou}, K. 2009,
  \href{http://adsabs.harvard.edu/abs/2009ApJ...700.1952H}{\JournalTitle{\apj},
  700, 1952}

\bibitem[{{Hansen} \& {Kawaler}(1994)}]{Hansen94}
{Hansen}, C.~J., \& {Kawaler}, S.~D. 1994, {Stellar Interiors. Physical
  Principles, Structure, and Evolution.}, 84

\bibitem[{{Heggie}(1975)}]{Heggie75}
{Heggie}, D.~C. 1975,
  \href{http://adsabs.harvard.edu/abs/1975MNRAS.173..729H}{\JournalTitle{\mnras},
  173, 729}

\bibitem[{{Heggie} {et~al.}(1996){Heggie}, {Hut}, \& {McMillan}}]{Heggie96}
{Heggie}, D.~C., {Hut}, P., \& {McMillan}, S. L.~W. 1996,
  \href{http://dx.doi.org/10.1086/177611}{\JournalTitle{\apj}, 467, 359}

\bibitem[{{Herrmann} {et~al.}(2007){Herrmann}, {Hinder}, {Shoemaker}, {Laguna},
  \& {Matzner}}]{Herrmann07}
{Herrmann}, F., {Hinder}, I., {Shoemaker}, D., {Laguna}, P., \& {Matzner},
  R.~A. 2007, \href{http://dx.doi.org/10.1086/513603}{\JournalTitle{\apj}, 661,
  430}

\bibitem[{{Hills}(1975)}]{Hills75}
{Hills}, J.~G. 1975,
  \href{http://adsabs.harvard.edu/abs/1975AJ.....80..809H}{\JournalTitle{AJ},
  80, 809}

\bibitem[{{Hinder} {et~al.}(2018){Hinder}, {Kidder}, \& {Pfeiffer}}]{Hinder18}
{Hinder}, I., {Kidder}, L.~E., \& {Pfeiffer}, H.~P. 2018,
  \href{http://adsabs.harvard.edu/abs/2018PhRvD..98d4015H}{\JournalTitle{Phys.
  Rev. D.}, 98, 044015}

\bibitem[{{Hoang} {et~al.}(2018{\natexlab{a}}){Hoang}, {Naoz}, {Kocsis},
  {Rasio}, \& {Dosopoulou}}]{Hoang2018}
{Hoang}, B.-M., {Naoz}, S., {Kocsis}, B., {Rasio}, F.~A., \& {Dosopoulou}, F.
  2018{\natexlab{a}},
  \href{http://dx.doi.org/10.3847/1538-4357/aaafce}{\JournalTitle{\apj}, 856,
  140}

\bibitem[{{Hoang} {et~al.}(2018{\natexlab{b}}){Hoang}, {Naoz}, {Kocsis},
  {Rasio}, \& F.}]{Hoang18}
{Hoang}, B.-M., {Naoz}, S., {Kocsis}, B., {Rasio}, F.~A., \& F., D.
  2018{\natexlab{b}},
  \href{http://adsabs.harvard.edu/abs/2018ApJ...856..140H}{\JournalTitle{\apj},
  856, 140}

\bibitem[{{Hopman} \& {Alexander}(2006)}]{Hopman06}
{Hopman}, C., \& {Alexander}, T. 2006,
  \href{http://adsabs.harvard.edu/abs/2006ApJ...645L.133H}{\JournalTitle{ApJL},
  645, L133}

\bibitem[{{Huerta} {et~al.}(2018){Huerta}, {Moore}, {Kumar},
  {et~al.}}]{Huerta18}
{Huerta}, E.~A., {Moore}, C.~J., {Kumar}, P., {et~al.} 2018,
  \href{http://adsabs.harvard.edu/abs/2018PhRvD..97b4031H}{\JournalTitle{Phys.
  Rev. D.}, 97, 024031}

\bibitem[{{Hughes} \& {Blandford}(2003)}]{Hughes03}
{Hughes}, S.~A., \& {Blandford}, R.~D. 2003,
  \href{http://adsabs.harvard.edu/abs/2003ApJ...585L.101H}{\JournalTitle{\apj},
  585, L101}

\bibitem[{{Ida} \& {Lin}(2004)}]{Ida04}
{Ida}, S., \& {Lin}, D.~N.~C. 2004,
  \href{http://adsabs.harvard.edu/abs/2004ApJ...616..567I}{\JournalTitle{\apj},
  616, 567}

\bibitem[{{Inayoshi} {et~al.}(2017{\natexlab{a}}){Inayoshi}, {Hirai},
  {Kinugawa}, \& {Hotokezaka}}]{Inayoshi17}
{Inayoshi}, K., {Hirai}, R., {Kinugawa}, T., \& {Hotokezaka}, K.
  2017{\natexlab{a}},
  \href{http://adsabs.harvard.edu/abs/2017MNRAS.468.5020I}{\JournalTitle{\mnras},
  468, 5020}

\bibitem[{{Inayoshi} {et~al.}(2018){Inayoshi}, {Ostriker}, {Haiman}, \&
  {Kuiper}}]{Inayoshi18}
{Inayoshi}, K., {Ostriker}, J.~P., {Haiman}, Z., \& {Kuiper}, R. 2018,
  \href{http://adsabs.harvard.edu/abs/2018MNRAS.476.1412I}{\JournalTitle{\mnras},
  476, 1412}

\bibitem[{{Inayoshi} {et~al.}(2017{\natexlab{b}}){Inayoshi}, {Tamanini},
  {Caprini}, \& {Haiman}}]{Inayoshi17b}
{Inayoshi}, K., {Tamanini}, N., {Caprini}, C., \& {Haiman}, Z.
  2017{\natexlab{b}},
  \href{http://adsabs.harvard.edu/abs/2017PhRvD..96f3014I}{\JournalTitle{Phys.
  Rev. D.}, 96, 063014}

\bibitem[{{Ivanov} {et~al.}(2015){Ivanov}, {Papaloizou}, {Paardekooper}, \&
  {Polnarev}}]{Ivanov15}
{Ivanov}, P.~B., {Papaloizou}, J.~C.~B., {Paardekooper}, S.~J., \& {Polnarev},
  A.~G. 2015,
  \href{http://dx.doi.org/10.1051/0004-6361/201424359}{\JournalTitle{\aap},
  576, A29}

\bibitem[{{Ivanov} {et~al.}(1999){Ivanov}, {Papaloizou}, \&
  {Polnarev}}]{Ivanov99}
{Ivanov}, P.~B., {Papaloizou}, J.~C.~B., \& {Polnarev}, A.~G. 1999,
  \href{http://adsabs.harvard.edu/abs/1999MNRAS.307...79I}{\JournalTitle{\mnras},
  307, 79}

\bibitem[{{Ivanova} {et~al.}(2013){Ivanova}, {Justham}, {Chen},
  {et~al.}}]{Ivanova13}
{Ivanova}, N., {Justham}, S., {Chen}, X., {et~al.} 2013,
  \href{http://adsabs.harvard.edu/abs/2013A%26ARv..21...59I}{\JournalTitle{The
  Astronomy and Astrophysics Review}, 21, 59}

\bibitem[{{Just} {et~al.}(2012){Just}, {Yurin}, {Makukov}, {Berczik}, {Omarov},
  {Spurzem}, \& {Vilkoviskij}}]{Just12}
{Just}, A., {Yurin}, D., {Makukov}, M., {et~al.} 2012,
  \href{http://adsabs.harvard.edu/abs/2012ApJ...758...51J}{\JournalTitle{\apj},
  758, 51}

\bibitem[{{Kanagawa} {et~al.}(2015){Kanagawa}, {Muto}, {Tanaka},
  {et~al.}}]{Kanagawa15}
{Kanagawa}, K.~D., {Muto}, T., {Tanaka}, H., {et~al.} 2015,
  \href{http://adsabs.harvard.edu/abs/2015ApJ...806L..15K}{\JournalTitle{ApJL},
  806, L15}

\bibitem[{{Kanagawa} {et~al.}(2018){Kanagawa}, {Tanaka}, \&
  {Szuszkiewicz}}]{Kanagawa18}
{Kanagawa}, K.~D., {Tanaka}, H., \& {Szuszkiewicz}, E. 2018,
  \href{http://adsabs.harvard.edu/abs/2018ApJ...861..140K}{\JournalTitle{\apj},
  861, 140}

\bibitem[{{Kelley} {et~al.}(2019){Kelley}, {Haiman}, {Sesana}, \&
  {Hernquist}}]{Kelley19}
{Kelley}, L.~Z., {Haiman}, Z., {Sesana}, A., \& {Hernquist}, L. 2019,
  \href{http://dx.doi.org/10.1093/mnras/stz150}{\JournalTitle{\mnras}, 485,
  1579}

\bibitem[{{Kelly} \& {Shen}(2013)}]{Kelly13}
{Kelly}, B.~C., \& {Shen}, Y. 2013,
  \href{http://adsabs.harvard.edu/abs/2013ApJ...764...45K}{\JournalTitle{\apj},
  764, 45}

\bibitem[{{Kennedy} {et~al.}(2016){Kennedy}, {Meiron}, {Shukirgaliyev},
  {Panamarev}, {Berczik}, {Just}, \& R.}]{Kennedy16}
{Kennedy}, G.~F., {Meiron}, Y., {Shukirgaliyev}, B., {et~al.} 2016,
  \href{http://adsabs.harvard.edu/abs/2016MNRAS.460..240K}{\JournalTitle{\mnras},
  460, 240}

\bibitem[{{Keshet} {et~al.}(2009){Keshet}, {Hopman}, \& {Alexander}}]{Keshet09}
{Keshet}, U., {Hopman}, C., \& {Alexander}, T. 2009,
  \href{http://adsabs.harvard.edu/abs/2009ApJ...698L..64K}{\JournalTitle{\apj},
  698, L64}

\bibitem[{{Kim} \& {Kim}(2007)}]{Kim07}
{Kim}, H., \& {Kim}, W.~T. 2007,
  \href{http://adsabs.harvard.edu/abs/2007ApJ...665..432K}{\JournalTitle{\apj},
  665, 432}

\bibitem[{{King} {et~al.}(2005){King}, {Lubow}, {Ogilvie}, \&
  {Pringle}}]{King05}
{King}, A.~R., {Lubow}, S.~H., {Ogilvie}, G.~I., \& {Pringle}, J.~E. 2005,
  \href{http://dx.doi.org/10.1111/j.1365-2966.2005.09378.x}{\JournalTitle{\mnras},
  363, 49}

\bibitem[{{King} {et~al.}(2007){King}, {Pringle}, \& {Livio}}]{King07}
{King}, A.~R., {Pringle}, J.~E., \& {Livio}, M. 2007,
  \href{http://dx.doi.org/10.1111/j.1365-2966.2007.11556.x}{\JournalTitle{\mnras},
  376, 1740}

\bibitem[{{Kinugawa} {et~al.}(2014){Kinugawa}, {Inayoshi}, {Hotokezaka},
  {Nakauchi}, \& T.}]{Kinugawa14}
{Kinugawa}, T., {Inayoshi}, K., {Hotokezaka}, K., {Nakauchi}, D., \& T., N.
  2014,
  \href{http://adsabs.harvard.edu/abs/2014MNRAS.442.2963K}{\JournalTitle{\mnras},
  442, 2963}

\bibitem[{{Kiseleva} {et~al.}(1998){Kiseleva}, {Eggleton}, \&
  {Mikkola}}]{Kiseleva98}
{Kiseleva}, L.~G., {Eggleton}, P.~P., \& {Mikkola}, S. 1998,
  \href{http://dx.doi.org/10.1046/j.1365-8711.1998.01903.x}{\JournalTitle{\mnras},
  300, 292}

\bibitem[{{Kissel} \& {Betzwieser}(2018)}]{Kissel18}
{Kissel}, J., \& {Betzwieser}, J. 2018,
  \href{https://dcc.ligo.org/LIGO-G1802164/public}{\JournalTitle{Ligo document
  ligo-g 1802164-v1, https://dcc.ligo.org/LIGO-G1802164/public}}

\bibitem[{{Klein} {et~al.}(2018){Klein}, {Boetzel}, {Gopakumar}, {Jetzer}, \&
  {de Vittori}}]{Klein18}
{Klein}, A., {Boetzel}, Y., {Gopakumar}, A., {Jetzer}, P., \& {de Vittori}, L.
  2018,
  \href{http://adsabs.harvard.edu/abs/2018PhRvD..98j4043K}{\JournalTitle{Phys.
  Rev. D.}, 98, 104043}

\bibitem[{{Kocsis}(2013)}]{Kocsis2013}
{Kocsis}, B. 2013,
  \href{http://dx.doi.org/10.1088/0004-637X/763/2/122}{\JournalTitle{\apj},
  763, 122}

\bibitem[{{Kocsis} \& {Levin}(2012)}]{KocsisLevin12}
{Kocsis}, B., \& {Levin}, J. 2012,
  \href{http://dx.doi.org/10.1103/PhysRevD.85.123005}{\JournalTitle{\prd}, 85,
  123005}

\bibitem[{{Kocsis} \& {Sesana}(2011)}]{KocsisSesana2011}
{Kocsis}, B., \& {Sesana}, A. 2011,
  \href{http://dx.doi.org/10.1111/j.1365-2966.2010.17782.x}{\JournalTitle{\mnras},
  411, 1467}

\bibitem[{{Kocsis} \& {Tremaine}(2011)}]{Kocsis11b}
{Kocsis}, B., \& {Tremaine}, S. 2011,
  \href{http://adsabs.harvard.edu/abs/2011MNRAS.412..187K}{\JournalTitle{\mnras},
  412, 187}

\bibitem[{{Kocsis} {et~al.}(2011){Kocsis}, {Yunes}, \& {Loeb}}]{Kocsis11}
{Kocsis}, B., {Yunes}, N., \& {Loeb}, A. 2011,
  \href{http://dx.doi.org/10.1103/PhysRevD.84.024032}{\JournalTitle{\prd}, 84,
  024032}

\bibitem[{{Koppitz} {et~al.}(2007){Koppitz}, {Pollney}, {Reisswig}, {Rezzolla},
  {Thornburg}, {Diener}, \& {Schnetter}}]{Koppitz07}
{Koppitz}, M., {Pollney}, D., {Reisswig}, C., {et~al.} 2007,
  \href{http://dx.doi.org/10.1103/PhysRevLett.99.041102}{\JournalTitle{\prl},
  99, 041102}

\bibitem[{{Kormendy} \& {Ho}(2013)}]{Kormendy13}
{Kormendy}, J., \& {Ho}, L.~C. 2013,
  \href{http://adsabs.harvard.edu/abs/2013ARA%26A..51..511K}{\JournalTitle{ARAA},
  51, 511}

\bibitem[{{Kumamoto} {et~al.}(2018){Kumamoto}, {Fujii}, \&
  {Tanikawa}}]{Kumamoto18}
{Kumamoto}, J., {Fujii}, M.~S., \& {Tanikawa}, A. 2018,
  \href{http://adsabs.harvard.edu/abs/2018arXiv181106726K}{\JournalTitle{arXiv
  e-prints}}, \href{http://arxiv.org/abs/1811.06726}{{\sffamily
  arXiv:1811.06726}}

\bibitem[{{Laughlin} {et~al.}(2004){Laughlin}, {Steinacker}, \&
  {Adams}}]{Laughlin2004}
{Laughlin}, G., {Steinacker}, A., \& {Adams}, F.~C. 2004,
  \href{http://dx.doi.org/10.1086/386316}{\JournalTitle{\apj}, 608, 489}

\bibitem[{{Leigh} {et~al.}(2018){Leigh}, {Geller}, {McKernan}, {Ford},
  {MacLow}, {Bellovary}, {Haiman}, {Lyra}, {Samsing}, {O'Dowd}, {Kocsis}, \&
  {Endlich}}]{Leigh18}
{Leigh}, N.~W.~C., {Geller}, A.~M., {McKernan}, B., {et~al.} 2018,
  \href{http://adsabs.harvard.edu/abs/2018MNRAS.474.5672L}{\JournalTitle{\mnras},
  474, 5672}

\bibitem[{{Levin}(2003)}]{Levin03}
{Levin}, Y. 2003, \JournalTitle{arXiv e-prints}, astro

\bibitem[{{Li} {et~al.}(2014){Li}, {Naoz}, {Kocsis}, \&
  {Loeb}}]{Li_Naoz_Kocsis2014}
{Li}, G., {Naoz}, S., {Kocsis}, B., \& {Loeb}, A. 2014,
  \href{http://dx.doi.org/10.1088/0004-637X/785/2/116}{\JournalTitle{\apj},
  785, 116}

\bibitem[{{Li} {et~al.}(2019){Li}, {Chang}, {Levin}, {Matzner}, \&
  {Armitage}}]{LiXinyu19}
{Li}, X., {Chang}, P., {Levin}, Y., {Matzner}, C.~D., \& {Armitage}, P.~J.
  2019, \JournalTitle{arXiv e-prints}, arXiv:1912.06864

\bibitem[{{Li}(2015)}]{Li15}
{Li}, X.-D. 2015,
  \href{http://adsabs.harvard.edu/abs/2015NewAR..64....1L}{\JournalTitle{New
  Astron. Rev.}, 64, 1}

\bibitem[{{Lin} \& {Papaloizou}(1986)}]{Lin86}
{Lin}, D.~N.~C., \& {Papaloizou}, J. 1986,
  \href{http://adsabs.harvard.edu/abs/1986ApJ...309..846L}{\JournalTitle{\apj},
  309, 846}

\bibitem[{{Liu} \& {Lai}(2017)}]{Liu17}
{Liu}, B., \& {Lai}, D. 2017,
  \href{http://adsabs.harvard.edu/abs/2017ApJ...846L..11L}{\JournalTitle{ApJL},
  846, L11}

\bibitem[{{Liu} \& {Lai}(2018)}]{Liu18}
---. 2018,
  \href{http://adsabs.harvard.edu/abs/2018ApJ...863...68L}{\JournalTitle{\apj},
  863, 68}

\bibitem[{{Liu} \& {Lai}(2019)}]{Liu18b}
---. 2019,
  \href{http://adsabs.harvard.edu/abs/2019MNRAS.483.4060L}{\JournalTitle{\mnras},
  483, 4060}

\bibitem[{{Lodato} \& {Gerosa}(2013)}]{Lodato13}
{Lodato}, G., \& {Gerosa}, D. 2013,
  \href{http://dx.doi.org/10.1093/mnrasl/sls018}{\JournalTitle{\mnras}, 429,
  L30}

\bibitem[{{Lower} {et~al.}(2018){Lower}, {Thrane}, {Lasky}, \&
  {Smith}}]{Lower18}
{Lower}, M.~E., {Thrane}, E., {Lasky}, P.~D., \& {Smith}, R. 2018,
  \href{http://dx.doi.org/10.1103/PhysRevD.98.083028}{\JournalTitle{\prd}, 98,
  083028}

\bibitem[{{Lu} {et~al.}(2013){Lu}, {Do}, {Ghez}, {Morris}, {Yelda}, \&
  {Matthews}}]{Lu13}
{Lu}, J.~R., {Do}, T., {Ghez}, A.~M., {et~al.} 2013,
  \href{http://adsabs.harvard.edu/abs/2013ApJ...764..155L}{\JournalTitle{\apj},
  764, 155}

\bibitem[{{Lubow} {et~al.}(1999){Lubow}, {Seibert}, \& {Artymowicz}}]{Lubow99}
{Lubow}, S.~H., {Seibert}, M., \& {Artymowicz}, P. 1999,
  \href{http://adsabs.harvard.edu/abs/1999ApJ...526.1001L}{\JournalTitle{\apj},
  526, 1001}

\bibitem[{{Mandel} \& {de Mink}(2016)}]{Mandel16}
{Mandel}, I., \& {de Mink}, S.~E. 2016,
  \href{http://adsabs.harvard.edu/abs/2016MNRAS.458.2634M}{\JournalTitle{\mnras},
  458, 2634}

\bibitem[{{Mapelli}(2016)}]{Mapelli16b}
{Mapelli}, M. 2016,
  \href{http://dx.doi.org/10.1093/mnras/stw869}{\JournalTitle{\mnras}, 459,
  3432}

\bibitem[{{Marchant} {et~al.}(2016){Marchant}, {Langer}, {Podsiadlowski},
  {Tauris}, \& {Moriya}}]{Marchant16}
{Marchant}, P., {Langer}, N., {Podsiadlowski}, P., {Tauris}, T., \& {Moriya},
  T. 2016,
  \href{http://adsabs.harvard.edu/abs/2016A%26A...588A..50M}{\JournalTitle{A\&A},
  588, A50}

\bibitem[{{Marconi} {et~al.}(2004){Marconi}, {Risaliti}, {Gilli}, {Hunt},
  {Maiolino}, \& {Salvati}}]{Marconi04}
{Marconi}, A., {Risaliti}, G., {Gilli}, R., {et~al.} 2004,
  \href{http://adsabs.harvard.edu/abs/2004MNRAS.351..169M}{\JournalTitle{\mnras},
  351, 169}

\bibitem[{{Mardling} \& {Aarseth}(2001)}]{Mardling01}
{Mardling}, R.~A., \& {Aarseth}, S.~J. 2001,
  \href{http://dx.doi.org/10.1046/j.1365-8711.2001.03974.x}{\JournalTitle{\mnras},
  321, 398}

\bibitem[{{Martini}(2004)}]{Martini04}
{Martini}, P. 2004,
  \href{http://adsabs.harvard.edu/abs/2004cbhg.symp..169M}{\JournalTitle{Coevolution
  of Black Holes and Galaxies}, 169}

\bibitem[{{Martini} \& {Weinberg}(2001)}]{Martini01}
{Martini}, P., \& {Weinberg}, D.~H. 2001,
  \href{http://adsabs.harvard.edu/abs/2001ApJ...547...12M}{\JournalTitle{\apj},
  547, 12}

\bibitem[{{McKernan} {et~al.}(2014){McKernan}, {Ford}, {Kocsis}, {Lyra}, \&
  {Winter}}]{McKernan14}
{McKernan}, B., {Ford}, K.~E.~S., {Kocsis}, B., {Lyra}, W., \& {Winter}, L.~M.
  2014, \href{http://dx.doi.org/10.1093/mnras/stu553}{\JournalTitle{\mnras},
  441, 900}

\bibitem[{{McKernan} {et~al.}(2012){McKernan}, {Ford}, {Lyra}, \&
  {Perets}}]{McKernan12}
{McKernan}, B., {Ford}, K.~E.~S., {Lyra}, W., \& {Perets}, H.~B. 2012,
  \href{http://dx.doi.org/10.1111/j.1365-2966.2012.21486.x}{\JournalTitle{\mnras},
  425, 460}

\bibitem[{{McKernan} {et~al.}(2019){McKernan}, {Ford}, {O'Shaughnessy}, \&
  {Wysocki}}]{McKernan19}
{McKernan}, B., {Ford}, K.~E.~S., {O'Shaughnessy}, R., \& {Wysocki}, D. 2019,
  \JournalTitle{arXiv e-prints}, arXiv:1907.04356

\bibitem[{{McKernan} {et~al.}(2018){McKernan}, {Ford}, {Bellovary}, {Leigh},
  {Haiman}, {Kocsis}, {Lyra}, {Mac Low}, {Metzger}, {O'Dowd}, {Endlich}, \&
  {Rosen}}]{McKernan17}
{McKernan}, B., {Ford}, K.~E.~S., {Bellovary}, J., {et~al.} 2018,
  \href{http://adsabs.harvard.edu/abs/2018ApJ...866...66M}{\JournalTitle{\apj},
  866, 66}

\bibitem[{{Meiron} {et~al.}(2017){Meiron}, {Kocsis}, \& {Loeb}}]{Meiron17}
{Meiron}, Y., {Kocsis}, B., \& {Loeb}, A. 2017,
  \href{http://adsabs.harvard.edu/abs/2017ApJ...834..200M}{\JournalTitle{\apj},
  834, 200}

\bibitem[{{Merritt}(2010)}]{Merritt10}
{Merritt}, D. 2010,
  \href{http://adsabs.harvard.edu/abs/2010ApJ...718..739M}{\JournalTitle{\apj},
  718, 739}

\bibitem[{{Michaely} \& {Perets}(2019)}]{Michaely19}
{Michaely}, E., \& {Perets}, H.~B. 2019,
  \href{http://dx.doi.org/10.3847/2041-8213/ab5b9b}{\JournalTitle{\apjl}, 887,
  L36}

\bibitem[{{Miller} {et~al.}(2005){Miller}, {Freitag}, {Hamilton}, \&
  {Lauburb}}]{Miller05}
{Miller}, M.~C., {Freitag}, M., {Hamilton}, D.~P., \& {Lauburb}, V.~M. 2005,
  \href{http://adsabs.harvard.edu/abs/2005ApJ...631L.117M}{\JournalTitle{ApJL},
  631, L117}

\bibitem[{{Miralda-Escude} \& {Gould}(2000)}]{MiraldaEscude00}
{Miralda-Escude}, J., \& {Gould}, A. 2000,
  \href{http://adsabs.harvard.edu/abs/2000ApJ...545..847M}{\JournalTitle{ApJ},
  545, 847}

\bibitem[{{Miranda} {et~al.}(2017){Miranda}, {Munoz}, \& {Lai}}]{Miranda17}
{Miranda}, R., {Munoz}, D.~J., \& {Lai}, D. 2017,
  \href{http://adsabs.harvard.edu/abs/2017MNRAS.466.1170M}{\JournalTitle{\mnras},
  466, 1170}

\bibitem[{{Moody} {et~al.}(2019){Moody}, {Shi}, \& {Stone}}]{Moody19}
{Moody}, M.~S.~L., {Shi}, J.-M., \& {Stone}, J.~M. 2019,
  \href{http://adsabs.harvard.edu/abs/2019arXiv190300008M}{\JournalTitle{arXiv
  e-prints}}, \href{http://arxiv.org/abs/1903.00008}{{\sffamily
  arXiv:1903.00008}}

\bibitem[{{Mu{\~n}oz} {et~al.}(2019){Mu{\~n}oz}, {Miranda}, \& {Lai}}]{Munoz19}
{Mu{\~n}oz}, D.~J., {Miranda}, R., \& {Lai}, D. 2019,
  \href{http://dx.doi.org/10.3847/1538-4357/aaf867}{\JournalTitle{\apj}, 871,
  84}

\bibitem[{{Muno} {et~al.}(2009){Muno}, {Bauer}, \& {Baganoff}}]{Muno09}
{Muno}, M.~P., {Bauer}, F.~E., \& {Baganoff}, F.~K. 2009,
  \href{http://adsabs.harvard.edu/abs/2009ApJS..181..110M}{\JournalTitle{ApJS},
  181, 110}

\bibitem[{{Namekata} \& {Umemura}(2016)}]{Namekata16}
{Namekata}, D., \& {Umemura}, M. 2016,
  \href{http://adsabs.harvard.edu/abs/2016MNRAS.460..980N}{\JournalTitle{\mnras},
  460, 980}

\bibitem[{{Naoz} {et~al.}(2013{\natexlab{a}}){Naoz}, {Farr}, {Lithwick},
  {Rasio}, \& {Teyssandier}}]{Naoz2013a}
{Naoz}, S., {Farr}, W.~M., {Lithwick}, Y., {Rasio}, F.~A., \& {Teyssandier}, J.
  2013{\natexlab{a}},
  \href{http://dx.doi.org/10.1093/mnras/stt302}{\JournalTitle{\mnras}, 431,
  2155}

\bibitem[{{Naoz} {et~al.}(2013{\natexlab{b}}){Naoz}, {Kocsis}, {Loeb}, \&
  {Yunes}}]{Naoz2013}
{Naoz}, S., {Kocsis}, B., {Loeb}, A., \& {Yunes}, N. 2013{\natexlab{b}},
  \href{http://dx.doi.org/10.1088/0004-637X/773/2/187}{\JournalTitle{\apj},
  773, 187}

\bibitem[{{Narayan} \& {McClintock}(2008)}]{Narayan08}
{Narayan}, R., \& {McClintock}, J.~E. 2008,
  \href{http://adsabs.harvard.edu/abs/2008NewAR..51..733N}{\JournalTitle{New
  Astronomy Reviews}, 51, 733}

\bibitem[{{Natarajan} \& {Pringle}(1998)}]{Natarajan98}
{Natarajan}, P., \& {Pringle}, J.~E. 1998,
  \href{http://adsabs.harvard.edu/abs/1998ApJ...506L..97N}{\JournalTitle{\apj},
  506, L97}

\bibitem[{{Nayakshin} {et~al.}(2007){Nayakshin}, {Cuadra}, \&
  {Springel}}]{Nayakshin07}
{Nayakshin}, S., {Cuadra}, J., \& {Springel}, V. 2007,
  \href{http://adsabs.harvard.edu/abs/2007MNRAS.379...21N}{\JournalTitle{\mnras},
  379, 21}

\bibitem[{{Nishizawa} {et~al.}(2016){Nishizawa}, {Berti}, {Klein}, \&
  {Sesana}}]{Nishizawa16}
{Nishizawa}, A., {Berti}, E., {Klein}, A., \& {Sesana}, A. 2016,
  \href{http://adsabs.harvard.edu/abs/2016PhRvD..94f4020N}{\JournalTitle{Phys.
  Rev. D}, 94, 064020}

\bibitem[{{Norris} {et~al.}(2014){Norris}, {Kannappan}, {Forbes}, {Romanowsky},
  {Brodie}, {Faifer}, {Huxor}, {Maraston}, {Moffett}, {Penny}, {Pota},
  {Smith-Castelli}, {Strader}, {Bradley}, {Eckert}, {Fohring}, {McBride},
  {Stark}, \& {Vaduvescu}}]{Norris+2014}
{Norris}, M.~A., {Kannappan}, S.~J., {Forbes}, D.~A., {et~al.} 2014,
  \href{http://dx.doi.org/10.1093/mnras/stu1186}{\JournalTitle{\mnras}, 443,
  1151}

\bibitem[{{Novak} {et~al.}(2019){Novak}, {Ba{\~n}ados}, {Decarli}, {Walter},
  {Venemans}, {Neeleman}, {Farina}, {Mazzucchelli}, {Carilli}, {Fan}, {Rix}, \&
  {Wang}}]{Novak19}
{Novak}, M., {Ba{\~n}ados}, E., {Decarli}, R., {et~al.} 2019,
  \href{http://dx.doi.org/10.3847/1538-4357/ab2beb}{\JournalTitle{\apj}, 881,
  63}

\bibitem[{{Ogilvie}(1999)}]{Ogilvie99}
{Ogilvie}, G.~I. 1999,
  \href{http://dx.doi.org/10.1046/j.1365-8711.1999.02340.x}{\JournalTitle{\mnras},
  304, 557}

\bibitem[{{O'Leary} {et~al.}(2009){O'Leary}, {Kocsis}, \& {Loeb}}]{OLeary09}
{O'Leary}, R.~M., {Kocsis}, B., \& {Loeb}, A. 2009,
  \href{http://adsabs.harvard.edu/abs/2009MNRAS.395.2127O}{\JournalTitle{\mnras},
  395, 2127}

\bibitem[{{O'Leary} {et~al.}(2016){O'Leary}, {Meiron}, \& {Kocsis}}]{OLeary16}
{O'Leary}, R.~M., {Meiron}, Y., \& {Kocsis}, B. 2016,
  \href{http://adsabs.harvard.edu/abs/2016ApJ...824L..12O}{\JournalTitle{ApJL},
  824, L12}

\bibitem[{{O'Leary} {et~al.}(2006){O'Leary}, {Rasio}, {Fregeau}, {Ivanova}, \&
  {O'Shaughnessy}}]{OLeary06}
{O'Leary}, R.~M., {Rasio}, F.~A., {Fregeau}, J.~M., {Ivanova}, N., \&
  {O'Shaughnessy}, R. 2006,
  \href{http://adsabs.harvard.edu/abs/2006ApJ...637..937O}{\JournalTitle{\apj},
  637, 937}

\bibitem[{{Ostriker}(1999)}]{Ostriker99}
{Ostriker}, E.~C. 1999,
  \href{http://adsabs.harvard.edu/abs/1999ApJ...513..252O}{\JournalTitle{\apj},
  513, 252}

\bibitem[{{Paardekooper} {et~al.}(2010){Paardekooper}, {Baruteau}, {Crida}, \&
  {Kley}}]{Paardekooper10}
{Paardekooper}, S.-J., {Baruteau}, C., {Crida}, A., \& {Kley}, W. 2010,
  \href{http://adsabs.harvard.edu/abs/2010MNRAS.401.1950P}{\JournalTitle{\mnras},
  401, 1950}

\bibitem[{{Paczynski}(1976)}]{Paczynski76}
{Paczynski}, B. 1976,
  \href{http://adsabs.harvard.edu/abs/1976IAUS...73...75P}{\JournalTitle{in IAU
  Symposium, Structure and Evolution of Close Binary Systems}, 73, 75}

\bibitem[{{Panamarev} {et~al.}(2018){Panamarev}, {Shukirgaliyev}, {Meiron},
  {Berczik}, {Just}, {Spurzem}, {Omarov}, \& {Vilkoviskij}}]{Panamarev18}
{Panamarev}, T., {Shukirgaliyev}, B., {Meiron}, Y., {et~al.} 2018,
  \href{http://adsabs.harvard.edu/abs/2018MNRAS.476.4224P}{\JournalTitle{\mnras},
  476, 4224}

\bibitem[{{Papaloizou} \& {Larwood}(2000)}]{Papaloizou00}
{Papaloizou}, J.~C.~B., \& {Larwood}, J.~D. 2000,
  \href{http://adsabs.harvard.edu/abs/2000MNRAS.315..823P}{\JournalTitle{\mnras},
  315, 823}

\bibitem[{{Park} \& {Bogdanovic}(2017)}]{Park17}
{Park}, K., \& {Bogdanovic}, T. 2017,
  \href{http://adsabs.harvard.edu/abs/2017ApJ...838..103P}{\JournalTitle{\apj},
  838, 103}

\bibitem[{{Park} \& {Ricotti}(2013)}]{Park13}
{Park}, K., \& {Ricotti}, M. 2013,
  \href{http://dx.doi.org/10.1088/0004-637X/767/2/163}{\JournalTitle{\apj},
  767, 163}

\bibitem[{{Pavlovskii} {et~al.}(2017){Pavlovskii}, {Ivanova}, {Belczynski}, \&
  {Van}}]{Pavlovskii17}
{Pavlovskii}, K., {Ivanova}, N., {Belczynski}, K., \& {Van}, K.~X. 2017,
  \href{http://adsabs.harvard.edu/abs/2017MNRAS.465.2092P}{\JournalTitle{\mnras},
  465, 2092}

\bibitem[{{P{\'e}rez-Torres} {et~al.}(2010){P{\'e}rez-Torres}, {Alberdi},
  {Romero-Ca{\~n}izales}, \& {Bondi}}]{PerezTorres10}
{P{\'e}rez-Torres}, M.~A., {Alberdi}, A., {Romero-Ca{\~n}izales}, C., \&
  {Bondi}, M. 2010,
  \href{http://dx.doi.org/10.1051/0004-6361/201015462}{\JournalTitle{\aap},
  519, L5}

\bibitem[{{Peters}(1964)}]{Peters64}
{Peters}, P.~C. 1964, \JournalTitle{Phys. Rev.}, 136, 1224

\bibitem[{{Pfuhl} {et~al.}(2014){Pfuhl}, {Alexander}, \& {Gillessen}}]{Pfuhl14}
{Pfuhl}, O., {Alexander}, T., \& {Gillessen}, S. 2014,
  \href{http://adsabs.harvard.edu/abs/2014ApJ...782..101P}{\JournalTitle{\apj},
  782, 101}

\bibitem[{{Podsiadlowski} {et~al.}(2003){Podsiadlowski}, {Rappaport}, \&
  {Han}}]{Podsiadlowski03}
{Podsiadlowski}, P., {Rappaport}, S., \& {Han}, Z. 2003,
  \href{http://adsabs.harvard.edu/abs/2003MNRAS.341..385P}{\JournalTitle{\mnras},
  341, 385}

\bibitem[{{Portegies Zwart} \& {McMillan}(2000)}]{PortegiesZwart00}
{Portegies Zwart}, S.~F., \& {McMillan}, S.~L.~W. 2000,
  \href{http://adsabs.harvard.edu/abs/2000ApJ...528L..17P}{\JournalTitle{\apj},
  528, L17}

\bibitem[{{Randall} \& {Xianyu}(2018)}]{Randall18}
{Randall}, L., \& {Xianyu}, Z.-Z. 2018,
  \href{http://dx.doi.org/10.3847/1538-4357/aad7fe}{\JournalTitle{\apj}, 864,
  134}

\bibitem[{{Rasskazov} \& {Kocsis}(2019)}]{Rasskazov19}
{Rasskazov}, A., \& {Kocsis}, B. 2019,
  \href{http://adsabs.harvard.edu/abs/2019arXiv190203242R}{\JournalTitle{arXiv
  e-prints}}, \href{http://arxiv.org/abs/1902.03242}{{\sffamily
  arXiv:1902.03242}}

\bibitem[{{Rastello} {et~al.}(2018){Rastello}, {Amaro-Seoane}, {Arca-Sedda},
  {Capuzzo-Dolcetta}, {Fragione}, \& I.}]{Rastello18}
{Rastello}, S., {Amaro-Seoane}, P., {Arca-Sedda}, M., {et~al.} 2018,
  \href{http://adsabs.harvard.edu/abs/2019MNRAS.483.1233R}{\JournalTitle{arXiv
  e-prints}}, \href{http://arxiv.org/abs/1811.10628}{{\sffamily
  arXiv:1811.10628}}

\bibitem[{{Regan} {et~al.}(2019){Regan}, {Downes}, {Volonteri},
  {et~al.}}]{Regan19}
{Regan}, J.~A., {Downes}, T.~P., {Volonteri}, M., {et~al.} 2019,
  \href{http://adsabs.harvard.edu/abs/2018arXiv181104953R}{\JournalTitle{arXiv
  e-prints}}, \href{http://arxiv.org/abs/1811.04953}{{\sffamily
  arXiv:1811.04953}}

\bibitem[{{R{\"o}dig} {et~al.}(2011){R{\"o}dig}, {Dotti}, {Sesana}, {Cuadra},
  \& {Colpi}}]{Rodig11}
{R{\"o}dig}, C., {Dotti}, M., {Sesana}, A., {Cuadra}, J., \& {Colpi}, M. 2011,
  \href{http://adsabs.harvard.edu/abs/2011MNRAS.415.3033R}{\JournalTitle{\mnras},
  415, 3033}

\bibitem[{{Rodriguez} {et~al.}(2018){Rodriguez}, {Amaro-Seoane}, {Chatterjee},
  {Kremer}, {Rasio}, {Samsing}, {Ye}, \& {Zevin}}]{Rodriguez18b}
{Rodriguez}, C.~L., {Amaro-Seoane}, P., {Chatterjee}, S., {et~al.} 2018,
  \href{http://adsabs.harvard.edu/abs/2018PhRvD..98l3005R}{\JournalTitle{Phys.
  Rev. D}, 98, 123005}

\bibitem[{{Rodriguez} {et~al.}(2016{\natexlab{a}}){Rodriguez}, {Chatterjee}, \&
  {Rasio}}]{Rodriguez16}
{Rodriguez}, C.~L., {Chatterjee}, S., \& {Rasio}, F.~A. 2016{\natexlab{a}},
  \href{http://adsabs.harvard.edu/abs/2016PhRvD..93h4029R}{\JournalTitle{Phys.
  Rev. D.}, 93, 084029}

\bibitem[{{Rodriguez} {et~al.}(2016{\natexlab{b}}){Rodriguez}, {Haster},
  {Chatterjee}, {Kalogera}, \& {Rasio}}]{Rodriguez16b}
{Rodriguez}, C.~L., {Haster}, C.-J., {Chatterjee}, S., {Kalogera}, V., \&
  {Rasio}, F.~A. 2016{\natexlab{b}},
  \href{http://adsabs.harvard.edu/abs/2016ApJ...824L...8R}{\JournalTitle{ApJL},
  824, L8}

\bibitem[{{Romero-Shaw} {et~al.}(2019){Romero-Shaw}, {Lasky}, \&
  {Thrane}}]{Romero-Shaw19}
{Romero-Shaw}, I.~M., {Lasky}, P.~D., \& {Thrane}, E. 2019, \JournalTitle{arXiv
  e-prints}, arXiv:1909.05466

\bibitem[{{Salaris} \& {Cassisi}(2005)}]{Salaris05}
{Salaris}, M., \& {Cassisi}, S. 2005, {Evolution of Stars and Stellar
  Populations}

\bibitem[{{Samsing} {et~al.}(2019){Samsing}, {D'Orazio}, {Kremer}, {Rodriguez},
  \& {Askar}}]{Samsing19}
{Samsing}, J., {D'Orazio}, D.~J., {Kremer}, K., {Rodriguez}, C.~L., \& {Askar},
  A. 2019, \JournalTitle{arXiv e-prints}, arXiv:1907.11231

\bibitem[{{Samsing} {et~al.}(2014){Samsing}, {MacLeod}, \&
  {Ramirez-Ruiz}}]{Samsing14}
{Samsing}, J., {MacLeod}, M., \& {Ramirez-Ruiz}, E. 2014,
  \href{http://dx.doi.org/10.1088/0004-637X/784/1/71}{\JournalTitle{\apj}, 784,
  71}

\bibitem[{{Sana} {et~al.}(2012){Sana}, {de Mink}, {de Koter},
  {et~al.}}]{Sana12}
{Sana}, H., {de Mink}, S.~E., {de Koter}, A., {et~al.} 2012,
  \href{http://adsabs.harvard.edu/abs/2012Sci...337..444S}{\JournalTitle{Science},
  337, 444}

\bibitem[{{S{\'a}nchez-Salcedo} {et~al.}(2018){S{\'a}nchez-Salcedo},
  {Chametla}, \& {Santill{\'a}n}}]{SanchezSalcedo18}
{S{\'a}nchez-Salcedo}, F.~J., {Chametla}, R.~O., \& {Santill{\'a}n}, A. 2018,
  \href{http://dx.doi.org/10.3847/1538-4357/aac494}{\JournalTitle{\apj}, 860,
  129}

\bibitem[{{Scheuer} \& {Feiler}(1996)}]{Scheuer96}
{Scheuer}, P.~A.~G., \& {Feiler}, R. 1996,
  \href{http://adsabs.harvard.edu/abs/1996MNRAS.282..291S}{\JournalTitle{\mnras},
  282, 291}

\bibitem[{{Sch{\"o}del} {et~al.}(2014){Sch{\"o}del}, {Feldmeier}, {Kunneriath},
  {Stolovy}, {Neumayer}, {Amaro-Seoane}, \& {Nishiyama}}]{Schodel14}
{Sch{\"o}del}, R., {Feldmeier}, A., {Kunneriath}, D., {et~al.} 2014,
  \href{http://dx.doi.org/10.1051/0004-6361/201423481}{\JournalTitle{\aap},
  566, A47}

\bibitem[{{Scott} \& {Graham}(2013)}]{Scott13}
{Scott}, N., \& {Graham}, A.~W. 2013,
  \href{http://adsabs.harvard.edu/abs/2013ApJ...763...76S}{\JournalTitle{\apj},
  763, 76}

\bibitem[{{Secunda} {et~al.}(2019){Secunda}, {Bellovary}, {Mac Low}, {Ford},
  {McKernan}, {Leigh}, {Lyra}, \& {S{\'a}ndor}}]{Secunda19}
{Secunda}, A., {Bellovary}, J., {Mac Low}, M.-M., {et~al.} 2019,
  \href{http://dx.doi.org/10.3847/1538-4357/ab20ca}{\JournalTitle{\apj}, 878,
  85}

\bibitem[{{Silsbee} \& {Tremaine}(2017)}]{Silsbee17}
{Silsbee}, K., \& {Tremaine}, S. 2017,
  \href{http://adsabs.harvard.edu/abs/2017ApJ...836...39S}{\JournalTitle{\apj},
  836, 39}

\bibitem[{{Spera} {et~al.}(2019){Spera}, {Mapelli}, {Giacobbo}, {Trani},
  {Bressan}, \& {Costa}}]{Spera19}
{Spera}, M., {Mapelli}, M., {Giacobbo}, N., {et~al.} 2019,
  \href{http://dx.doi.org/10.1093/mnras/stz359}{\JournalTitle{\mnras}, 485,
  889}

\bibitem[{{Stephan} {et~al.}(2016){Stephan}, {Naoz}, {Ghez}, {Witzel},
  {Sitarski}, {Do}, \& {Kocsis}}]{Stephan16}
{Stephan}, A.~P., {Naoz}, S., {Ghez}, A.~M., {et~al.} 2016,
  \href{http://dx.doi.org/10.1093/mnras/stw1220}{\JournalTitle{\mnras}, 460,
  3494}

\bibitem[{{Stephan} {et~al.}(2019){Stephan}, {Naoz}, {Ghez}, {Morris},
  {Ciurlo}, {Do}, {Breivik}, {Coughlin}, \& {Rodriguez}}]{Stephan19}
---. 2019,
  \href{http://adsabs.harvard.edu/abs/2019arXiv190300010S}{\JournalTitle{arXiv
  e-prints}}, \href{http://arxiv.org/abs/1903.00010}{{\sffamily
  arXiv:1903.00010}}

\bibitem[{{Stone} {et~al.}(2017){Stone}, {Metzger}, \& {Haiman}}]{Stone17}
{Stone}, N.~C., {Metzger}, B.~D., \& {Haiman}, Z. 2017,
  \href{http://adsabs.harvard.edu/abs/2017MNRAS.464..946S}{\JournalTitle{\mnras},
  464, 946}

\bibitem[{{Su} {et~al.}(2010){Su}, {Slatyer}, \& {Finkbeiner}}]{Su10}
{Su}, M., {Slatyer}, T.~R., \& {Finkbeiner}, D.~P. 2010,
  \href{http://dx.doi.org/10.1088/0004-637X/724/2/1044}{\JournalTitle{\apj},
  724, 1044}

\bibitem[{{Sugimura} {et~al.}(2018){Sugimura}, {Hosokawa}, {Yajima},
  {Inayoshi}, \& {Omukai}}]{Sugimura18}
{Sugimura}, K., {Hosokawa}, T., {Yajima}, H., {Inayoshi}, K., \& {Omukai}, K.
  2018,
  \href{http://adsabs.harvard.edu/abs/2018MNRAS.478.3961S}{\JournalTitle{\mnras},
  478, 3961}

\bibitem[{{Szolgyen} \& {Kocsis}(2018)}]{Szolgyen18}
{Szolgyen}, A., \& {Kocsis}, B. 2018,
  \href{http://adsabs.harvard.edu/abs/2018PhRvL.121j1101S}{\JournalTitle{Phys.
  Rev. Lett.}, 121, 101101}

\bibitem[{{Tagawa} {et~al.}(2020){Tagawa}, {Haiman}, {Bartos}, \&
  {Kocsis}}]{Tagawa20}
{Tagawa}, H., {Haiman}, Z., {Bartos}, I., \& {Kocsis}, B. 2020,
  \JournalTitle{arXiv e-prints}, arXiv:2004.11914

\bibitem[{{Tagawa} {et~al.}(2018){Tagawa}, {Kocsis}, \& {Saitoh}}]{Tagawa18}
{Tagawa}, H., {Kocsis}, B., \& {Saitoh}, R.~T. 2018,
  \href{http://adsabs.harvard.edu/abs/2018PhRvL.120z1101T}{\JournalTitle{Phys.
  Rev. Lett.}, 120, 261101}

\bibitem[{{Tagawa} \& {Umemura}(2018)}]{Tagawa18b}
{Tagawa}, H., \& {Umemura}, M. 2018,
  \href{http://adsabs.harvard.edu/abs/2018ApJ...856...47T}{\JournalTitle{\apj},
  856, 47}

\bibitem[{{Takeo} {et~al.}(2020){Takeo}, {Inayoshi}, \& {Mineshige}}]{Takeo20}
{Takeo}, E., {Inayoshi}, K., \& {Mineshige}, S. 2020, \JournalTitle{arXiv
  e-prints}, arXiv:2002.07187

\bibitem[{{Tanaka} {et~al.}(2002){Tanaka}, {Takeuchi}, \& {Ward}}]{Tanaka02}
{Tanaka}, H., {Takeuchi}, T., \& {Ward}, W.~R. 2002,
  \href{http://adsabs.harvard.edu/abs/2002ApJ...565.1257T}{\JournalTitle{\apj},
  565, 1257}

\bibitem[{{Tang} {et~al.}(2017){Tang}, {MacFadyen}, \& Z.}]{Tang17}
{Tang}, Y., {MacFadyen}, A., \& Z., H. 2017,
  \href{http://adsabs.harvard.edu/abs/2017MNRAS.469.4258T}{\JournalTitle{\mnras},
  469, 4258}

\bibitem[{{The LIGO Scientific Collaboration} \& {The Virgo
  Collaboration}(2012)}]{TheLIGO12}
{The LIGO Scientific Collaboration}, \& {The Virgo Collaboration}. 2012,
  \JournalTitle{arXiv e-prints}, arXiv:1203.2674

\bibitem[{{The LIGO Scientific Collaboration} \& {the Virgo
  Collaboration}(2020)}]{LIGO20_GW190412}
{The LIGO Scientific Collaboration}, \& {the Virgo Collaboration}. 2020,
  \JournalTitle{arXiv e-prints}, arXiv:2004.08342

\bibitem[{{Thompson} {et~al.}(2005){Thompson}, {Quataert}, \&
  {Murray}}]{Thompson05}
{Thompson}, T.~A., {Quataert}, E., \& {Murray}, N. 2005,
  \href{http://adsabs.harvard.edu/abs/2005ApJ...630..167T}{\JournalTitle{\apj},
  630, 167}

\bibitem[{{Tichy} \& {Marronetti}(2008)}]{Tichy08}
{Tichy}, W., \& {Marronetti}, P. 2008,
  \href{http://dx.doi.org/10.1103/PhysRevD.78.081501}{\JournalTitle{\prd}, 78,
  081501}

\bibitem[{{Tiede} {et~al.}(2020){Tiede}, {Zrake}, {MacFadyen}, \&
  Zoltan}]{Tiede+20}
{Tiede}, C., {Zrake}, J., {MacFadyen}, A., \& Zoltan, H. 2020,
  \JournalTitle{\mnras, submitted}

\bibitem[{{Torres} {et~al.}(2010){Torres}, {Andersen}, \& {Gimenez}}]{Torres10}
{Torres}, G., {Andersen}, J., \& {Gimenez}, A. 2010,
  \href{http://adsabs.harvard.edu/abs/2010A%26ARv..18...67T}{\JournalTitle{ARAA},
  18, 67}

\bibitem[{{Toyouchi} {et~al.}(2020){Toyouchi}, {Hosokawa}, {Sugimura}, \&
  {Kuiper}}]{Toyouchi20}
{Toyouchi}, D., {Hosokawa}, T., {Sugimura}, K., \& {Kuiper}, R. 2020,
  \JournalTitle{arXiv e-prints}, arXiv:2002.08017

\bibitem[{{Toyouchi} {et~al.}(2019){Toyouchi}, {Hosokawa}, {Sugimura},
  {Nakatani}, \& {Kuiper}}]{Toyouchi19}
{Toyouchi}, D., {Hosokawa}, T., {Sugimura}, K., {Nakatani}, R., \& {Kuiper}, R.
  2019,
  \href{http://adsabs.harvard.edu/abs/2019MNRAS.483.2031T}{\JournalTitle{\mnras},
  483, 2031}

\bibitem[{{Trani} {et~al.}(2018){Trani}, {Fujii}, \& {Spera}}]{Trani18}
{Trani}, A.~A., {Fujii}, M.~S., \& {Spera}, M. 2018,
  \href{http://adsabs.harvard.edu/abs/2018arXiv180907339T}{\JournalTitle{arXiv
  e-prints}}, \href{http://arxiv.org/abs/1809.07339}{{\sffamily
  arXiv:1809.07339}}

\bibitem[{{Trippe} {et~al.}(2008){Trippe}, {Gillessen}, {Gerhard},
  {et~al.}}]{Trippe08}
{Trippe}, S., {Gillessen}, S., {Gerhard}, O.~E., {et~al.} 2008,
  \href{http://adsabs.harvard.edu/abs/2008A%26A...492..419T}{\JournalTitle{A$\&$
  A}, 492, 419}

\bibitem[{{Udall} {et~al.}(2019){Udall}, {Jani}, {Lange}, {O'Shaughnessy},
  {Clark}, {Cadonati}, {Shoemaker}, \& {Holley-Bockelmann}}]{Udall19}
{Udall}, R., {Jani}, K., {Lange}, J., {et~al.} 2019, \JournalTitle{arXiv
  e-prints}, arXiv:1912.10533

\bibitem[{{Valtonen} \& {Karttunen}(2006)}]{Valtonen06}
{Valtonen}, M., \& {Karttunen}, H. 2006,
  \href{http://adsabs.harvard.edu/abs/2006tbp..book.....V}{\JournalTitle{The
  Three-Body Problem (Cambridge: Cambridge University Press)}}

\bibitem[{{van den Heuvel} {et~al.}(2017){van den Heuvel}, {Portegies Zwart},
  \& {de Mink}}]{vandenHeuvel17}
{van den Heuvel}, E.~P.~J., {Portegies Zwart}, S.~F., \& {de Mink}, S.~E. 2017,
  \href{http://adsabs.harvard.edu/abs/2017MNRAS.471.4256V}{\JournalTitle{\mnras},
  471, 4256}

\bibitem[{{Venumadhav} {et~al.}(2019){Venumadhav}, {Zackay}, {Roulet}, {Dai},
  \& {Zaldarriaga}}]{Venumadhav19}
{Venumadhav}, T., {Zackay}, B., {Roulet}, J., {Dai}, L., \& {Zaldarriaga}, M.
  2019,
  \href{http://adsabs.harvard.edu/abs/2019arXiv190407214V}{\JournalTitle{arXiv
  e-prints}}, \href{http://arxiv.org/abs/1904.07214}{{\sffamily
  arXiv:1904.07214}}

\bibitem[{{Vokrouhlicky} \& {Karas}(1998)}]{Vokrouhlicky98}
{Vokrouhlicky}, D., \& {Karas}, V. 1998,
  \href{http://dx.doi.org/10.1046/j.1365-8711.1998.01564.x}{\JournalTitle{\mnras},
  298, 53}

\bibitem[{{Volonteri} {et~al.}(2007){Volonteri}, {Sikora}, \&
  {Lasota}}]{Volonteri07}
{Volonteri}, M., {Sikora}, M., \& {Lasota}, J.-P. 2007,
  \href{http://dx.doi.org/10.1086/521186}{\JournalTitle{\apj}, 667, 704}

\bibitem[{{{\v{S}}ubr} \& {Karas}(2005)}]{Subr05}
{{\v{S}}ubr}, L., \& {Karas}, V. 2005,
  \href{http://dx.doi.org/10.1051/0004-6361:20042089}{\JournalTitle{\aap}, 433,
  405}

\bibitem[{{Wada} {et~al.}(2016){Wada}, {Schartmann}, \& {Meijerink}}]{Wada16}
{Wada}, K., {Schartmann}, M., \& {Meijerink}, R. 2016,
  \href{http://dx.doi.org/10.3847/2041-8205/828/2/L19}{\JournalTitle{\apjl},
  828, L19}

\bibitem[{{Walcher} {et~al.}(2005){Walcher}, {van der Marel}, {McLaughlin},
  {Rix}, {Boker}, {Haring}, {Ho}, {Sarzi}, \& {Shields}}]{Walcher05}
{Walcher}, C.~J., {van der Marel}, R.~P., {McLaughlin}, D., {et~al.} 2005,
  \href{http://adsabs.harvard.edu/abs/2005ApJ...618..237W}{\JournalTitle{\apj},
  618, 237}

\bibitem[{{Wang} {et~al.}(2016){Wang}, {Jia}, \& {Li}}]{Wang16}
{Wang}, C., {Jia}, K., \& {Li}, X.-D. 2016,
  \href{http://adsabs.harvard.edu/abs/2016MNRAS.457.1015W}{\JournalTitle{\mnras},
  457, 1015}

\bibitem[{{Ward}(1997)}]{Ward97}
{Ward}, W.~R. 1997,
  \href{http://adsabs.harvard.edu/abs/1997Icar..126..261W}{\JournalTitle{Icarus},
  126, 261}

\bibitem[{{Wardle} \& {Yusef-Zadeh}(2008)}]{Wardle08}
{Wardle}, M., \& {Yusef-Zadeh}, F. 2008,
  \href{http://dx.doi.org/10.1086/591471}{\JournalTitle{\apjl}, 683, L37}

\bibitem[{{Wen}(2003)}]{Wen2003}
{Wen}, L. 2003, \href{http://dx.doi.org/10.1086/378794}{\JournalTitle{\apj},
  598, 419}

\bibitem[{{Wiktorowicz} {et~al.}(2014){Wiktorowicz}, {Belczynski}, \&
  {Maccarone}}]{Wiktorowicz14}
{Wiktorowicz}, G., {Belczynski}, K., \& {Maccarone}, T. 2014,
  \href{http://adsabs.harvard.edu/abs/2014bsee.confE..37W}{\JournalTitle{in
  Binary Systems, their Evolution and Environments}, 37}

\bibitem[{{Williamson} {et~al.}(2019){Williamson}, {H{\"o}nig}, \&
  {Venanzi}}]{Williamson19}
{Williamson}, D., {H{\"o}nig}, S., \& {Venanzi}, M. 2019,
  \href{http://dx.doi.org/10.3847/1538-4357/ab17d5}{\JournalTitle{\apj}, 876,
  137}

\bibitem[{{Wong} {et~al.}(2018){Wong}, {Baibhav}, \& {Berti}}]{Wong19}
{Wong}, K.~W.~K., {Baibhav}, V., \& {Berti}, E. 2018,
  \href{http://adsabs.harvard.edu/abs/2019arXiv190201402W}{\JournalTitle{arXiv
  e-prints}}, \href{http://arxiv.org/abs/1902.01402}{{\sffamily
  arXiv:1902.01402}}

\bibitem[{{Xu} {et~al.}(2018){Xu}, {Bian}, {Shen}, {Zuo}, {Fan}, \&
  {Zhu}}]{Xu18}
{Xu}, F., {Bian}, F., {Shen}, Y., {et~al.} 2018,
  \href{http://dx.doi.org/10.1093/mnras/sty1763}{\JournalTitle{\mnras}, 480,
  345}

\bibitem[{{Yang} {et~al.}(2019{\natexlab{a}}){Yang}, {Bartos}, {Haiman},
  {Kocsis}, {M{\'a}rka}, {Stone}, \& {M{\'a}rka}}]{Yang19a}
{Yang}, Y., {Bartos}, I., {Haiman}, Z., {et~al.} 2019{\natexlab{a}},
  \href{http://dx.doi.org/10.3847/1538-4357/ab16e3}{\JournalTitle{\apj}, 876,
  122}

\bibitem[{{Yang} {et~al.}(2019{\natexlab{b}}){Yang}, {Bartos}, {Gayathri},
  {Ford}, {Haiman}, {Klimenko}, {Kocsis}, {M{\'a}rka}, {M{\'a}rka}, \&
  {McKernan}}]{Yang19b}
{Yang}, Y., {Bartos}, I., {Gayathri}, V., {et~al.} 2019{\natexlab{b}},
  \JournalTitle{arXiv e-prints}, arXiv:1906.09281

\bibitem[{{Yelda} {et~al.}(2014){Yelda}, {Ghez}, {Lu}, {Do}, {Meyer}, {Morris},
  \& {Matthews}}]{Yelda14}
{Yelda}, S., {Ghez}, A.~M., {Lu}, J.~R., {et~al.} 2014,
  \href{http://adsabs.harvard.edu/abs/2014ApJ...783..131Y}{\JournalTitle{\apj},
  783, 131}

\bibitem[{{Yi} {et~al.}(2018){Yi}, {Cheng}, \& {Taam}}]{Yi18}
{Yi}, S.-X., {Cheng}, K.~S., \& {Taam}, R.~E. 2018,
  \href{http://dx.doi.org/10.3847/2041-8213/aac649}{\JournalTitle{\apjl}, 859,
  L25}

\bibitem[{{Yunes} {et~al.}(2011){Yunes}, {Kocsis}, {Loeb}, \&
  {Haiman}}]{Yunes2011}
{Yunes}, N., {Kocsis}, B., {Loeb}, A., \& {Haiman}, Z. 2011,
  \href{http://dx.doi.org/10.1103/PhysRevLett.107.171103}{\JournalTitle{\prl},
  107, 171103}

\bibitem[{{Zackay} {et~al.}(2019{\natexlab{a}}){Zackay}, {Dai}, {Venumadhav},
  {Roulet}, \& {Zaldarriaga}}]{Zackay19b}
{Zackay}, B., {Dai}, L., {Venumadhav}, T., {Roulet}, J., \& {Zaldarriaga}, M.
  2019{\natexlab{a}}, \JournalTitle{arXiv e-prints}, arXiv:1910.09528

\bibitem[{{Zackay} {et~al.}(2019{\natexlab{b}}){Zackay}, {Venumadhav}, {Dai},
  {Roulet}, \& {Zaldarriaga}}]{Zackay19}
{Zackay}, B., {Venumadhav}, T., {Dai}, L., {Roulet}, J., \& {Zaldarriaga}, M.
  2019{\natexlab{b}},
  \href{http://adsabs.harvard.edu/abs/2019arXiv190210331Z}{\JournalTitle{arXiv
  e-prints}}, \href{http://arxiv.org/abs/1902.10331}{{\sffamily
  arXiv:1902.10331}}

\bibitem[{{Zevin} {et~al.}(2019){Zevin}, {Samsing}, {Rodriguez}, {Haster}, \&
  {Ramirez-Ruiz}}]{Zevin19}
{Zevin}, M., {Samsing}, J., {Rodriguez}, C., {Haster}, C.-J., \&
  {Ramirez-Ruiz}, E. 2019,
  \href{http://adsabs.harvard.edu/abs/2019ApJ...871...91Z}{\JournalTitle{\apj},
  871, 91}

\bibitem[{{Zhang} {et~al.}(2019){Zhang}, {Shao}, \& {Zhu}}]{Zhang19}
{Zhang}, F., {Shao}, L., \& {Zhu}, W. 2019,
  \href{http://adsabs.harvard.edu/abs/2019arXiv190302685Z}{\JournalTitle{arXiv
  e-prints}}, \href{http://arxiv.org/abs/1903.02685}{{\sffamily
  arXiv:1903.02685}}

\bibitem[{{Ziosi} {et~al.}(2014){Ziosi}, {Mapelli}, {Branchesi}, \&
  {Tormen}}]{Ziosi14}
{Ziosi}, B.~M., {Mapelli}, M., {Branchesi}, M., \& {Tormen}, G. 2014,
  \href{http://dx.doi.org/10.1093/mnras/stu824}{\JournalTitle{\mnras}, 441,
  3703}

\bibitem[{{Zubovas} {et~al.}(2011){Zubovas}, {King}, \&
  {Nayakshin}}]{Zubovas11}
{Zubovas}, K., {King}, A.~R., \& {Nayakshin}, S. 2011,
  \href{https://ui.adsabs.harvard.edu/abs/2011MNRAS.415L..21Z/abstract}{\JournalTitle{MNRAS
  Letters}, 415, L21}

\end{thebibliography}

\appendix

\section{Conversion efficiency}
\label{sec:conversion_eff}

In the disk model of \citet{Thompson05}, the outer regions of the AGN disk are stabilized by radiation pressure and supernovae from in-situ formed stars. 
The model introduces a parameter $\epsilon$ defined as the conversion efficiency with which star formation converts mass into radiation, and uses this parameter to calculate the stellar contribution of radiation pressure and the radiative flux.
We modify the calcalation of $\epsilon$ by accounting for the stellar and the AGN lifetime. 
The radiation flux of main sequence stars formed in the disk averaged over the two disk faces is
\begin{align}
\label{eq:flux_luminosity}
    F_\mathrm{AGN,*}=  \frac{\sum_i^{N_\mathrm{form}}L(m_{\mathrm{star},i})\min[t_\mathrm{AGN}, t_{\mathrm{star}}(m_{\mathrm{star},i})]}{2At_\mathrm{AGN}},
\end{align}
where $A=\pi r^2$ is the disk's area of radius $r$, 
$t_{\mathrm{star}} = 10\,\mathrm{Gyr}(m_\mathrm{star}/\Msun)(L_\mathrm{star}/\Lsun)^{-1}$ is the lifetime of a star \citep{Hansen94}, $t_\mathrm{AGN}=100\,\rm Myr$ is the typical lifetime of AGN disks, and $N_\mathrm{form}$ is the number of stars formed during the AGN phase,
$L(m_\mathrm{star})$ is the luminosity of a star of mass $m_\mathrm{star}$:
\begin{align}\label{eq:mass_luminosity}
L(m_{\mathrm{star}})
=
\left\{
\begin{array}{l}
0.27\,\Lsun (m_{\mathrm{star}}/\Msun)^{2.6}
~~\mathrm{for}~m_{\mathrm{star}}<0.5\,\Msun, \\
\Lsun (m_{\mathrm{star}}/\Msun)^{4.5}
~~~\mathrm{for}~0.5\,\Msun<m_{\mathrm{star}}<2\,\Msun, \\
1.9\,\Lsun (m_{\mathrm{star}}/\Msun)^{3.6} ~~\mathrm{for}~2\,\Msun<m_{\mathrm{star}}<42\,\Msun, \\
32000\,\Lsun (m_{\mathrm{star}}/\Msun)
~~~~\mathrm{for}~42\,\Msun<m_{\mathrm{star}}<140\,\Msun, 
\end{array}
\right.
\end{align}
\citep{Salaris05}. 
In Eq.~\eqref{eq:flux_luminosity}, we introduced the limitation of the stellar lifetime during an AGN episode. 
The conversion efficiency $\epsilon$ is defined as
\begin{align}\label{eq:epsilondef}
    F_\mathrm{AGN,*} = \frac{1}{2}\epsilon {\dot \Sigma}_* c^2, 
\end{align}
where the ${\dot \Sigma}_*$ is the star formation rate surface density, which is given by 
\begin{align}\label{eq:dotsigma*}
    {\dot \Sigma}_*=\frac{ \sum_i^{N_\mathrm{form}} m_{\mathrm{star},i}}{At_\mathrm{AGN}}.
\end{align}
Combining Eqs.~\eqref{eq:flux_luminosity}, \eqref{eq:epsilondef}, and \eqref{eq:dotsigma*}, and that $m_{\mathrm{star},i}$ are drawn from the initial mass function, $\epsilon$ is expressed as 
\begin{equation}
    \epsilon = \dfrac{\int L(m_{\mathrm{star}}) \min[t_\mathrm{AGN}, t_{\mathrm{star}}(m_{\mathrm{star}})] \, \frac{{\rm d}N}{{\rm d}m} {\rm d}m_{\rm star}}{\int m_{\rm star} c^2 \frac{{\rm d}N}{{\rm d}m}{\rm d}m_{\rm star}}.
\end{equation}
The result lies between $\epsilon=1.5\times 10^{-4}$ and $7.7\times 10^{-4}$ depending on the IMF exponent between $dN/dm \propto m^{-2.35}$ and $m^{-1.7}$.
The limitation due to the stellar and AGN lifetimes reduces $\epsilon$ by a factor of $\sim 4$. 

\section{Scattering in two-dimensions}

\label{sec:scattering_twod}
We utilize the following approximations to evaluate the first and second order diffusion coefficients during gravitational scattering encounters in two dimensions, Eqs.~\eqref{eq:dv_3}--\eqref{eq:dvdv_3} in the main text
\begin{align}
\label{eq:dv_app_12}
D[\Delta v_{\parallel}] 
&= -G\Sigma_{\rm c}\times 
\left\{
\begin{array}{cl}
      2^{1/2}\pi^{3/2}   e^{-x}(x^{1/2} + \frac12 x^{3/2} +\frac14 x^{5/2})
 & ~~~\mathrm{for}~~v<2.3\sigma_{\rm c}\,, \\[2ex]
    2\pi   & ~~~\mathrm{for}~~v>2.3\sigma_{\rm c}\,, 
\end{array}
\right.
\\
\label{eq:dv_app_34}
D[\Delta v_{\perp}^2] 
&= D[\Delta v_{\parallel}^2] =G \Sigma_{\rm c}\times 
\left\{
\begin{array}{cl}
   (2\pi)^{3/2}  \frac{ \sigma_{\rm c} m_{\rm c}}{m+m_{\rm c}}
      e^{-x}
    (\frac12 + x + \frac58 x^2 + \frac14 x^3)
    &\mathrm{for}~~v<2.3\sigma_{\rm c}\,,\\[2ex]
  \pi \frac{\sigma_{\rm c} m_{\rm c}}{m+m_{\rm c}}
    \left[
     \left(4x+1\right)/\sqrt{x} 
    \right]
    &\mathrm{for}~~v>2.3\sigma_{\rm c}\,.
\end{array}
\right.
\end{align}
where we use $I_0(x)\approx 1 + \frac14 x^2$ and $I_1(x) \approx \frac12 x$ for small $x$, $I_0(x)\approx I_1(x) \approx e^x/\sqrt{2\pi x}$ for large $x$. The boundary between the two cases is adjusted to give comparable errors for the two formulae, and the maximum error of the approximated formulae is $\sim10\%$ at the boundary. 
The acceleration is calculated as
\begin{align}
\bm{a}_{\mathrm{WS},k}=
p_{\mathrm{c},k}\left\{D[\Delta v_{\parallel}] \bm{\hat{v}}_{k} + 
\left(\frac{D[\Delta v_{\perp}^2]+D[\Delta v_{\parallel}^2] }{\Delta t} \right)^{1/2}\bm{\hat{n}}_{xy}\right\}
\end{align}
where $\bm{n}_{xy}$ is a unit vector in a random direction in the $xy$ (AGN) plane. In the simulation we adopt an acceleration in unit time $\Delta t$.

\end{document}